\documentclass[reprint,prb,twocolumn,showpacs,superscriptaddress,aps,longbibliography]{revtex4-2}

\usepackage[colorlinks=true,citecolor=blue]{hyperref} 
\usepackage{graphicx}
\usepackage{bm}
\usepackage{amsmath}
\usepackage{amssymb}
\usepackage{comment}

\DeclareGraphicsExtensions{.png,.jpg,.eps}
\usepackage{xcolor}

\newcommand{\beq}{\begin{equation}}
\newcommand{\eeq}{\end{equation}}

\newcommand{\Hscf}{\mathcal{H}_{\text{MF}}}

\begin{document}

\author{Adolfo O. Fumega}
\thanks{These authors contributed equally.}
\affiliation{Department of Applied Physics, Aalto University, 02150 Espoo, Finland}
 
\author{Marcel Niedermeier} 
\thanks{These authors contributed equally.}
\affiliation{Department of Applied Physics, Aalto University, 02150 Espoo, Finland}
 
\author{Jose L. Lado}
\thanks{Corresponding author: jose.lado@aalto.fi}
\affiliation{Department of Applied Physics, Aalto University, 02150 Espoo, Finland}

\title{Correlated states in super-moir\'e materials with a kernel polynomial quantics tensor cross interpolation algorithm}

\begin{abstract}
Super-moir\'e materials represent a novel playground to engineer states of matter beyond the possibilities of conventional moir\'e materials. However, from the computational point of view, understanding correlated matter in these systems requires solving models with several millions of atoms, a formidable task for state-of-the-art methods.  Conventional wavefunction methods for correlated matter scale with a cubic power with the number of sites, a major challenge for super-moir\'e materials.  Here, we introduce a methodology capable of solving correlated states in super-moir\'e materials by combining a kernel polynomial method with a quantics tensor cross interpolation matrix product state algorithm. This strategy leverages a mapping of the super-moir\'e structure to a many-body Hilbert space, that is efficiently sampled with tensor cross interpolation with matrix product states, where individual evaluations are performed with a Chebyshev kernel polynomial algorithm.  We demonstrate this approach with interacting super-moir\'e systems with up to several millions of atoms, showing its ability to capture correlated states in moir\'e-of-moir\'e systems and domain walls between different moir\'e systems.  Our manuscript puts forward a widely applicable methodology to study correlated matter in ultra-long length scales, enabling rationalizing correlated super-moir\'e phenomena. 
\end{abstract}

\date{\today}

\maketitle

\section{Introduction}

Twisted moir\'e materials\cite{Andrei2021} provide a unique playground to engineer artificial states of matter, 
including topological states\cite{Zeng2023,Serlin2020,Sharpe2019,Rickhaus2018,Cai2023}, 
correlated phases\cite{Cao2018corr,Burg2022,Kerelsky2019,Kim2023,Lu2019,Zhao2023}, and superconductivity\cite{Cao2018,Yankowitz2019,Park2021,Oh2021,Park2022}.
Moir\'e patterns arise due to the lattice mismatch between two or more van der Waals layers, leading
to several coexisting length scales. This can naturally occur when two layers of different van der Waals materials with distinct lattice parameters are stacked together, or when layers of the same material are twisted or strained.
Interestingly, when three or more different layers are stacked, the moir\'e pattern itself can feature a long-range modulation, giving rise to a super-moir\'e pattern\cite{Devakul2023,Kapfer2023,Li2022,Turkel2022}. 
Among super-moir\'e patterns, for generic twist angles quasiperiodic patterns emerge,
where recent experiments have demonstrated even more exotic states\cite{Li2024,Ahn2018,Uri2023}, 
including competing correlated mosaic orders and quasiperiodic 
correlated phases, as well as superconductivity\cite{Uri2023}.
From a theoretical point of view, understanding the electronic structure of moir\'e patterns microscopically 
at the atomistic level 
requires treating interacting systems with 
tens of thousands of atoms\cite{PhysRevB.82.121407,PhysRevLett.119.107201,Long2022,PhysRevB.101.060505,PhysRevB.92.075402,PhysRevLett.127.026401,PhysRevB.108.125141,Ramzan2023,PhysRevLett.121.146801,PhysRevB.107.125423}, a task 
that pushes the limits of conventional methods\cite{Carr2020}.
Modeling super-moir\'e patterns requires solving systems with millions of atoms
and incorporating electronic interactions in a selfconsistent manner, 
a task challenging beyond 
current atomistic electronic structure methods.

\begin{figure}[t!]
\centering
\includegraphics[width=\linewidth]{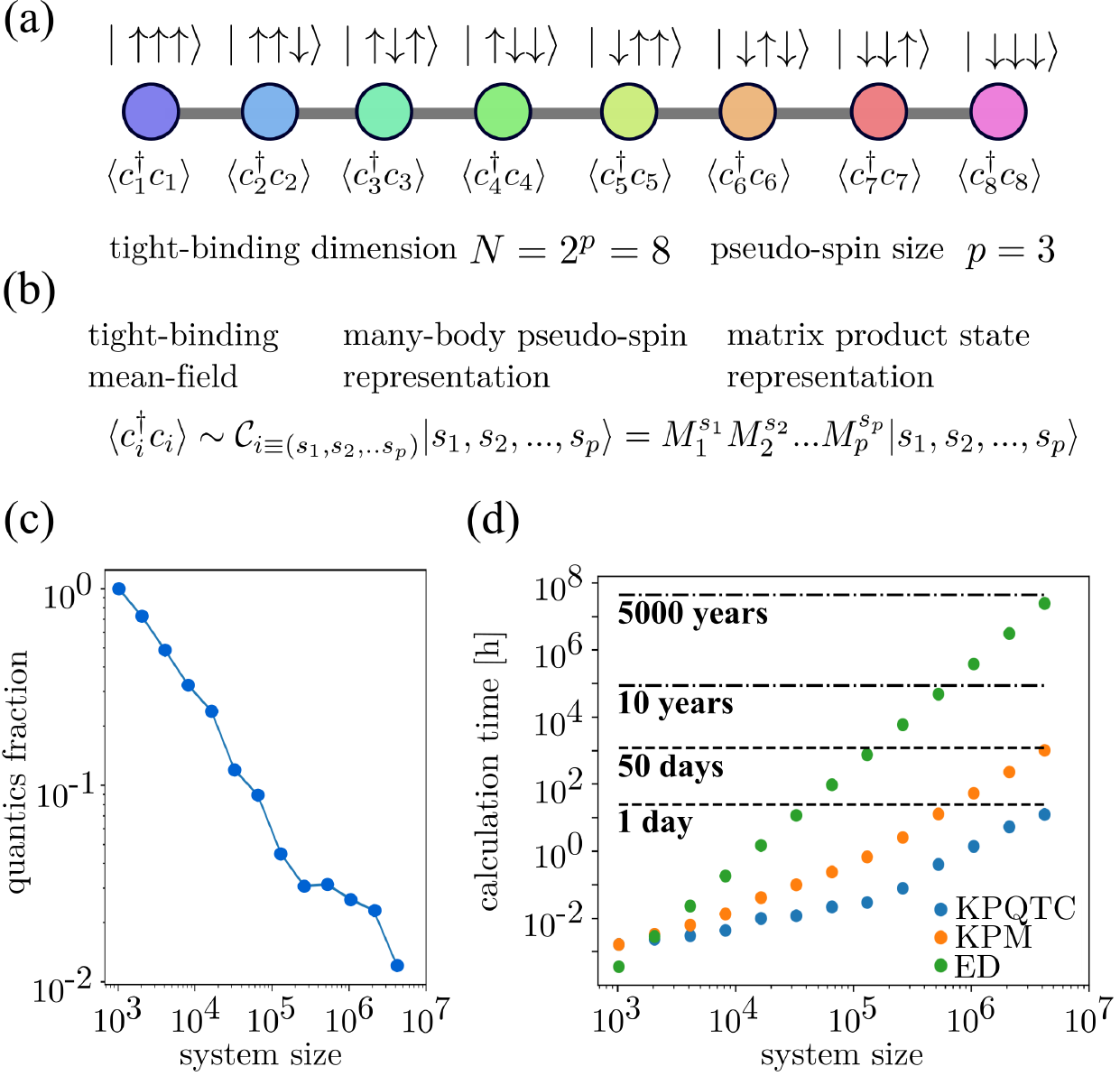}
\caption{
(a) Schematic of the mapping between an interacting super-moir\'e
and an auxiliary many-body spin model. A many-body pseudo spin model
with $p$ spins allows us to encode the mean-field of a 
tight binding model with $2^p$ sites.
(b) The interacting mean-field
is encoded as the amplitude of a
many-body spin model as a matrix product state.
Panel (c) shows the compression of the mean-field
for different system sizes achieved by the algorithm.
Panel (d) shows a comparison between our algorithm (KPQTC), a pure KPM method and
exact diagonalization (ED).
}
\label{fig:schematic}
\end{figure}

The problem of dealing with very high-dimensional objects is well-known in physics, in particular in the case of quantum many-body calculations\cite{Ors2014,Carleo2017,Cerezo2021}.
For a quantum many-body system with $L$ sites, the Hilbert space has a dimension of $2^L$, making quantum many-body calculations extremely challenging even for moderate system sizes\cite{Arovas2022,Qin2022}. A very successful strategy to deal with this problem is to use variational tensor network states to parametrize quantum many-body wavefunctions\cite{PhysRevLett.69.2863,Ors2019,Schollwck2011,PhysRevB.94.165116,RevModPhys.77.259,RevModPhys.93.045003,Ors2014,10.21468/SciPostPhysCodeb.4}.
This approach allows us to solve with nearly arbitrary precision one-dimensional models,
and it has provided the most accurate solutions for paradigmatic two-dimensional
models such as the doped Hubbard and frustrated Heisenberg models\cite{Zheng_Science_2017,10.21468/SciPostPhys.15.6.236,PhysRevX.13.011039}.
In recent years, it has been realized that the power of tensor networks
parametrizing very high-dimensional objects can be applied beyond the realm of quantum many-body physics. This has lead to applications in tensor networks for machine learning\cite{NIPS2016_5314b967,Stoudenmire2018,PhysRevResearch.4.043007,PhysRevX.8.031012,Bradley2020,PhysRevB.99.155131}, quantum computing\cite{PhysRevX.10.041038,PRXQuantum.5.010308,2024arXiv240406048N,PhysRevLett.129.090502,PhysRevLett.128.030501,PhysRevResearch.6.033325,Zhang2023,pastaq}, and parsimonious function representation\cite{PhysRevLett.132.056501,PhysRevX.13.021015,2024arXiv241022975R,2024arXiv240512730S}. In particular, tensor networks can be used to efficiently compress and numerically represent functions that exhibit internal structures. 
This suggests that this methodology may enable addressing
super-moir\'e systems, whose spatially dependent electronic structure
gives rise to phenomena occurring at different length scales.

Here, we demonstrate a technique capable of solving interacting super-moir\'e structures with several millions of atoms. Our method combines a kernel polynomial method with a quantics tensor cross interpolation (KPQTC) with matrix product states. The methodology maps the super-moir\'e structure to a many-body Hilbert space, whose mean-field Hamiltonian is compressed in a matrix product state. This tensor network representation of the mean-field Hamiltonian is learned by applying a tensor cross-interpolation algorithm, which greatly reduces the number of real-space correlators which have to be evaluated with the (expensive) kernel polynomial method (KPM).
With this methodology, we show that interacting electronic models in real space for systems with millions of atoms can be solved,
allowing us to compute interaction-induced symmetry-broken states in those systems while treating interactions in a self-consistent manner.
In particular, we show that this technique allows us to efficiently solve interacting
super-moir\'e models in one and two dimensions, and even in the presence of super-moir\'e domain boundaries.
Our results establish a methodology capable of dealing with interacting problems well beyond
conventional wavefunction methods, providing a technique capable of addressing correlated
physics in super-moir\'e systems from microscopic models.

The paper is organized as follows. We first introduce and describe the KPQTC methodology that we have developed in this work. In the next two sections, we apply the KPQTC to the study of super-moir\'e 1D models and 2D materials. Finally, we provide the discussion and conclusion sections highlighting the results and the broad applicability of the KPQTC method.

\section{Methods}

In the following, we elaborate on the methodology to solve interacting super-moir\'e systems.
We will focus on interacting fermionic models solved at the mean-field level,
where the individual mean-field parameters can be computed with a kernel
polynomial algorithm. The tensor cross interpolation algorithm allows us to reconstruct the whole
mean-field Hamiltonian by iteratively selecting the mean field parameters to be computed.

\subsection{Interactions in super-moir\'e}

The Hamiltonian of a super-moir\'e system in the presence of electronic interactions takes the form
\begin{equation}
H = \sum_{ijs} t_{ij} c^\dagger_{i,s} c_{j,s} + 
\sum_{ijss'} V_{ij} c^\dagger_{i,s} c_{i,s} c^\dagger_{j,s'} c_{j,s'}
\end{equation}
where $t_{ij}$ are the hopping parameters in the system and $V_{ij}$ parametrizes the electronic interactions. To
find the ground state of the Hamiltonian above, 
the interacting term can be decoupled with a mean-field ansatz of the form
$V_{ij} c^\dagger_{i,s} c_{i,s} c^\dagger_{j,s'} c_{j,s'} \approx 
V_{ij} \langle c^\dagger_{i,s} c_{i,s} \rangle c^\dagger_{j,s'} c_{j,s'}
+ ...
$ where $...$ is a shorthand for the remaining Wick contractions.
The previous decoupling gives rise to a mean-field Hamiltonian of the form
\begin{equation}
H_{MF} = \sum_{ijs} t_{ij} c^\dagger_{i,s} c_{j,s} + 
\sum_{ijss'} \chi_{ijss'} c^\dagger_{i,s} c_{j,s'}
\label{eq:hmf}
\end{equation}
with $\chi_{ijss'} \equiv \chi_{ijss'} (V_{ij},|GS\rangle) $  the mean-field parameters, where 
$|GS\rangle$ is the variational many-body ground state 
$H_{MF}|GS \rangle = E_{GS} |GS\rangle$. The variational ground state
is taken as a product state of the form $|GS\rangle = \prod_\alpha \psi^\dagger_\alpha |\Omega\rangle$,
with $\psi^\dagger_\alpha$ variational single-particle states and $|\Omega\rangle$ the vacuum state, 
$\psi_\alpha |\Omega\rangle = 0$. As $H_{MF}$ depends on the $|GS\rangle$, and $|GS\rangle$ depends 
on $H_{MF}$, the previous problem can be solved with a conventional iterative self-consistent algorithm.
From the computational point of view, the most demanding step consists of computing the
mean-field parameters $\chi_{ijss'}$ at each step of the self-consistent procedure.
In particular, for a super-moir\'e system with $N$ sites, conventional algorithms
based on matrix diagonalization
as implemented in electronic structure codes scale as $N^3$,
whereas a full Chebyshev expansion reduces the computational cost to $N^2$. This sets the maximum number of sites computable 
with typical computational resources with diagonalization
on the order of $N = 10^4$ atoms, 
and with Chebyshev expansion in $N = 10^6$ atoms.
Modeling interacting states in super-moir\'e materials requires solving problems
with several millions of atoms, well above the
capabilities of the previous two methods.

\subsection{Interactions with a kernel polynomial expansion}

We now address how a Chebyshev kernel polynomial expansion can be used to solve
self-consistent mean-field systems.
In each iteration of the self-consistent procedure, the variational
parameters of the mean-field Hamiltonian $\chi_{ijss'}$ can be calculated
once the correlators $\langle c^\dagger_{i,s} c_{j,s'} \rangle \equiv \langle GS | c^\dagger_{i,s} c_{j,s'} | GS\rangle$ 
are known.
These correlators can be computed as
\begin{equation}
\langle c^\dagger_{i,s} c_{j,s'} \rangle 
= 
\int_{-\infty}^{\epsilon_F}
\langle \Omega | c_{j,s'} \delta(\omega - \Hscf) c^\dagger_{i,s} |\Omega \rangle
d\omega
\end{equation}
where $\epsilon_F$ is the single particle Fermi energy and $\delta(\omega - \Hscf)$ is the Dirac-delta function operator.
Taking $ \langle c^\dagger_{i,s} c_{j,s'} \rangle = \int_{-\infty}^{\epsilon_F} g_{ijss'}(\omega) d\omega$
with $g_{ijss'}(\omega) = \langle \Omega | c_{j,s'} \delta(\omega - \Hscf) c^\dagger_{i,s} |\Omega \rangle$,
where $g_{ijss'}(\omega)$ is the dynamical correlator between sites $i$ and $j$ with spin $s$ and $s'$,
and $| \Omega \rangle$ the empty many-body state.
For the sake of concreteness, we now take that the Hamiltonian $\Hscf$ has a single particle spectrum bounded in the
interval $(-1,1)$.
The function $g_{ijss'}(\omega)$ can be efficiently computed with a Chebyshev kernel polynomial expansion\cite{RevModPhys.78.275} of the
form $g_{ijss'}(\omega)= \frac{1}{\pi\sqrt{1-\omega^2}}[\gamma_0T_0(\omega) + 2\sum_{n>0} \gamma_n T_n(\omega)]$,
where $T_n(\omega)$ are the Chebyshev polynomials and $\gamma_n$ are the coefficients of the expansion.
Thanks to the Chebyshev recursion relation, the moments of the expansion can be computed as
$\gamma_n = \langle \Omega | c_{j,s'} | v_n \rangle$, where $| v_{n+1} \rangle = 2 \Hscf |v_n\rangle - |v_{n-1}\rangle$,
with $|v_1\rangle = \Hscf |v_0 \rangle $ and $|v_0\rangle = c^\dagger_{i,s}|\Omega\rangle $.
The calculation of a single correlator $\langle c^\dagger_{i,s} c_{j,s'} \rangle$ scales linearly with the number of atoms $N$.
To compute the self-consistent Hamiltonian, the number of correlators that have to be evaluated is proportional
to the number of atoms $N$, so that a Chebyshev kernel polynomial
expansion allows us to perform self-consistent calculations\cite{Nagai2012} with a scaling $N^2$, in contrast the scaling $N^3$
for exact diagonalization.

\subsection{Quantics tensor-network representation of the mean-field Hamiltonian}

The most expensive part of the algorithm is to evaluate the variational parameters $\chi_{ijss'}$ of the mean-field Hamiltonian. The $\chi_{ijss'}$'s depend on the correlators $\langle c^\dagger_{i,s} c_{j,s'} \rangle = \langle GS | c^\dagger_{i,s} c_{j,s'} | GS\rangle$, and
where each $\langle c^\dagger_{i,s} c_{j,s'}  \rangle$ has to be found by performing a full run of the KPM algorithm. Therefore, to perform the full mean-field calculation, a large number of KPM's has to be executed. In the following, we will apply the tensor cross interpolation algorithm to construct an approximation of the function $\chi_{ijss'} (\{\langle c^\dagger_{i,s} c_{j,s'}  \rangle \})$ as a matrix product state:
\begin{equation}
    \chi_{ijss'} \approx M_1^{s_1} M_2^{s_2} ... M_p^{s_p}.
\end{equation}
The main benefit of this method is, that it allows us to construct a high-fidelity approximation of $\chi_{ijss'}$, while only requiring an exact evaluation for a very small number of arguments $\langle c^\dagger_{i,s} c_{j,s'}  \rangle $. All other correlators, that are not called during the construction of $ {\chi}_{ijss'}$, do not need to be calculated in the first place, which greatly reduces the number of individual
KPM runs, the most expensive part of the algorithm. A schematic of the mapping used by the KPQTC is shown in Fig.~\ref{fig:schematic}.

\begin{figure}[t!]
\centering
\includegraphics[width=\linewidth]{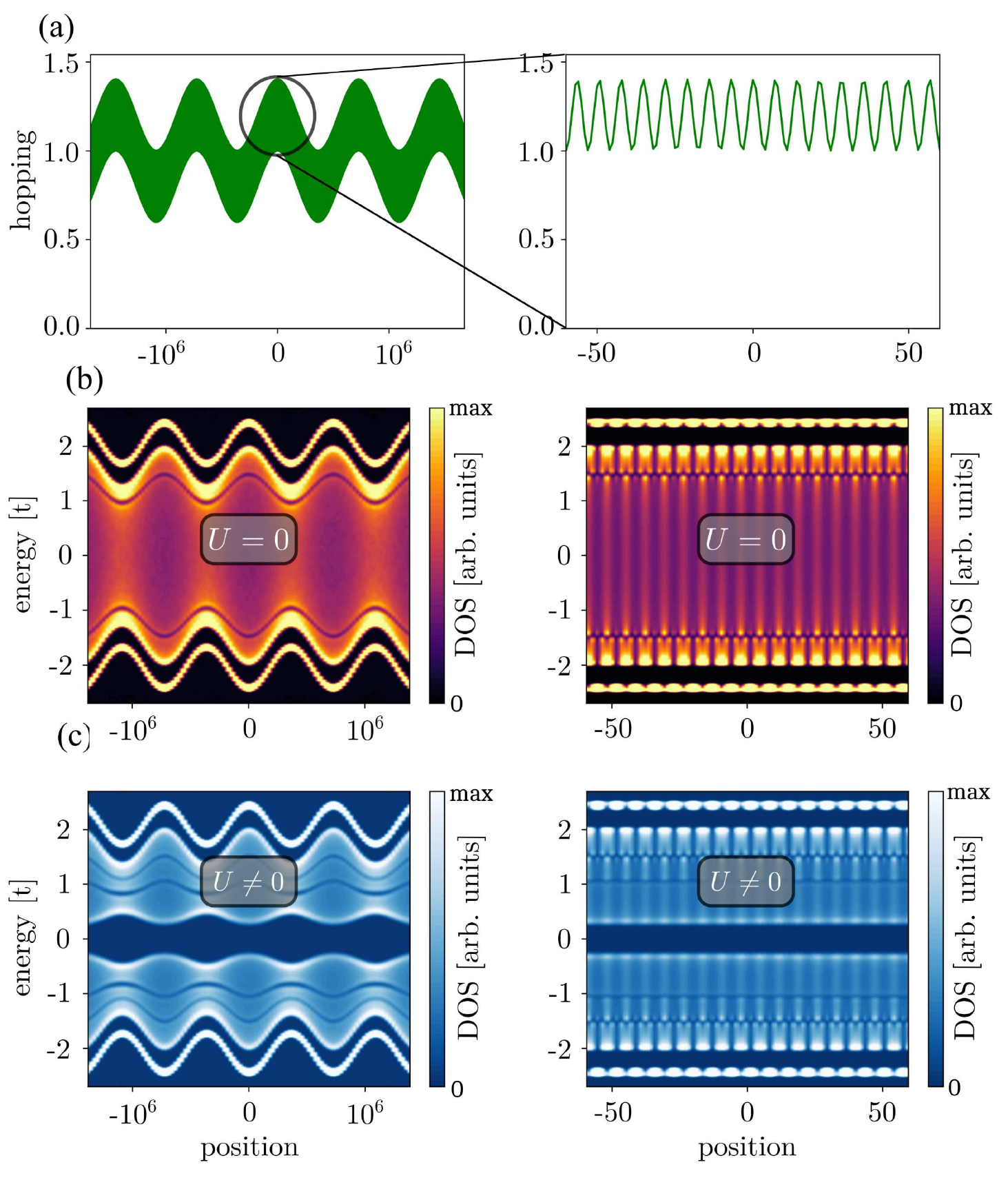}
\caption{
	\textbf{Correlated super-moir\'e for $L=2^{22}$ sites} (above 4 million sites).
Panel (a) shows the super-moir\'e modulation of the Hamiltonian,
featuring a moir\'e pattern at short scales, and another at long length scales.
Panel (b) shows the density of states in the system at long and short length
scales, showing how its spectral properties are modulated.
Panel (c) shows the spectral function of the interacting
super-lattice solved selfconsistently with KPQTC.
It is observed that the interaction-induced gap in the spectral function
is modulated in the long length scale, whereas its intensity
is modulated at the short length scale.
}
\label{fig:supermoire1d}
\end{figure}

The kernel polynomial tensor cross interpolation method relies on exploiting the 
natural structure and length scales of the mean-field Hamiltonian in a super-moir\'e system.
The large number of components of $\chi_{ijss'} (\{\langle c^\dagger_{i,s} c_{j,s'} \rangle \})$ 
can be reformulated as a rank-$R$ tensor $\chi^{\sigma_1 ... \sigma_R}$, with
$R \propto \log (\text{number of correlators})$. This tensor can be
re-expressed as a much cheaper matrix product state, using the tensor cross
interpolation algorithm, which learns a quasi-optimal approximation of
$\chi^{\sigma_1 ... \sigma_R}$ by evaluating it exactly for only a small subset
of its entries\cite{PhysRevX.12.041018,2024arXiv240702454N,PhysRevLett.132.056501,PhysRevB.107.245135,PhysRevB.110.035124,2023arXiv230803508J,PhysRevB.109.165135}. Therefore, only a small subset of the full list $[\langle
c^\dagger_1 c_1 \rangle, ..., \langle c^\dagger_{2^p} c_{2^p} \rangle]$ of
real-space correlators has to be calculated in practice.
Furthermore, the architecture of the underlying matrix product state and the update strategy is dynamically optimized during the self-consistent loop,
where at each iteration we optimize for the strategy that requires the least amount of evaluations of the mean-field
of the previous iteration. The dynamically optimized parameters include the matrix product state bond dimension, the number
of orbitals for which independent matrix product states are created, the initial pivot,
the choice of a rook or accumulative optimization method, and the number and location of global pivots. 
The convergence of the MPS representation of the mean-field Hamiltonian is controlled
by two main parameters: the maximum bond dimension and a per-tensor singular value decomposition-compression
error. Our methodology sets a value of the compression error and lets the algorithm
make evaluations until the matrix product state constructed has enough accuracy. This is thus an iterative procedure
that is only halted when the mean-field is accurate enough. The threshold we set in the matrix product state
construction is taken as one order of magnitude below the target error of the selfconsistency loop, guaranteeing that
the mean-field features high enough accuracy.

\section{Interactions in super-moir\'e 1D models}

We now use the KPQTC method to address one-dimensional models. First, we will focus on a system that features two incommensurate moir\'e patterns, also incommensurate with the lattice. Second, we will address a system featuring an interface between two moir\'e patterns, to show that the methodology is able to deal with inhomogeneous problems.
While 1D models are not the most relevant use case for van der Waals materials, they provide an excellent testing ground for the KPQTC algorithm to show that calculations with millions of atoms can be performed.

\subsection{Incommensurate 1D super-moir\'e}
We first consider a model featuring two moir\'e patterns, incommensurate with each other and with the
original lattice. Super-moir\'e models can be realized in artificial platforms including engineering optical resonators and cold atom systems\cite{Wang2019,Fu2020}. Furthermore, within van der Waals materials, 
the electronic properties of multi-walled nanotubes
are effectively described by a one-dimensional super-moir\'e Hamiltonian\cite{2024arXiv240514967C}.
We take a Hamiltonian of the form

\begin{equation}
\mathcal{H} 
= \sum_{ij,s} t_{ij} c^\dagger_{i,s} c_{j,s} + U
\sum_{i} 
\left [ c^\dagger_{i,\uparrow} c_{i,\uparrow} - \frac{1}{2} \right ]
\left [ c^\dagger_{i,\downarrow} c_{i,\downarrow} - \frac{1}{2} \right ],
\end{equation}
where the hopping is modulated
by two incommensurate moir\'e patterns as
\begin{equation}
t_{n,n+1} = t_0 + t_1 \cos{(k_1 X_{n,n+1})} + t_2 \cos{(k_2 X_{n,n+1})}, 
\end{equation}
with 
$X_{n,n+1} = (x_n + x_{n+1})/2$,
$k_1$ and $k_2$ the wavevectors of the two moir\'es
and $x_n$ is the location of site $n$.
We take 
$k_1 = 2 \pi /5\sqrt{2}$
and 
$k_2 = 2\pi 5/2^{p-1}\sqrt3$,
which leads to two incommensurate modulations,
also incommensurate with the lattice. 
The modulation in the local hopping gives rise to a different competition between electronic interactions and kinetic energy
in different regions in the system.
We solve the model with the QTCI for a system with $L=2^{22}$ sites, approximately
4 million atoms.
The results of our calculation are shown in Fig.~\ref{fig:supermoire1d},
where we show the moir\'e modulation of the Hamiltonian at the two length scales,
together with the non-interacting (Fig.~\ref{fig:supermoire1d}b) and interacting (Fig.~\ref{fig:supermoire1d}c)
local spectral function.
In particular, we observe that interactions give rise to a spatially dependent gap opening, as shown by comparing the density of states
in the absence (Fig.~\ref{fig:supermoire1d}b) and presence (Fig.~\ref{fig:supermoire1d}c) of electronic interactions.
Interestingly, on the largest length scale, the gap opening fully follows the moir\'e
length scale, whereas, on the shorter length scale, the spectrum shows a modulation of the spectral
weight. This stems from the fact that at the smaller moir\'e length scale, the correlation length
associated with the electronic order is of the same order as the moir\'e length scale, which gives rise to
the lowest electronic excitation to extend in the whole moir\'e. In contrast, for the biggest moir\'e length scale,
the correlation length associated with the order is much smaller than the length scale of the moir\'e, which leads to the spectral gap being modulated exactly following the moir\'e.

The performance of the KPQTC method\cite{pyqula,qtcipy} as compared to the current techniques
is highlighted in Fig.~\ref{fig:schematic}. Figure \ref{fig:schematic}c shows
the compression of the mean-field components introduced by the KPQTC. The
fraction of real-space correlators required as compared to the pure KPM is
plotted as a function of the system size. We can clearly observe the advantage
introduced by the KPQTC method for large systems above $10^5$ sites, 
where the fraction of
correlators required decreases as the system size grows.
The convergence time of the self-consistent mean-field calculation for the
KPQTC method is reported in Fig.~\ref{fig:schematic}d. We
show the estimated
calculation times for the pure KPM and the ED methods are shown for comparison.
We can observe that our KPQTC allows solving systems of millions of atoms in
less than one day, while the traditional KPM requires around 50 days.
KPM and KPQTC become faster than exact diagonalization once the system size goes above $10^3$.
sites, and KPQTC becomes substantially faster than KPM for systems above $10^5$
sites.
Therefore, these results demonstrate the advantage of the KPQTC method to study
correlated states in super-moir\'e materials composed of millions of atoms.

\begin{figure}[t!]
\centering
\includegraphics[width=\columnwidth]{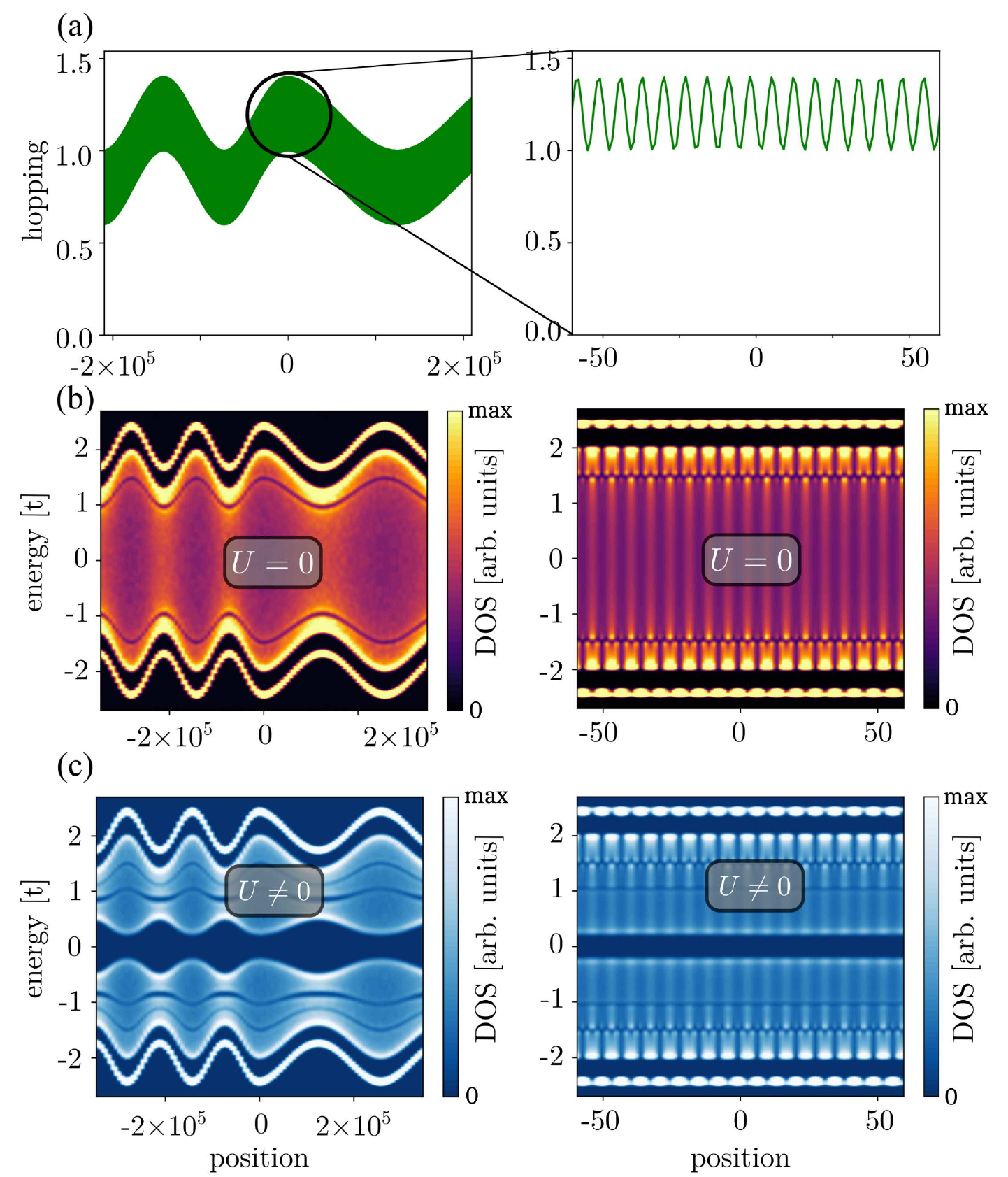}
\caption{
\textbf{
	Correlated super-moir\'e domain wall for $L=2^{20}$ sites} (above 1 million sites).
Panel (a) shows the super-moir\'e 
modulation featuring a domain wall,
featuring different super-moir\'e modulations
at the left and right boundary.
Panel (b) shows the density of states in the system at long and short length
scales, showing how its spectral properties are modulated
according to each domain.
Panel (c) shows the spectral function of the interacting
system solved selfconsistently with KPQTC.
The interaction-induced gap in the spectral function
is modulated in the long length scale following
the super-moir\'e at each domain, 
whereas at the domain wall its intensity
is modulated following the local moir\'e.
}
\label{fig:domain1d}
\end{figure}

\subsection{Super-moir\'e domain wall in 1D}

An alternative situation that appears in super-moir\'e systems is a domain wall
between different length scales. This emerges in situations where structural relaxations 
strongly prefer specific stacking or moir\'e length scales, a phenomenon that gives rise to
domain walls appearing between different regions.
We will here consider a system where the shortest moir\'e length scale
is the same in the whole system, whereas the biggest one features
two domains.
We take the hopping $t_{n,n+1}$ to be modulated in space by two wavelengths $k_1$ and $\bar k_2$, one of them spatially dependent

\begin{equation}
t_{n,n+1} = t_0 + t_1 \cos{(k_1 X_{n,n+1})} + t_2 \cos{(\bar k_2(X_{n,n+1}) X_{n,n+1})}, 
\end{equation}
with the wavelength of the modulation featuring a domain wall:
\begin{equation}
\bar k_2(X_{n,n+1}) = \bar k_2(1 + \delta \tanh{(X_{n,n+1}/W)}).
\end{equation}

Here, $W$ describes the width of the domain wall and $\delta$ parametrizes the
mismatch between the length scales of the moir\'e modulations in the two domains,
that in the asymptotic limit are $k_2(1-\delta)$ and $k_2(1+\delta)$. 
In our calculations, we have used $\delta=0.2$, $W=2^p/40$, $k_2 = 2\pi 5/2^{p-1}\sqrt3$ and $k_1 = 2 \pi /5\sqrt{2}$.
This Hamiltonian therefore describes a system breaking translational symmetry regardless of the moir\'e length.
In Fig.~\ref{fig:domain1d}, we show the solution of the interacting model for $2^{21}$ sites, or approximately 2 million atoms. The impact of the 
electronic interactions can be observed
by comparing the non-interacting (Fig.~\ref{fig:domain1d}b) and interacting (Fig.~\ref{fig:domain1d}c) spectral functions.
The gap in the spectral function follows the moir\'e
pattern both in the left and right domains at the longest length scale. 
As in the case studied previously, this phenomenology stems from the fact that
both moir\'e length scales are much longer than the localization
length associated with the correlated state, which leads to a spectral gap
reflecting the large-scale moir\'e modulation. At shorter scales,
the spectrum is modulated according to the moir\'e, but leading to a spectral gap that
is uniform due to the comparable correlation and moir\'e length scales.

\section{Interactions in super-moir\'e 2D materials}

We now consider interacting super-moir\'e materials in two dimensions,
which is the most physically relevant scenario for van der Waals materials.
We will focus on the Hamiltonian of a purely two-dimensional system of super-moir\'e graphene
monolayer. moir\'e patterns in monolayer graphene can emerge from
periodically modulated strain from buckling\cite{Mao2020,Manesco2021,PhysRevLett.131.096401,PhysRevLett.130.216401,PhysRevLett.128.176406}, 
or from a moir\'e pattern with boron nitride.
For the sake of concreteness, we will focus on the case of periodically buckled monolayer
graphene, which has been demonstrated to lead to a variety of correlated states.
The Hamiltonian of the system thus takes the form
\begin{equation}
\mathcal{H} 
= \sum_{\langle ij \rangle,s} t_{ij} c^\dagger_{i,s} c_{j,s} + 
U \sum_{i} 
\left [ c^\dagger_{i,\uparrow} c_{i,\uparrow} - \frac{1}{2} \right ]
\left [ c^\dagger_{i,\downarrow} c_{i,\downarrow} - \frac{1}{2} \right ],
\end{equation}
where the sites $ij$ form a honeycomb lattice and $\langle ij \rangle$ runs over the first neighbors in the graphene
honeycomb lattice. In the presence of buckling, the hopping parameters $t_{ij}$ of graphene are modified as\cite{Manesco2021,PhysRevLett.128.176406,PhysRevLett.130.216401}

\begin{equation}
t_{ij} = t^0_{ij}(1 + \delta \sin{(\Omega \mathbf u_{ij} \cdot \mathbf R_{ij})} ),
\end{equation}
where
$\mathbf R_{ij} = (\mathbf r_i + \mathbf r_j)/2$ is the location of the bond,
$u_{ij}$ is the vector linking sites $i$ and $j$, and $\Omega$ parametrizes the frequency of the buckling.
The previous modulation gives rise to a direction-dependent modulation
of the hopping of wavevector $\Omega$.
The previous buckling modulation gives rise to pseudo-Landau levels due to the emergence
of a non-uniform artificial gauge field. The pseudo-Landau levels get localized in an emergent honeycomb
lattice structure due to the modulation of the gauge field. In the presence of interactions, those
localized modes give rise to a correlated state.

We will study two cases where super-moir\'e physics emerges in this system. First, we will
consider the case where two different bucklings at different length scales emerge, with relative frequencies
$\Omega_M$ and $\Omega_{SM}$ and strengths $\delta_M$ and $\delta_{SM}$. Afterwards, we will consider
an interface between two buckling modulations.

\begin{figure*}[t!]
\centering
\includegraphics[width=\textwidth]{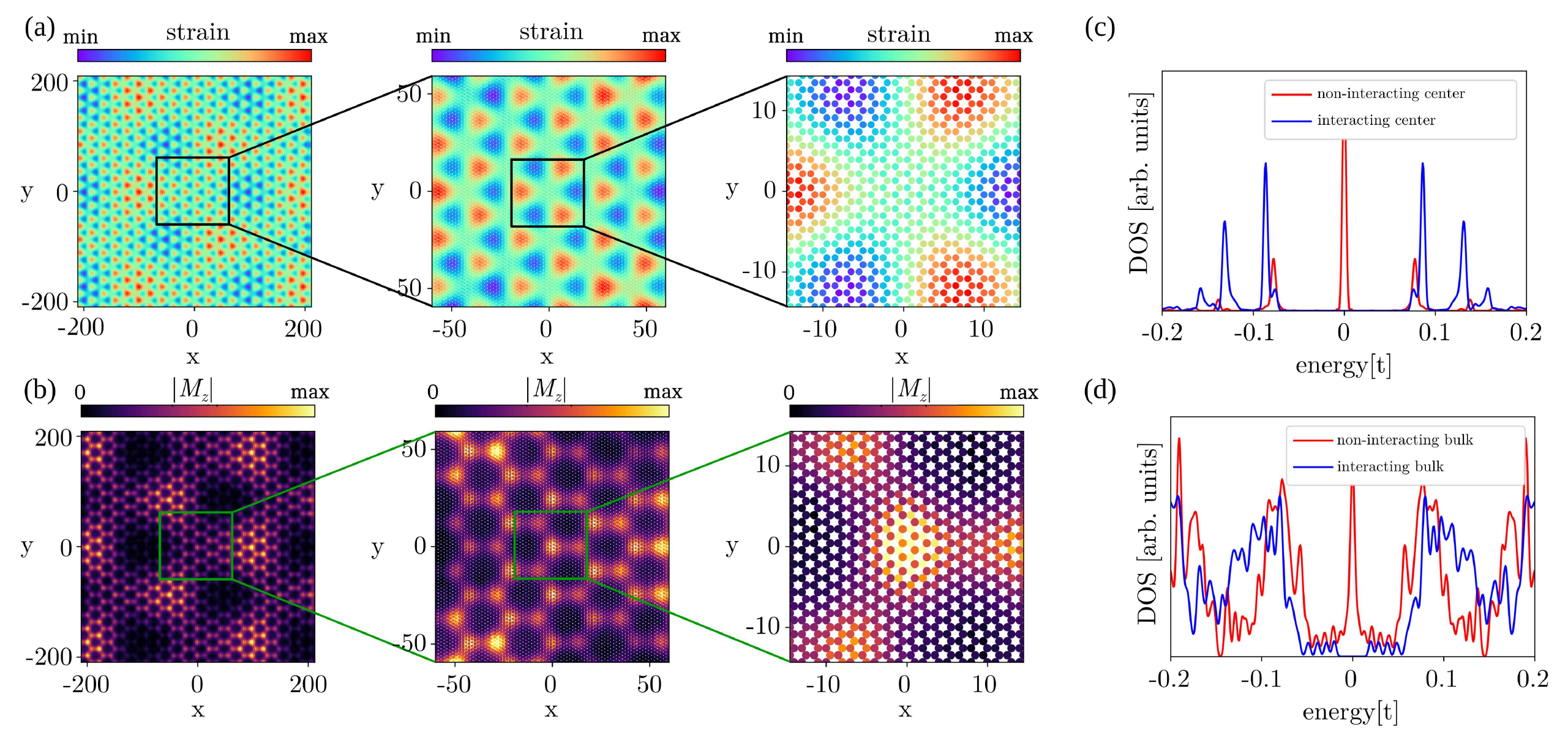}
\caption{
\textbf{2D super-moir\'e:}  Panel (a) shows the profile of strain in the super-moir\'e system,
showing a modulation of the strain both at a large length scale and at a shorter one.
Panel (b) shows the self-consistent magnetization computed with KPQTC algorithm. We observe
that the order parameter is modulated both at the super-moir\'e and moir\'e length scales, giving rise
to localized regions of space with electronic order. Panels (c,d) show the spectral function in the absence and
presence of electronic interactions, both at the center of the super-moir\'e pattern (c) and averaged over the super-moir\'e pattern (d).
We observe that interactions give rise to a gap in the spectral function, associated with a correlated insulating
state in the buckled super-moir\'e.
Self-consistent calculations are performed in systems with more than 200000 sites ($N=2^{18}$).
}
\label{fig:supermoire2d}
\end{figure*}

\subsection{Correlations in a 2D super-moir\'e}

We start first with the moir\'e of moir\'e buckling. In this scenario, the hoppings of the graphene monolayer are modulated as
\begin{equation}
\begin{split}
    t_{ij} = t^0_{ij}
&(1 + \delta_M \sin{(\Omega_M \mathbf u_{ij} \cdot \mathbf R_{ij})} ) \\
&(1 + \delta_{SM} \sin{(\Omega_{SM} \mathbf u_{ij} \cdot \mathbf R_{ij})} ),
\end{split}
\end{equation}
where $\delta_M$ and $\delta_{SM}$ correspond to the strength of the buckling at the moir\'e and super-moir\'e length scales,
and $\Omega_M$, $\Omega_{SM}$ are the wavevectors of the moir\'e and super-moir\'e buckling. 
We take $\Omega_M = 2\pi\sqrt{2}/30$, $\Omega_{SM} = \Omega_M/7$, $U=2t$, $\delta_M = 0.2$ and $\delta_{SM}=0.1$.
In Fig.~\ref{fig:supermoire2d}a, we show 
the strength of the local strain field at each point in space, defined as the average value of the neighboring
hopping $s(\mathbf r_i) \sim \sum_j t_{ij} $. We solve a system with $2^{18}$, or approximately $200000$, sites, where the self-consistent
magnetization is shown in Fig.~\ref{fig:supermoire2d}b. We observe that the symmetry broken order clearly follows both moir\'e patterns,
giving rise to an emergent honeycomb lattice modulated at the super-moir\'e length scale of magnetic order.
We can now compare the spectral function of the system both with and without interactions, as shown in Fig.~\ref{fig:supermoire2d}cd.
In particular, in Fig.~\ref{fig:supermoire2d}c we show the local spectral function right at the center of the super-moir\'e pattern.
In the absence of interactions, a zero energy peak emerges, which
in the presence of interactions gives rise to a gap (Fig.~\ref{fig:supermoire2d}c). Such zero energy mode is precisely the one responsible for the
spatially localized magnetic order as shown in the shortest length scale of Fig.~\ref{fig:supermoire2d}b.
Figure \ref{fig:supermoire2d}d shows the comparison of the spectral function computed in the whole length scale of the
super-moir\'e pattern. We observe that in the absence of correlations, the system features a gapless electron gas with a van Hove singularity at charge neutrality, whereas in the presence of interactions a full band gap opens up (Fig.~\ref{fig:supermoire2d}d). Such a van Hove singularity corresponds to the localized modes in specific regions of the moir\'e pattern, which
in the interacting regime give rise to a correlated insulator in the full system.

\begin{figure*}[t!]
\centering
\includegraphics[width=\textwidth]{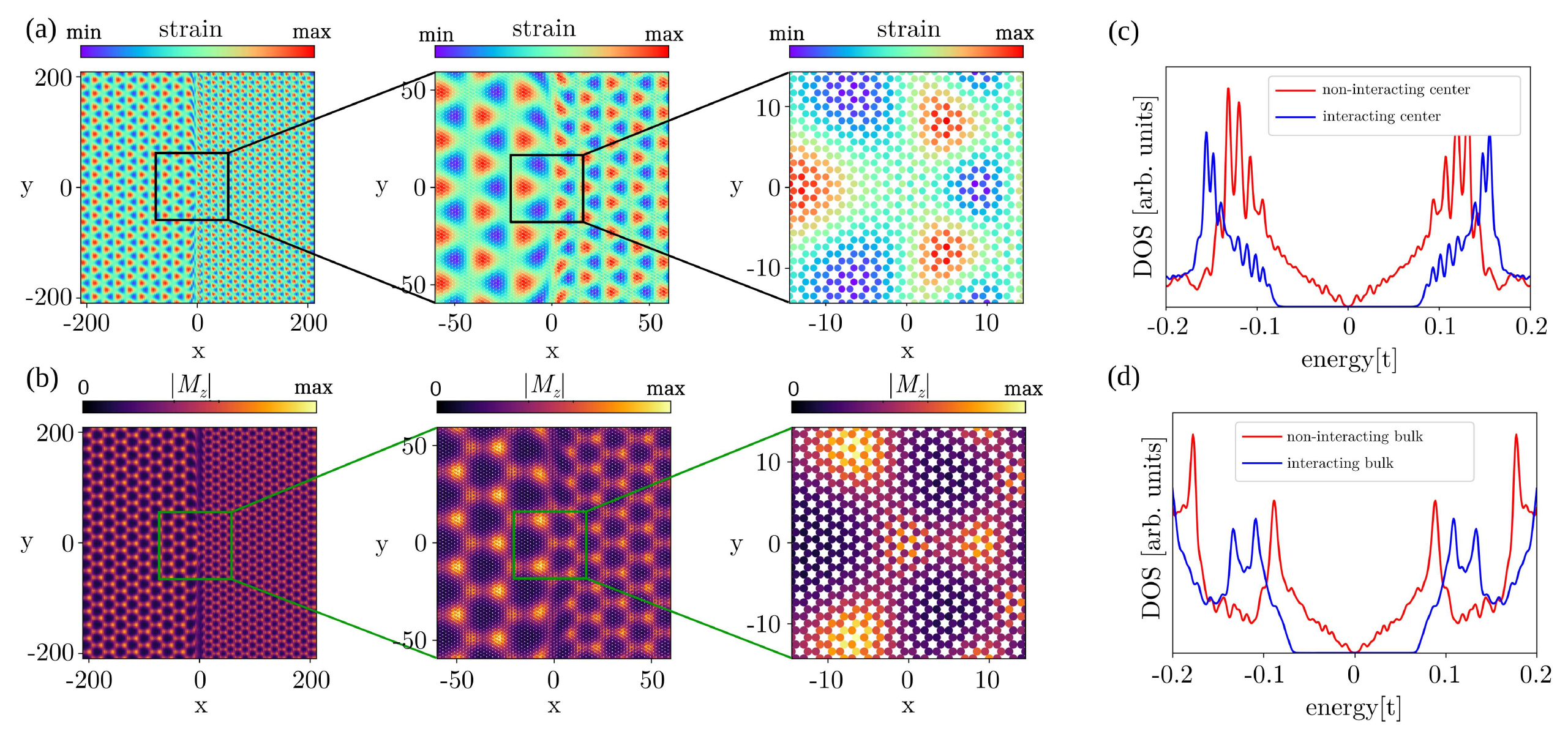}
\caption{
\textbf{2D moir\'e domain wall:} Panel (a) shows the profile of strain in a domain wall
in the moir\'e system, featuring a different
moir\'e on the left and on the right.
Panel (b) shows the self-consistent magnetization computed with KPQTC algorithm. The order parameter follows the moir\'e at the two domains,
while also featuring a modulation at the domain wall. Panels (c,d) show the spectral function in the absence and
presence of electronic interactions, both at the center of the domain wall (c), and averaged over a length
scale around the domain wall on the order
of the super-moir\'e length scale (d).
Interactions give rise to a gap in the spectral function
at the domain wall, that coexists with the correlated state in the two domains.
Self-consistent calculations are performed in systems with more than 200000 sites ($N=2^{18}$).
}
\label{fig:domain2d}
\end{figure*}

\subsection{Correlations between moir\'e domains}
In the presence of two moir\'e patterns stemming from several twisted two-dimensional materials,
structural relaxations can lead to relatively uniform moir\'e regions separated from each other by domain walls\cite{Turkel2022,PhysRevX.13.041007,Craig2024,PhysRevB.107.125413}.
This phenomenology observed experimentally requires that computational methods be capable of dealing with geometries lacking translational symmetry
and hosting a very large amount of atoms. We now show that the KPQTC methodology
is able to capture correlated states emerging in the presence of moir\'e domains.
For the sake of concreteness, we will focus on studying a domain wall between two different
bucklings in graphene. In this scenario, the hopping of the graphene system will be modulated as

\begin{equation}
t_{ij} = t^0_{ij}(1 + \delta \sin{(\Omega (\mathbf R_{ij}) \mathbf u_{ij} \cdot \mathbf R_{ij})} ),
\end{equation}
where now the buckling wavevector changes in space and features a domain wall of the form

\begin{equation}
\Omega (\mathbf R_{ij}) = \Omega_0 (1 + \gamma \tanh{(X_{ij}/W)}).
\end{equation}
Here $W$ parametrizes the width of the moir\'e domain wall and $\gamma$ controls the difference between the two moir\'e buckling frequencies,
which asymptotically inside the domain become
$ \Omega_0 (1-\gamma)$ and $\Omega_0(1+\gamma)$. 
We take $\Omega_0= 2\pi\sqrt{2}/23$, $\gamma=0.3$, $U=2t$, and $\delta = 0.2$.

The spatial modulation of the strain is shown in Fig.~\ref{fig:domain2d}(a) at different length scales. At the right and left domains the
strain profile corresponds to a buckled monolayer, yet featuring different modulation length scales. When introducing electronic interactions (Fig.~\ref{fig:domain2d}b), a moir\'e correlated state emerges, featuring an order parameter following the modulation of the strain in the
left and right domains. Interestingly, certain regions develop a correlated order at the interface between the two domains, whereas
in other regions of the domain wall, the correlated order is quenched (Fig.~\ref{fig:domain2d}b). This phenomenology is due to the mismatch in the non-uniform strain between the two domains in different regions of the domain wall.
With the correlated state, the spectral function of the system can be computed both at the
center of the domain wall (Fig.~\ref{fig:domain2d}c), or averaged over a large length scale (Fig.~\ref{fig:domain2d}d).
In the absence of interactions, the system remains gapless, featuring a linear dispersion close to
the domain wall (Fig.~\ref{fig:domain2d}c). In the presence of interactions, the whole system develops a correlated insulating state,
including the domain wall, thus giving rise to a full spectral gap in the system (Fig.~\ref{fig:domain2d}d).

\section{Discussion}

We have shown that the kernel polynomial tensor cross algorithm enables us to solve interacting super-moir\'e
models with several millions of atoms. While our calculations focus on Hubbard models,
our methodology can be used to compress and infer a generic mean-field Hamiltonian,
including in the presence of non-local interactions. In particular, beyond the magnetic orders we have considered
in our manuscript, it is worth noting that a similar strategy can be used to capture charge order,
superconducting or valley coherent states\cite{Kim2023}, which are especially relevant for twisted graphene multilayers.
The energy resolution of our algorithm is set by the number of Chebyshev polynomials. Symmetry-broken states with very small energy scales, such as superconducting order, require a higher number of Chebyshev polynomials.
A key step is that the models we considered have a certain structure, both in the single-particle Hamiltonian
and its mean-field. The tensor cross interpolation algorithm relies on the existence of a compressibility in the mean-field Hamiltonian.

While this is true for generic twisted van der Waals materials, even in the presence of domain walls, systems with strong disorder may represent a challenge for our algorithm.
This stems from the fact that in the presence of strong disorder, the compressibility of the mean-field is lost due to randomness, thus substantially increasing the required number of evaluations to reconstruct the mean-field.
This is however not a limitation in the presence of a finite dilute amount of impurities, and thus
the KPQTC methodology would allow us to tackle super-moir\'e systems with a small amount of impurities.
It is important to note that
the choice of pivots and update strategy of tensors can strongly influence the required number of evaluations
required to converge the tensor network. As work in quantics tensor cross interpolation methods is progressing rapidly, we foresee that further improvements in these algorithms will enable us to address more complex and bigger interacting super-moir\'e systems.

In our manuscript, we have used a matrix product state representation of the mean-field terms, yet the Hamiltonian is still stored as a sparse matrix. With this algorithm, the maximum system size is 
determined by the required kernel polynomial expansion with vector for $N$ sites, 
which requires storing vectors of size $N$, where memory becomes the bottleneck. 
A potential step in the future is to store the Hamiltonian itself as a matrix product operator, such that
the kernel polynomial expansion is done directly with tensor networks. This would allow us to reach system sizes beyond $N=10^8$. It is also worth noting that since the evaluations of each individual correlator
with the kernel polynomial algorithm are fully independent, our approach can be massively parallelized to thousands
of cores almost with linear scaling\cite{Kreutzer2015}. This should be contrasted with electronic structure methods based on diagonalization,
where diagonalization tasks show a worse than linear scaling with parallelization.
While our calculations have focused on correlated states in tight binding models, a similar approach to the
one presented here can be implemented in conventional Hartree-Fock quantum chemistry codes
and density functional theory, in particular, those based on describing the electronic density
and Hamiltonian on real space grids\cite{Kronik2006,Prentice2020,Ivanov2021,Mortensen2024,Soler2002}.

While our demonstration above focuses on a super-moir\'e pattern in a graphene monolayer,
our methodology can be readily extended to twisted
multilayers. Experiments in moir\'e systems are rapidly developing\cite{Park_2024_2D,Son20202D},
with notable demonstrations including super-moir\'e
twisted multilayer graphene and other van der Waals heterostructures \cite{Devakul2023,Kapfer2023,Li2022,Turkel2022}. 
In particular, twisted graphene trilayers provide an exceptional platform to observe the effects of super-moir\'e patterns
in the case where different twisting angles are taken between the top and bottom layers\cite{Uri2023}.
Spectroscopic measurements with scanning tunneling microscopy
have been performed in a variety of correlated twisted graphene
heterostructures\cite{Nuckolls2023,Nuckolls2024,Oh2021,Kim2022},
and it is expected that future experiments will enable probing the unique physics
of super-moir\'e patterns in real-space.
Scanning probe experiments are particularly 
promising for directly validating symmetry broken states,
as a comparison of the local reconstruction of the spectral function
in different regions of the super-moir\'e provides a highly non-trivial test
for a correlated state\cite{Oh2021,Kim2022}.
This can be especially interesting to characterize incommensurate Kekule symmetry broken states in twisted graphene
multilayers\cite{Kim2023}.
We expect that our methodology will enable understanding of a variety of symmetry-broken
correlated states in super-moir\'e systems,
whose system sizes are well beyond the capabilities of conventional methods.
We finally note that our 
methodology has no fundamental limitation to considering other systems that can be captured with mean-field theory,
and it is not limited to van der Waals materials.
In particular, our method can be used for generic symmetry-broken states
that have very long length scales, such as incommensurate charge density waves
in high temperature superconductors\cite{Frano2016,PhysRevX.9.031042} or
incommensurate charge density waves in transition metal
chalcogenides\cite{PhysRevLett.45.576,PhysRevLett.118.106405,PhysRevLett.99.046401}. 

\section{Conclusion}

Solving interacting models in super-moir\'e materials
represents a formidable theory challenge to understanding emerging phenomena in van der Waals materials,
due to the unprecedented system sizes
required to capture their physics.
Here, we have presented a kernel polynomial tensor cross interpolation algorithm, that can solve interacting models with several millions of atoms, considerably outperforming the current state-of-the-art.
Our strategy relies on mapping the mean-field Hamiltonian
of a large electronic model to an auxiliary many-body Hilbert space that is compressed using a many-body tensor network. The tensor network is constructed
with a tensor cross interpolation algorithm, which greatly reduces the number of individual evaluations performed with a kernel polynomial method. This demonstrates how a quantum-inspired
methodology enables massive speed-up of the calculation of
mean-field interacting ground states of tight-binding models.
We have applied our algorithm to both one-
and two-dimensional models, showing that this approach allows us to deal with interacting problems with multiple long-range modulations and domain walls. In particular,
we have demonstrated that this methodology can describe correlated states in super-moir\'e buckled graphene, capturing both the electronic reconstructions and symmetry-breaking at the moir\'e and super-moir\'e length scales. Our methodology enables tackling the interacting models with a number of sites required to rationalize the physics of a whole new family
of artificial materials based on twisted van der Waals heterostructures.
In particular, it can be readily extended to account for
charge order, bond-ordered, topological and superconducting states,
providing the required computational tool to study super-moir\'e quantum matter.

\textbf{Acknowledgements:}
We acknowledge financial support from
the Academy of Finland Projects Nos. 331342, 358088, and 349696,
InstituteQ, the Jane and Aatos Erkko Foundation, and the Finnish Quantum Flagship.
We thank X. Waintal, C. Groth, A. Moulinas, N. Jolly, T. Louvet, M. Srdinsek, A. Manesco, O. Zilberberg,
B. Amorim, E. Castro, P. San-Jose and C. Flindt for useful discussions.
We acknowledge the computational resources provided by the Aalto Science-IT project.

\bibliography{biblio}

\begin{thebibliography}{105}%
\makeatletter
\providecommand \@ifxundefined [1]{%
 \@ifx{#1\undefined}
}%
\providecommand \@ifnum [1]{%
 \ifnum #1\expandafter \@firstoftwo
 \else \expandafter \@secondoftwo
 \fi
}%
\providecommand \@ifx [1]{%
 \ifx #1\expandafter \@firstoftwo
 \else \expandafter \@secondoftwo
 \fi
}%
\providecommand \natexlab [1]{#1}%
\providecommand \enquote  [1]{``#1''}%
\providecommand \bibnamefont  [1]{#1}%
\providecommand \bibfnamefont [1]{#1}%
\providecommand \citenamefont [1]{#1}%
\providecommand \href@noop [0]{\@secondoftwo}%
\providecommand \href [0]{\begingroup \@sanitize@url \@href}%
\providecommand \@href[1]{\@@startlink{#1}\@@href}%
\providecommand \@@href[1]{\endgroup#1\@@endlink}%
\providecommand \@sanitize@url [0]{\catcode `\\12\catcode `\$12\catcode
  `\&12\catcode `\#12\catcode `\^12\catcode `\_12\catcode `\%12\relax}%
\providecommand \@@startlink[1]{}%
\providecommand \@@endlink[0]{}%
\providecommand \url  [0]{\begingroup\@sanitize@url \@url }%
\providecommand \@url [1]{\endgroup\@href {#1}{\urlprefix }}%
\providecommand \urlprefix  [0]{URL }%
\providecommand \Eprint [0]{\href }%
\providecommand \doibase [0]{http://dx.doi.org/}%
\providecommand \selectlanguage [0]{\@gobble}%
\providecommand \bibinfo  [0]{\@secondoftwo}%
\providecommand \bibfield  [0]{\@secondoftwo}%
\providecommand \translation [1]{[#1]}%
\providecommand \BibitemOpen [0]{}%
\providecommand \bibitemStop [0]{}%
\providecommand \bibitemNoStop [0]{.\EOS\space}%
\providecommand \EOS [0]{\spacefactor3000\relax}%
\providecommand \BibitemShut  [1]{\csname bibitem#1\endcsname}%
\let\auto@bib@innerbib\@empty
\bibitem [{\citenamefont {Andrei}\ \emph {et~al.}(2021)\citenamefont {Andrei},
  \citenamefont {Efetov}, \citenamefont {Jarillo-Herrero}, \citenamefont
  {MacDonald}, \citenamefont {Mak}, \citenamefont {Senthil}, \citenamefont
  {Tutuc}, \citenamefont {Yazdani},\ and\ \citenamefont {Young}}]{Andrei2021}%
  \BibitemOpen
  \bibfield  {author} {\bibinfo {author} {\bibfnamefont {Eva~Y.}\ \bibnamefont
  {Andrei}}, \bibinfo {author} {\bibfnamefont {Dmitri~K.}\ \bibnamefont
  {Efetov}}, \bibinfo {author} {\bibfnamefont {Pablo}\ \bibnamefont
  {Jarillo-Herrero}}, \bibinfo {author} {\bibfnamefont {Allan~H.}\ \bibnamefont
  {MacDonald}}, \bibinfo {author} {\bibfnamefont {Kin~Fai}\ \bibnamefont
  {Mak}}, \bibinfo {author} {\bibfnamefont {T.}~\bibnamefont {Senthil}},
  \bibinfo {author} {\bibfnamefont {Emanuel}\ \bibnamefont {Tutuc}}, \bibinfo
  {author} {\bibfnamefont {Ali}\ \bibnamefont {Yazdani}}, \ and\ \bibinfo
  {author} {\bibfnamefont {Andrea~F.}\ \bibnamefont {Young}},\ }\bibfield
  {title} {\enquote {\bibinfo {title} {The marvels of moiré materials},}\
  }\href {\doibase 10.1038/s41578-021-00284-1} {\bibfield  {journal} {\bibinfo
  {journal} {Nature Reviews Materials}\ }\textbf {\bibinfo {volume} {6}},\
  \bibinfo {pages} {201–206} (\bibinfo {year} {2021})}\BibitemShut {NoStop}%
\bibitem [{\citenamefont {Zeng}\ \emph {et~al.}(2023)\citenamefont {Zeng},
  \citenamefont {Xia}, \citenamefont {Kang}, \citenamefont {Zhu}, \citenamefont
  {Kn\"{u}ppel}, \citenamefont {Vaswani}, \citenamefont {Watanabe},
  \citenamefont {Taniguchi}, \citenamefont {Mak},\ and\ \citenamefont
  {Shan}}]{Zeng2023}%
  \BibitemOpen
  \bibfield  {author} {\bibinfo {author} {\bibfnamefont {Yihang}\ \bibnamefont
  {Zeng}}, \bibinfo {author} {\bibfnamefont {Zhengchao}\ \bibnamefont {Xia}},
  \bibinfo {author} {\bibfnamefont {Kaifei}\ \bibnamefont {Kang}}, \bibinfo
  {author} {\bibfnamefont {Jiacheng}\ \bibnamefont {Zhu}}, \bibinfo {author}
  {\bibfnamefont {Patrick}\ \bibnamefont {Kn\"{u}ppel}}, \bibinfo {author}
  {\bibfnamefont {Chirag}\ \bibnamefont {Vaswani}}, \bibinfo {author}
  {\bibfnamefont {Kenji}\ \bibnamefont {Watanabe}}, \bibinfo {author}
  {\bibfnamefont {Takashi}\ \bibnamefont {Taniguchi}}, \bibinfo {author}
  {\bibfnamefont {Kin~Fai}\ \bibnamefont {Mak}}, \ and\ \bibinfo {author}
  {\bibfnamefont {Jie}\ \bibnamefont {Shan}},\ }\bibfield  {title} {\enquote
  {\bibinfo {title} {Thermodynamic evidence of fractional chern insulator in
  moiré mote2},}\ }\href {\doibase 10.1038/s41586-023-06452-3} {\bibfield
  {journal} {\bibinfo  {journal} {Nature}\ }\textbf {\bibinfo {volume} {622}},\
  \bibinfo {pages} {69–73} (\bibinfo {year} {2023})}\BibitemShut {NoStop}%
\bibitem [{\citenamefont {Serlin}\ \emph {et~al.}(2020)\citenamefont {Serlin},
  \citenamefont {Tschirhart}, \citenamefont {Polshyn}, \citenamefont {Zhang},
  \citenamefont {Zhu}, \citenamefont {Watanabe}, \citenamefont {Taniguchi},
  \citenamefont {Balents},\ and\ \citenamefont {Young}}]{Serlin2020}%
  \BibitemOpen
  \bibfield  {author} {\bibinfo {author} {\bibfnamefont {M.}~\bibnamefont
  {Serlin}}, \bibinfo {author} {\bibfnamefont {C.~L.}\ \bibnamefont
  {Tschirhart}}, \bibinfo {author} {\bibfnamefont {H.}~\bibnamefont {Polshyn}},
  \bibinfo {author} {\bibfnamefont {Y.}~\bibnamefont {Zhang}}, \bibinfo
  {author} {\bibfnamefont {J.}~\bibnamefont {Zhu}}, \bibinfo {author}
  {\bibfnamefont {K.}~\bibnamefont {Watanabe}}, \bibinfo {author}
  {\bibfnamefont {T.}~\bibnamefont {Taniguchi}}, \bibinfo {author}
  {\bibfnamefont {L.}~\bibnamefont {Balents}}, \ and\ \bibinfo {author}
  {\bibfnamefont {A.~F.}\ \bibnamefont {Young}},\ }\bibfield  {title} {\enquote
  {\bibinfo {title} {Intrinsic quantized anomalous hall effect in a moiré
  heterostructure},}\ }\href {\doibase 10.1126/science.aay5533} {\bibfield
  {journal} {\bibinfo  {journal} {Science}\ }\textbf {\bibinfo {volume}
  {367}},\ \bibinfo {pages} {900–903} (\bibinfo {year} {2020})}\BibitemShut
  {NoStop}%
\bibitem [{\citenamefont {Sharpe}\ \emph {et~al.}(2019)\citenamefont {Sharpe},
  \citenamefont {Fox}, \citenamefont {Barnard}, \citenamefont {Finney},
  \citenamefont {Watanabe}, \citenamefont {Taniguchi}, \citenamefont
  {Kastner},\ and\ \citenamefont {Goldhaber-Gordon}}]{Sharpe2019}%
  \BibitemOpen
  \bibfield  {author} {\bibinfo {author} {\bibfnamefont {Aaron~L.}\
  \bibnamefont {Sharpe}}, \bibinfo {author} {\bibfnamefont {Eli~J.}\
  \bibnamefont {Fox}}, \bibinfo {author} {\bibfnamefont {Arthur~W.}\
  \bibnamefont {Barnard}}, \bibinfo {author} {\bibfnamefont {Joe}\ \bibnamefont
  {Finney}}, \bibinfo {author} {\bibfnamefont {Kenji}\ \bibnamefont
  {Watanabe}}, \bibinfo {author} {\bibfnamefont {Takashi}\ \bibnamefont
  {Taniguchi}}, \bibinfo {author} {\bibfnamefont {M.~A.}\ \bibnamefont
  {Kastner}}, \ and\ \bibinfo {author} {\bibfnamefont {David}\ \bibnamefont
  {Goldhaber-Gordon}},\ }\bibfield  {title} {\enquote {\bibinfo {title}
  {Emergent ferromagnetism near three-quarters filling in twisted bilayer
  graphene},}\ }\href {\doibase 10.1126/science.aaw3780} {\bibfield  {journal}
  {\bibinfo  {journal} {Science}\ }\textbf {\bibinfo {volume} {365}},\ \bibinfo
  {pages} {605–608} (\bibinfo {year} {2019})}\BibitemShut {NoStop}%
\bibitem [{\citenamefont {Rickhaus}\ \emph {et~al.}(2018)\citenamefont
  {Rickhaus}, \citenamefont {Wallbank}, \citenamefont {Slizovskiy},
  \citenamefont {Pisoni}, \citenamefont {Overweg}, \citenamefont {Lee},
  \citenamefont {Eich}, \citenamefont {Liu}, \citenamefont {Watanabe},
  \citenamefont {Taniguchi}, \citenamefont {Ihn},\ and\ \citenamefont
  {Ensslin}}]{Rickhaus2018}%
  \BibitemOpen
  \bibfield  {author} {\bibinfo {author} {\bibfnamefont {Peter}\ \bibnamefont
  {Rickhaus}}, \bibinfo {author} {\bibfnamefont {John}\ \bibnamefont
  {Wallbank}}, \bibinfo {author} {\bibfnamefont {Sergey}\ \bibnamefont
  {Slizovskiy}}, \bibinfo {author} {\bibfnamefont {Riccardo}\ \bibnamefont
  {Pisoni}}, \bibinfo {author} {\bibfnamefont {Hiske}\ \bibnamefont {Overweg}},
  \bibinfo {author} {\bibfnamefont {Yongjin}\ \bibnamefont {Lee}}, \bibinfo
  {author} {\bibfnamefont {Marius}\ \bibnamefont {Eich}}, \bibinfo {author}
  {\bibfnamefont {Ming-Hao}\ \bibnamefont {Liu}}, \bibinfo {author}
  {\bibfnamefont {Kenji}\ \bibnamefont {Watanabe}}, \bibinfo {author}
  {\bibfnamefont {Takashi}\ \bibnamefont {Taniguchi}}, \bibinfo {author}
  {\bibfnamefont {Thomas}\ \bibnamefont {Ihn}}, \ and\ \bibinfo {author}
  {\bibfnamefont {Klaus}\ \bibnamefont {Ensslin}},\ }\bibfield  {title}
  {\enquote {\bibinfo {title} {Transport through a network of topological
  channels in twisted bilayer graphene},}\ }\href {\doibase
  10.1021/acs.nanolett.8b02387} {\bibfield  {journal} {\bibinfo  {journal}
  {Nano Letters}\ }\textbf {\bibinfo {volume} {18}},\ \bibinfo {pages}
  {6725–6730} (\bibinfo {year} {2018})}\BibitemShut {NoStop}%
\bibitem [{\citenamefont {Cai}\ \emph {et~al.}(2023)\citenamefont {Cai},
  \citenamefont {Anderson}, \citenamefont {Wang}, \citenamefont {Zhang},
  \citenamefont {Liu}, \citenamefont {Holtzmann}, \citenamefont {Zhang},
  \citenamefont {Fan}, \citenamefont {Taniguchi}, \citenamefont {Watanabe},
  \citenamefont {Ran}, \citenamefont {Cao}, \citenamefont {Fu}, \citenamefont
  {Xiao}, \citenamefont {Yao},\ and\ \citenamefont {Xu}}]{Cai2023}%
  \BibitemOpen
  \bibfield  {author} {\bibinfo {author} {\bibfnamefont {Jiaqi}\ \bibnamefont
  {Cai}}, \bibinfo {author} {\bibfnamefont {Eric}\ \bibnamefont {Anderson}},
  \bibinfo {author} {\bibfnamefont {Chong}\ \bibnamefont {Wang}}, \bibinfo
  {author} {\bibfnamefont {Xiaowei}\ \bibnamefont {Zhang}}, \bibinfo {author}
  {\bibfnamefont {Xiaoyu}\ \bibnamefont {Liu}}, \bibinfo {author}
  {\bibfnamefont {William}\ \bibnamefont {Holtzmann}}, \bibinfo {author}
  {\bibfnamefont {Yinong}\ \bibnamefont {Zhang}}, \bibinfo {author}
  {\bibfnamefont {Fengren}\ \bibnamefont {Fan}}, \bibinfo {author}
  {\bibfnamefont {Takashi}\ \bibnamefont {Taniguchi}}, \bibinfo {author}
  {\bibfnamefont {Kenji}\ \bibnamefont {Watanabe}}, \bibinfo {author}
  {\bibfnamefont {Ying}\ \bibnamefont {Ran}}, \bibinfo {author} {\bibfnamefont
  {Ting}\ \bibnamefont {Cao}}, \bibinfo {author} {\bibfnamefont {Liang}\
  \bibnamefont {Fu}}, \bibinfo {author} {\bibfnamefont {Di}~\bibnamefont
  {Xiao}}, \bibinfo {author} {\bibfnamefont {Wang}\ \bibnamefont {Yao}}, \ and\
  \bibinfo {author} {\bibfnamefont {Xiaodong}\ \bibnamefont {Xu}},\ }\bibfield
  {title} {\enquote {\bibinfo {title} {Signatures of fractional quantum
  anomalous hall states in twisted mote2},}\ }\href {\doibase
  10.1038/s41586-023-06289-w} {\bibfield  {journal} {\bibinfo  {journal}
  {Nature}\ }\textbf {\bibinfo {volume} {622}},\ \bibinfo {pages} {63–68}
  (\bibinfo {year} {2023})}\BibitemShut {NoStop}%
\bibitem [{\citenamefont {Cao}\ \emph {et~al.}(2018{\natexlab{a}})\citenamefont
  {Cao}, \citenamefont {Fatemi}, \citenamefont {Demir}, \citenamefont {Fang},
  \citenamefont {Tomarken}, \citenamefont {Luo}, \citenamefont
  {Sanchez-Yamagishi}, \citenamefont {Watanabe}, \citenamefont {Taniguchi},
  \citenamefont {Kaxiras}, \citenamefont {Ashoori},\ and\ \citenamefont
  {Jarillo-Herrero}}]{Cao2018corr}%
  \BibitemOpen
  \bibfield  {author} {\bibinfo {author} {\bibfnamefont {Yuan}\ \bibnamefont
  {Cao}}, \bibinfo {author} {\bibfnamefont {Valla}\ \bibnamefont {Fatemi}},
  \bibinfo {author} {\bibfnamefont {Ahmet}\ \bibnamefont {Demir}}, \bibinfo
  {author} {\bibfnamefont {Shiang}\ \bibnamefont {Fang}}, \bibinfo {author}
  {\bibfnamefont {Spencer~L.}\ \bibnamefont {Tomarken}}, \bibinfo {author}
  {\bibfnamefont {Jason~Y.}\ \bibnamefont {Luo}}, \bibinfo {author}
  {\bibfnamefont {Javier~D.}\ \bibnamefont {Sanchez-Yamagishi}}, \bibinfo
  {author} {\bibfnamefont {Kenji}\ \bibnamefont {Watanabe}}, \bibinfo {author}
  {\bibfnamefont {Takashi}\ \bibnamefont {Taniguchi}}, \bibinfo {author}
  {\bibfnamefont {Efthimios}\ \bibnamefont {Kaxiras}}, \bibinfo {author}
  {\bibfnamefont {Ray~C.}\ \bibnamefont {Ashoori}}, \ and\ \bibinfo {author}
  {\bibfnamefont {Pablo}\ \bibnamefont {Jarillo-Herrero}},\ }\bibfield  {title}
  {\enquote {\bibinfo {title} {Correlated insulator behaviour at half-filling
  in magic-angle graphene superlattices},}\ }\href {\doibase
  10.1038/nature26154} {\bibfield  {journal} {\bibinfo  {journal} {Nature}\
  }\textbf {\bibinfo {volume} {556}},\ \bibinfo {pages} {80–84} (\bibinfo
  {year} {2018}{\natexlab{a}})}\BibitemShut {NoStop}%
\bibitem [{\citenamefont {Burg}\ \emph {et~al.}(2022)\citenamefont {Burg},
  \citenamefont {Khalaf}, \citenamefont {Wang}, \citenamefont {Watanabe},
  \citenamefont {Taniguchi},\ and\ \citenamefont {Tutuc}}]{Burg2022}%
  \BibitemOpen
  \bibfield  {author} {\bibinfo {author} {\bibfnamefont {G.~William}\
  \bibnamefont {Burg}}, \bibinfo {author} {\bibfnamefont {Eslam}\ \bibnamefont
  {Khalaf}}, \bibinfo {author} {\bibfnamefont {Yimeng}\ \bibnamefont {Wang}},
  \bibinfo {author} {\bibfnamefont {Kenji}\ \bibnamefont {Watanabe}}, \bibinfo
  {author} {\bibfnamefont {Takashi}\ \bibnamefont {Taniguchi}}, \ and\ \bibinfo
  {author} {\bibfnamefont {Emanuel}\ \bibnamefont {Tutuc}},\ }\bibfield
  {title} {\enquote {\bibinfo {title} {Emergence of correlations in alternating
  twist quadrilayer graphene},}\ }\href {\doibase 10.1038/s41563-022-01286-2}
  {\bibfield  {journal} {\bibinfo  {journal} {Nature Materials}\ }\textbf
  {\bibinfo {volume} {21}},\ \bibinfo {pages} {884–889} (\bibinfo {year}
  {2022})}\BibitemShut {NoStop}%
\bibitem [{\citenamefont {Kerelsky}\ \emph {et~al.}(2019)\citenamefont
  {Kerelsky}, \citenamefont {McGilly}, \citenamefont {Kennes}, \citenamefont
  {Xian}, \citenamefont {Yankowitz}, \citenamefont {Chen}, \citenamefont
  {Watanabe}, \citenamefont {Taniguchi}, \citenamefont {Hone}, \citenamefont
  {Dean}, \citenamefont {Rubio},\ and\ \citenamefont
  {Pasupathy}}]{Kerelsky2019}%
  \BibitemOpen
  \bibfield  {author} {\bibinfo {author} {\bibfnamefont {Alexander}\
  \bibnamefont {Kerelsky}}, \bibinfo {author} {\bibfnamefont {Leo~J.}\
  \bibnamefont {McGilly}}, \bibinfo {author} {\bibfnamefont {Dante~M.}\
  \bibnamefont {Kennes}}, \bibinfo {author} {\bibfnamefont {Lede}\ \bibnamefont
  {Xian}}, \bibinfo {author} {\bibfnamefont {Matthew}\ \bibnamefont
  {Yankowitz}}, \bibinfo {author} {\bibfnamefont {Shaowen}\ \bibnamefont
  {Chen}}, \bibinfo {author} {\bibfnamefont {K.}~\bibnamefont {Watanabe}},
  \bibinfo {author} {\bibfnamefont {T.}~\bibnamefont {Taniguchi}}, \bibinfo
  {author} {\bibfnamefont {James}\ \bibnamefont {Hone}}, \bibinfo {author}
  {\bibfnamefont {Cory}\ \bibnamefont {Dean}}, \bibinfo {author} {\bibfnamefont
  {Angel}\ \bibnamefont {Rubio}}, \ and\ \bibinfo {author} {\bibfnamefont
  {Abhay~N.}\ \bibnamefont {Pasupathy}},\ }\bibfield  {title} {\enquote
  {\bibinfo {title} {Maximized electron interactions at the magic angle in
  twisted bilayer graphene},}\ }\href {\doibase 10.1038/s41586-019-1431-9}
  {\bibfield  {journal} {\bibinfo  {journal} {Nature}\ }\textbf {\bibinfo
  {volume} {572}},\ \bibinfo {pages} {95–100} (\bibinfo {year}
  {2019})}\BibitemShut {NoStop}%
\bibitem [{\citenamefont {Kim}\ \emph {et~al.}(2023)\citenamefont {Kim},
  \citenamefont {Choi}, \citenamefont {Lantagne-Hurtubise}, \citenamefont
  {Lewandowski}, \citenamefont {Thomson}, \citenamefont {Kong}, \citenamefont
  {Zhou}, \citenamefont {Baum}, \citenamefont {Zhang}, \citenamefont {Holleis},
  \citenamefont {Watanabe}, \citenamefont {Taniguchi}, \citenamefont {Young},
  \citenamefont {Alicea},\ and\ \citenamefont {Nadj-Perge}}]{Kim2023}%
  \BibitemOpen
  \bibfield  {author} {\bibinfo {author} {\bibfnamefont {Hyunjin}\ \bibnamefont
  {Kim}}, \bibinfo {author} {\bibfnamefont {Youngjoon}\ \bibnamefont {Choi}},
  \bibinfo {author} {\bibfnamefont {Etienne}\ \bibnamefont
  {Lantagne-Hurtubise}}, \bibinfo {author} {\bibfnamefont {Cyprian}\
  \bibnamefont {Lewandowski}}, \bibinfo {author} {\bibfnamefont {Alex}\
  \bibnamefont {Thomson}}, \bibinfo {author} {\bibfnamefont {Lingyuan}\
  \bibnamefont {Kong}}, \bibinfo {author} {\bibfnamefont {Haoxin}\ \bibnamefont
  {Zhou}}, \bibinfo {author} {\bibfnamefont {Eli}\ \bibnamefont {Baum}},
  \bibinfo {author} {\bibfnamefont {Yiran}\ \bibnamefont {Zhang}}, \bibinfo
  {author} {\bibfnamefont {Ludwig}\ \bibnamefont {Holleis}}, \bibinfo {author}
  {\bibfnamefont {Kenji}\ \bibnamefont {Watanabe}}, \bibinfo {author}
  {\bibfnamefont {Takashi}\ \bibnamefont {Taniguchi}}, \bibinfo {author}
  {\bibfnamefont {Andrea~F.}\ \bibnamefont {Young}}, \bibinfo {author}
  {\bibfnamefont {Jason}\ \bibnamefont {Alicea}}, \ and\ \bibinfo {author}
  {\bibfnamefont {Stevan}\ \bibnamefont {Nadj-Perge}},\ }\bibfield  {title}
  {\enquote {\bibinfo {title} {Imaging inter-valley coherent order in
  magic-angle twisted trilayer graphene},}\ }\href {\doibase
  10.1038/s41586-023-06663-8} {\bibfield  {journal} {\bibinfo  {journal}
  {Nature}\ }\textbf {\bibinfo {volume} {623}},\ \bibinfo {pages} {942–948}
  (\bibinfo {year} {2023})}\BibitemShut {NoStop}%
\bibitem [{\citenamefont {Lu}\ \emph {et~al.}(2019)\citenamefont {Lu},
  \citenamefont {Stepanov}, \citenamefont {Yang}, \citenamefont {Xie},
  \citenamefont {Aamir}, \citenamefont {Das}, \citenamefont {Urgell},
  \citenamefont {Watanabe}, \citenamefont {Taniguchi}, \citenamefont {Zhang},
  \citenamefont {Bachtold}, \citenamefont {MacDonald},\ and\ \citenamefont
  {Efetov}}]{Lu2019}%
  \BibitemOpen
  \bibfield  {author} {\bibinfo {author} {\bibfnamefont {Xiaobo}\ \bibnamefont
  {Lu}}, \bibinfo {author} {\bibfnamefont {Petr}\ \bibnamefont {Stepanov}},
  \bibinfo {author} {\bibfnamefont {Wei}\ \bibnamefont {Yang}}, \bibinfo
  {author} {\bibfnamefont {Ming}\ \bibnamefont {Xie}}, \bibinfo {author}
  {\bibfnamefont {Mohammed~Ali}\ \bibnamefont {Aamir}}, \bibinfo {author}
  {\bibfnamefont {Ipsita}\ \bibnamefont {Das}}, \bibinfo {author}
  {\bibfnamefont {Carles}\ \bibnamefont {Urgell}}, \bibinfo {author}
  {\bibfnamefont {Kenji}\ \bibnamefont {Watanabe}}, \bibinfo {author}
  {\bibfnamefont {Takashi}\ \bibnamefont {Taniguchi}}, \bibinfo {author}
  {\bibfnamefont {Guangyu}\ \bibnamefont {Zhang}}, \bibinfo {author}
  {\bibfnamefont {Adrian}\ \bibnamefont {Bachtold}}, \bibinfo {author}
  {\bibfnamefont {Allan~H.}\ \bibnamefont {MacDonald}}, \ and\ \bibinfo
  {author} {\bibfnamefont {Dmitri~K.}\ \bibnamefont {Efetov}},\ }\bibfield
  {title} {\enquote {\bibinfo {title} {Superconductors, orbital magnets and
  correlated states in magic-angle bilayer graphene},}\ }\href {\doibase
  10.1038/s41586-019-1695-0} {\bibfield  {journal} {\bibinfo  {journal}
  {Nature}\ }\textbf {\bibinfo {volume} {574}},\ \bibinfo {pages} {653–657}
  (\bibinfo {year} {2019})}\BibitemShut {NoStop}%
\bibitem [{\citenamefont {Zhao}\ \emph {et~al.}(2023)\citenamefont {Zhao},
  \citenamefont {Shen}, \citenamefont {Tao}, \citenamefont {Han}, \citenamefont
  {Kang}, \citenamefont {Watanabe}, \citenamefont {Taniguchi}, \citenamefont
  {Mak},\ and\ \citenamefont {Shan}}]{Zhao2023}%
  \BibitemOpen
  \bibfield  {author} {\bibinfo {author} {\bibfnamefont {Wenjin}\ \bibnamefont
  {Zhao}}, \bibinfo {author} {\bibfnamefont {Bowen}\ \bibnamefont {Shen}},
  \bibinfo {author} {\bibfnamefont {Zui}\ \bibnamefont {Tao}}, \bibinfo
  {author} {\bibfnamefont {Zhongdong}\ \bibnamefont {Han}}, \bibinfo {author}
  {\bibfnamefont {Kaifei}\ \bibnamefont {Kang}}, \bibinfo {author}
  {\bibfnamefont {Kenji}\ \bibnamefont {Watanabe}}, \bibinfo {author}
  {\bibfnamefont {Takashi}\ \bibnamefont {Taniguchi}}, \bibinfo {author}
  {\bibfnamefont {Kin~Fai}\ \bibnamefont {Mak}}, \ and\ \bibinfo {author}
  {\bibfnamefont {Jie}\ \bibnamefont {Shan}},\ }\bibfield  {title} {\enquote
  {\bibinfo {title} {Gate-tunable heavy fermions in a moiré kondo lattice},}\
  }\href {\doibase 10.1038/s41586-023-05800-7} {\bibfield  {journal} {\bibinfo
  {journal} {Nature}\ }\textbf {\bibinfo {volume} {616}},\ \bibinfo {pages}
  {61–65} (\bibinfo {year} {2023})}\BibitemShut {NoStop}%
\bibitem [{\citenamefont {Cao}\ \emph {et~al.}(2018{\natexlab{b}})\citenamefont
  {Cao}, \citenamefont {Fatemi}, \citenamefont {Fang}, \citenamefont
  {Watanabe}, \citenamefont {Taniguchi}, \citenamefont {Kaxiras},\ and\
  \citenamefont {Jarillo-Herrero}}]{Cao2018}%
  \BibitemOpen
  \bibfield  {author} {\bibinfo {author} {\bibfnamefont {Yuan}\ \bibnamefont
  {Cao}}, \bibinfo {author} {\bibfnamefont {Valla}\ \bibnamefont {Fatemi}},
  \bibinfo {author} {\bibfnamefont {Shiang}\ \bibnamefont {Fang}}, \bibinfo
  {author} {\bibfnamefont {Kenji}\ \bibnamefont {Watanabe}}, \bibinfo {author}
  {\bibfnamefont {Takashi}\ \bibnamefont {Taniguchi}}, \bibinfo {author}
  {\bibfnamefont {Efthimios}\ \bibnamefont {Kaxiras}}, \ and\ \bibinfo {author}
  {\bibfnamefont {Pablo}\ \bibnamefont {Jarillo-Herrero}},\ }\bibfield  {title}
  {\enquote {\bibinfo {title} {Unconventional superconductivity in magic-angle
  graphene superlattices},}\ }\href {\doibase 10.1038/nature26160} {\bibfield
  {journal} {\bibinfo  {journal} {Nature}\ }\textbf {\bibinfo {volume} {556}},\
  \bibinfo {pages} {43--50} (\bibinfo {year} {2018}{\natexlab{b}})}\BibitemShut
  {NoStop}%
\bibitem [{\citenamefont {Yankowitz}\ \emph {et~al.}(2019)\citenamefont
  {Yankowitz}, \citenamefont {Chen}, \citenamefont {Polshyn}, \citenamefont
  {Zhang}, \citenamefont {Watanabe}, \citenamefont {Taniguchi}, \citenamefont
  {Graf}, \citenamefont {Young},\ and\ \citenamefont {Dean}}]{Yankowitz2019}%
  \BibitemOpen
  \bibfield  {author} {\bibinfo {author} {\bibfnamefont {Matthew}\ \bibnamefont
  {Yankowitz}}, \bibinfo {author} {\bibfnamefont {Shaowen}\ \bibnamefont
  {Chen}}, \bibinfo {author} {\bibfnamefont {Hryhoriy}\ \bibnamefont
  {Polshyn}}, \bibinfo {author} {\bibfnamefont {Yuxuan}\ \bibnamefont {Zhang}},
  \bibinfo {author} {\bibfnamefont {K.}~\bibnamefont {Watanabe}}, \bibinfo
  {author} {\bibfnamefont {T.}~\bibnamefont {Taniguchi}}, \bibinfo {author}
  {\bibfnamefont {David}\ \bibnamefont {Graf}}, \bibinfo {author}
  {\bibfnamefont {Andrea~F.}\ \bibnamefont {Young}}, \ and\ \bibinfo {author}
  {\bibfnamefont {Cory~R.}\ \bibnamefont {Dean}},\ }\bibfield  {title}
  {\enquote {\bibinfo {title} {Tuning superconductivity in twisted bilayer
  graphene},}\ }\href {\doibase 10.1126/science.aav1910} {\bibfield  {journal}
  {\bibinfo  {journal} {Science}\ }\textbf {\bibinfo {volume} {363}},\ \bibinfo
  {pages} {1059–1064} (\bibinfo {year} {2019})}\BibitemShut {NoStop}%
\bibitem [{\citenamefont {Park}\ \emph {et~al.}(2021)\citenamefont {Park},
  \citenamefont {Cao}, \citenamefont {Watanabe}, \citenamefont {Taniguchi},\
  and\ \citenamefont {Jarillo-Herrero}}]{Park2021}%
  \BibitemOpen
  \bibfield  {author} {\bibinfo {author} {\bibfnamefont {Jeong~Min}\
  \bibnamefont {Park}}, \bibinfo {author} {\bibfnamefont {Yuan}\ \bibnamefont
  {Cao}}, \bibinfo {author} {\bibfnamefont {Kenji}\ \bibnamefont {Watanabe}},
  \bibinfo {author} {\bibfnamefont {Takashi}\ \bibnamefont {Taniguchi}}, \ and\
  \bibinfo {author} {\bibfnamefont {Pablo}\ \bibnamefont {Jarillo-Herrero}},\
  }\bibfield  {title} {\enquote {\bibinfo {title} {Tunable strongly coupled
  superconductivity in magic-angle twisted trilayer graphene},}\ }\href
  {\doibase 10.1038/s41586-021-03192-0} {\bibfield  {journal} {\bibinfo
  {journal} {Nature}\ }\textbf {\bibinfo {volume} {590}},\ \bibinfo {pages}
  {249–255} (\bibinfo {year} {2021})}\BibitemShut {NoStop}%
\bibitem [{\citenamefont {Oh}\ \emph {et~al.}(2021)\citenamefont {Oh},
  \citenamefont {Nuckolls}, \citenamefont {Wong}, \citenamefont {Lee},
  \citenamefont {Liu}, \citenamefont {Watanabe}, \citenamefont {Taniguchi},\
  and\ \citenamefont {Yazdani}}]{Oh2021}%
  \BibitemOpen
  \bibfield  {author} {\bibinfo {author} {\bibfnamefont {Myungchul}\
  \bibnamefont {Oh}}, \bibinfo {author} {\bibfnamefont {Kevin~P.}\ \bibnamefont
  {Nuckolls}}, \bibinfo {author} {\bibfnamefont {Dillon}\ \bibnamefont {Wong}},
  \bibinfo {author} {\bibfnamefont {Ryan~L.}\ \bibnamefont {Lee}}, \bibinfo
  {author} {\bibfnamefont {Xiaomeng}\ \bibnamefont {Liu}}, \bibinfo {author}
  {\bibfnamefont {Kenji}\ \bibnamefont {Watanabe}}, \bibinfo {author}
  {\bibfnamefont {Takashi}\ \bibnamefont {Taniguchi}}, \ and\ \bibinfo {author}
  {\bibfnamefont {Ali}\ \bibnamefont {Yazdani}},\ }\bibfield  {title} {\enquote
  {\bibinfo {title} {Evidence for unconventional superconductivity in twisted
  bilayer graphene},}\ }\href {\doibase 10.1038/s41586-021-04121-x} {\bibfield
  {journal} {\bibinfo  {journal} {Nature}\ }\textbf {\bibinfo {volume} {600}},\
  \bibinfo {pages} {240–245} (\bibinfo {year} {2021})}\BibitemShut {NoStop}%
\bibitem [{\citenamefont {Park}\ \emph {et~al.}(2022)\citenamefont {Park},
  \citenamefont {Cao}, \citenamefont {Xia}, \citenamefont {Sun}, \citenamefont
  {Watanabe}, \citenamefont {Taniguchi},\ and\ \citenamefont
  {Jarillo-Herrero}}]{Park2022}%
  \BibitemOpen
  \bibfield  {author} {\bibinfo {author} {\bibfnamefont {Jeong~Min}\
  \bibnamefont {Park}}, \bibinfo {author} {\bibfnamefont {Yuan}\ \bibnamefont
  {Cao}}, \bibinfo {author} {\bibfnamefont {Li-Qiao}\ \bibnamefont {Xia}},
  \bibinfo {author} {\bibfnamefont {Shuwen}\ \bibnamefont {Sun}}, \bibinfo
  {author} {\bibfnamefont {Kenji}\ \bibnamefont {Watanabe}}, \bibinfo {author}
  {\bibfnamefont {Takashi}\ \bibnamefont {Taniguchi}}, \ and\ \bibinfo {author}
  {\bibfnamefont {Pablo}\ \bibnamefont {Jarillo-Herrero}},\ }\bibfield  {title}
  {\enquote {\bibinfo {title} {Robust superconductivity in magic-angle
  multilayer graphene family},}\ }\href {\doibase 10.1038/s41563-022-01287-1}
  {\bibfield  {journal} {\bibinfo  {journal} {Nature Materials}\ }\textbf
  {\bibinfo {volume} {21}},\ \bibinfo {pages} {877–883} (\bibinfo {year}
  {2022})}\BibitemShut {NoStop}%
\bibitem [{\citenamefont {Devakul}\ \emph {et~al.}(2023)\citenamefont
  {Devakul}, \citenamefont {Ledwith}, \citenamefont {Xia}, \citenamefont {Uri},
  \citenamefont {de~la Barrera}, \citenamefont {Jarillo-Herrero},\ and\
  \citenamefont {Fu}}]{Devakul2023}%
  \BibitemOpen
  \bibfield  {author} {\bibinfo {author} {\bibfnamefont {Trithep}\ \bibnamefont
  {Devakul}}, \bibinfo {author} {\bibfnamefont {Patrick~J.}\ \bibnamefont
  {Ledwith}}, \bibinfo {author} {\bibfnamefont {Li-Qiao}\ \bibnamefont {Xia}},
  \bibinfo {author} {\bibfnamefont {Aviram}\ \bibnamefont {Uri}}, \bibinfo
  {author} {\bibfnamefont {Sergio~C.}\ \bibnamefont {de~la Barrera}}, \bibinfo
  {author} {\bibfnamefont {Pablo}\ \bibnamefont {Jarillo-Herrero}}, \ and\
  \bibinfo {author} {\bibfnamefont {Liang}\ \bibnamefont {Fu}},\ }\bibfield
  {title} {\enquote {\bibinfo {title} {Magic-angle helical trilayer
  graphene},}\ }\href {\doibase 10.1126/sciadv.adi6063} {\bibfield  {journal}
  {\bibinfo  {journal} {Science Advances}\ }\textbf {\bibinfo {volume} {9}}
  (\bibinfo {year} {2023}),\ 10.1126/sciadv.adi6063}\BibitemShut {NoStop}%
\bibitem [{\citenamefont {Kapfer}\ \emph {et~al.}(2023)\citenamefont {Kapfer},
  \citenamefont {Jessen}, \citenamefont {Eisele}, \citenamefont {Fu},
  \citenamefont {Danielsen}, \citenamefont {Darlington}, \citenamefont {Moore},
  \citenamefont {Finney}, \citenamefont {Marchese}, \citenamefont {Hsieh},
  \citenamefont {Majchrzak}, \citenamefont {Jiang}, \citenamefont {Biswas},
  \citenamefont {Dudin}, \citenamefont {Avila}, \citenamefont {Watanabe},
  \citenamefont {Taniguchi}, \citenamefont {Ulstrup}, \citenamefont {Bøggild},
  \citenamefont {Schuck}, \citenamefont {Basov}, \citenamefont {Hone},\ and\
  \citenamefont {Dean}}]{Kapfer2023}%
  \BibitemOpen
  \bibfield  {author} {\bibinfo {author} {\bibfnamefont {Maëlle}\ \bibnamefont
  {Kapfer}}, \bibinfo {author} {\bibfnamefont {Bjarke~S.}\ \bibnamefont
  {Jessen}}, \bibinfo {author} {\bibfnamefont {Megan~E.}\ \bibnamefont
  {Eisele}}, \bibinfo {author} {\bibfnamefont {Matthew}\ \bibnamefont {Fu}},
  \bibinfo {author} {\bibfnamefont {Dorte~R.}\ \bibnamefont {Danielsen}},
  \bibinfo {author} {\bibfnamefont {Thomas~P.}\ \bibnamefont {Darlington}},
  \bibinfo {author} {\bibfnamefont {Samuel~L.}\ \bibnamefont {Moore}}, \bibinfo
  {author} {\bibfnamefont {Nathan~R.}\ \bibnamefont {Finney}}, \bibinfo
  {author} {\bibfnamefont {Ariane}\ \bibnamefont {Marchese}}, \bibinfo {author}
  {\bibfnamefont {Valerie}\ \bibnamefont {Hsieh}}, \bibinfo {author}
  {\bibfnamefont {Paulina}\ \bibnamefont {Majchrzak}}, \bibinfo {author}
  {\bibfnamefont {Zhihao}\ \bibnamefont {Jiang}}, \bibinfo {author}
  {\bibfnamefont {Deepnarayan}\ \bibnamefont {Biswas}}, \bibinfo {author}
  {\bibfnamefont {Pavel}\ \bibnamefont {Dudin}}, \bibinfo {author}
  {\bibfnamefont {José}\ \bibnamefont {Avila}}, \bibinfo {author}
  {\bibfnamefont {Kenji}\ \bibnamefont {Watanabe}}, \bibinfo {author}
  {\bibfnamefont {Takashi}\ \bibnamefont {Taniguchi}}, \bibinfo {author}
  {\bibfnamefont {Søren}\ \bibnamefont {Ulstrup}}, \bibinfo {author}
  {\bibfnamefont {Peter}\ \bibnamefont {Bøggild}}, \bibinfo {author}
  {\bibfnamefont {P.~J.}\ \bibnamefont {Schuck}}, \bibinfo {author}
  {\bibfnamefont {Dmitri~N.}\ \bibnamefont {Basov}}, \bibinfo {author}
  {\bibfnamefont {James}\ \bibnamefont {Hone}}, \ and\ \bibinfo {author}
  {\bibfnamefont {Cory~R.}\ \bibnamefont {Dean}},\ }\bibfield  {title}
  {\enquote {\bibinfo {title} {Programming twist angle and strain profiles in
  2d materials},}\ }\href {\doibase 10.1126/science.ade9995} {\bibfield
  {journal} {\bibinfo  {journal} {Science}\ }\textbf {\bibinfo {volume}
  {381}},\ \bibinfo {pages} {677–681} (\bibinfo {year} {2023})}\BibitemShut
  {NoStop}%
\bibitem [{\citenamefont {Li}\ \emph {et~al.}(2022)\citenamefont {Li},
  \citenamefont {Xue}, \citenamefont {Fan}, \citenamefont {Gao}, \citenamefont
  {Shi}, \citenamefont {Liu}, \citenamefont {Watanabe}, \citenamefont
  {Tanguchi}, \citenamefont {Zhao}, \citenamefont {Wu}, \citenamefont {Wang},
  \citenamefont {Shi}, \citenamefont {Guo}, \citenamefont {Zhang},
  \citenamefont {Fei},\ and\ \citenamefont {Li}}]{Li2022}%
  \BibitemOpen
  \bibfield  {author} {\bibinfo {author} {\bibfnamefont {Yuhao}\ \bibnamefont
  {Li}}, \bibinfo {author} {\bibfnamefont {Minmin}\ \bibnamefont {Xue}},
  \bibinfo {author} {\bibfnamefont {Hua}\ \bibnamefont {Fan}}, \bibinfo
  {author} {\bibfnamefont {Cun-Fa}\ \bibnamefont {Gao}}, \bibinfo {author}
  {\bibfnamefont {Yan}\ \bibnamefont {Shi}}, \bibinfo {author} {\bibfnamefont
  {Yang}\ \bibnamefont {Liu}}, \bibinfo {author} {\bibfnamefont {Kenji}\
  \bibnamefont {Watanabe}}, \bibinfo {author} {\bibfnamefont {Takashi}\
  \bibnamefont {Tanguchi}}, \bibinfo {author} {\bibfnamefont {Yue}\
  \bibnamefont {Zhao}}, \bibinfo {author} {\bibfnamefont {Fengcheng}\
  \bibnamefont {Wu}}, \bibinfo {author} {\bibfnamefont {Xinran}\ \bibnamefont
  {Wang}}, \bibinfo {author} {\bibfnamefont {Yi}~\bibnamefont {Shi}}, \bibinfo
  {author} {\bibfnamefont {Wanlin}\ \bibnamefont {Guo}}, \bibinfo {author}
  {\bibfnamefont {Zhuhua}\ \bibnamefont {Zhang}}, \bibinfo {author}
  {\bibfnamefont {Zaiyao}\ \bibnamefont {Fei}}, \ and\ \bibinfo {author}
  {\bibfnamefont {Jiangyu}\ \bibnamefont {Li}},\ }\bibfield  {title} {\enquote
  {\bibinfo {title} {Symmetry breaking and anomalous conductivity in a
  double-moiré superlattice},}\ }\href {\doibase 10.1021/acs.nanolett.2c01710}
  {\bibfield  {journal} {\bibinfo  {journal} {Nano Letters}\ }\textbf {\bibinfo
  {volume} {22}},\ \bibinfo {pages} {6215–6222} (\bibinfo {year}
  {2022})}\BibitemShut {NoStop}%
\bibitem [{\citenamefont {Turkel}\ \emph {et~al.}(2022)\citenamefont {Turkel},
  \citenamefont {Swann}, \citenamefont {Zhu}, \citenamefont {Christos},
  \citenamefont {Watanabe}, \citenamefont {Taniguchi}, \citenamefont {Sachdev},
  \citenamefont {Scheurer}, \citenamefont {Kaxiras}, \citenamefont {Dean},\
  and\ \citenamefont {Pasupathy}}]{Turkel2022}%
  \BibitemOpen
  \bibfield  {author} {\bibinfo {author} {\bibfnamefont {Simon}\ \bibnamefont
  {Turkel}}, \bibinfo {author} {\bibfnamefont {Joshua}\ \bibnamefont {Swann}},
  \bibinfo {author} {\bibfnamefont {Ziyan}\ \bibnamefont {Zhu}}, \bibinfo
  {author} {\bibfnamefont {Maine}\ \bibnamefont {Christos}}, \bibinfo {author}
  {\bibfnamefont {K.}~\bibnamefont {Watanabe}}, \bibinfo {author}
  {\bibfnamefont {T.}~\bibnamefont {Taniguchi}}, \bibinfo {author}
  {\bibfnamefont {Subir}\ \bibnamefont {Sachdev}}, \bibinfo {author}
  {\bibfnamefont {Mathias~S.}\ \bibnamefont {Scheurer}}, \bibinfo {author}
  {\bibfnamefont {Efthimios}\ \bibnamefont {Kaxiras}}, \bibinfo {author}
  {\bibfnamefont {Cory~R.}\ \bibnamefont {Dean}}, \ and\ \bibinfo {author}
  {\bibfnamefont {Abhay~N.}\ \bibnamefont {Pasupathy}},\ }\bibfield  {title}
  {\enquote {\bibinfo {title} {Orderly disorder in magic-angle twisted trilayer
  graphene},}\ }\href {\doibase 10.1126/science.abk1895} {\bibfield  {journal}
  {\bibinfo  {journal} {Science}\ }\textbf {\bibinfo {volume} {376}},\ \bibinfo
  {pages} {193–199} (\bibinfo {year} {2022})}\BibitemShut {NoStop}%
\bibitem [{\citenamefont {Li}\ \emph {et~al.}(2024)\citenamefont {Li},
  \citenamefont {Zhang}, \citenamefont {Ha}, \citenamefont {Lin}, \citenamefont
  {Dong}, \citenamefont {Gao}, \citenamefont {Liu}, \citenamefont {Liu},
  \citenamefont {Ryu}, \citenamefont {Kim}, \citenamefont {Jozwiak},
  \citenamefont {Bostwick}, \citenamefont {Watanabe}, \citenamefont
  {Taniguchi}, \citenamefont {Kousa}, \citenamefont {Li}, \citenamefont
  {Rotenberg}, \citenamefont {Khalaf}, \citenamefont {Robinson}, \citenamefont
  {Giustino},\ and\ \citenamefont {Shih}}]{Li2024}%
  \BibitemOpen
  \bibfield  {author} {\bibinfo {author} {\bibfnamefont {Yanxing}\ \bibnamefont
  {Li}}, \bibinfo {author} {\bibfnamefont {Fan}\ \bibnamefont {Zhang}},
  \bibinfo {author} {\bibfnamefont {Viet-Anh}\ \bibnamefont {Ha}}, \bibinfo
  {author} {\bibfnamefont {Yu-Chuan}\ \bibnamefont {Lin}}, \bibinfo {author}
  {\bibfnamefont {Chengye}\ \bibnamefont {Dong}}, \bibinfo {author}
  {\bibfnamefont {Qiang}\ \bibnamefont {Gao}}, \bibinfo {author} {\bibfnamefont
  {Zhida}\ \bibnamefont {Liu}}, \bibinfo {author} {\bibfnamefont {Xiaohui}\
  \bibnamefont {Liu}}, \bibinfo {author} {\bibfnamefont {Sae~Hee}\ \bibnamefont
  {Ryu}}, \bibinfo {author} {\bibfnamefont {Hyunsue}\ \bibnamefont {Kim}},
  \bibinfo {author} {\bibfnamefont {Chris}\ \bibnamefont {Jozwiak}}, \bibinfo
  {author} {\bibfnamefont {Aaron}\ \bibnamefont {Bostwick}}, \bibinfo {author}
  {\bibfnamefont {Kenji}\ \bibnamefont {Watanabe}}, \bibinfo {author}
  {\bibfnamefont {Takashi}\ \bibnamefont {Taniguchi}}, \bibinfo {author}
  {\bibfnamefont {Bishoy}\ \bibnamefont {Kousa}}, \bibinfo {author}
  {\bibfnamefont {Xiaoqin}\ \bibnamefont {Li}}, \bibinfo {author}
  {\bibfnamefont {Eli}\ \bibnamefont {Rotenberg}}, \bibinfo {author}
  {\bibfnamefont {Eslam}\ \bibnamefont {Khalaf}}, \bibinfo {author}
  {\bibfnamefont {Joshua~A.}\ \bibnamefont {Robinson}}, \bibinfo {author}
  {\bibfnamefont {Feliciano}\ \bibnamefont {Giustino}}, \ and\ \bibinfo
  {author} {\bibfnamefont {Chih-Kang}\ \bibnamefont {Shih}},\ }\bibfield
  {title} {\enquote {\bibinfo {title} {Tuning commensurability in twisted van
  der waals bilayers},}\ }\href {\doibase 10.1038/s41586-023-06904-w}
  {\bibfield  {journal} {\bibinfo  {journal} {Nature}\ }\textbf {\bibinfo
  {volume} {625}},\ \bibinfo {pages} {494–499} (\bibinfo {year}
  {2024})}\BibitemShut {NoStop}%
\bibitem [{\citenamefont {Ahn}\ \emph {et~al.}(2018)\citenamefont {Ahn},
  \citenamefont {Moon}, \citenamefont {Kim}, \citenamefont {Kim}, \citenamefont
  {Shin}, \citenamefont {Kim}, \citenamefont {Cha}, \citenamefont {Kahng},
  \citenamefont {Kim}, \citenamefont {Koshino}, \citenamefont {Son},
  \citenamefont {Yang},\ and\ \citenamefont {Ahn}}]{Ahn2018}%
  \BibitemOpen
  \bibfield  {author} {\bibinfo {author} {\bibfnamefont {Sung~Joon}\
  \bibnamefont {Ahn}}, \bibinfo {author} {\bibfnamefont {Pilkyung}\
  \bibnamefont {Moon}}, \bibinfo {author} {\bibfnamefont {Tae-Hoon}\
  \bibnamefont {Kim}}, \bibinfo {author} {\bibfnamefont {Hyun-Woo}\
  \bibnamefont {Kim}}, \bibinfo {author} {\bibfnamefont {Ha-Chul}\ \bibnamefont
  {Shin}}, \bibinfo {author} {\bibfnamefont {Eun~Hye}\ \bibnamefont {Kim}},
  \bibinfo {author} {\bibfnamefont {Hyun~Woo}\ \bibnamefont {Cha}}, \bibinfo
  {author} {\bibfnamefont {Se-Jong}\ \bibnamefont {Kahng}}, \bibinfo {author}
  {\bibfnamefont {Philip}\ \bibnamefont {Kim}}, \bibinfo {author}
  {\bibfnamefont {Mikito}\ \bibnamefont {Koshino}}, \bibinfo {author}
  {\bibfnamefont {Young-Woo}\ \bibnamefont {Son}}, \bibinfo {author}
  {\bibfnamefont {Cheol-Woong}\ \bibnamefont {Yang}}, \ and\ \bibinfo {author}
  {\bibfnamefont {Joung~Real}\ \bibnamefont {Ahn}},\ }\bibfield  {title}
  {\enquote {\bibinfo {title} {Dirac electrons in a dodecagonal graphene
  quasicrystal},}\ }\href {\doibase 10.1126/science.aar8412} {\bibfield
  {journal} {\bibinfo  {journal} {Science}\ }\textbf {\bibinfo {volume}
  {361}},\ \bibinfo {pages} {782–786} (\bibinfo {year} {2018})}\BibitemShut
  {NoStop}%
\bibitem [{\citenamefont {Uri}\ \emph {et~al.}(2023)\citenamefont {Uri},
  \citenamefont {de~la Barrera}, \citenamefont {Randeria}, \citenamefont
  {Rodan-Legrain}, \citenamefont {Devakul}, \citenamefont {Crowley},
  \citenamefont {Paul}, \citenamefont {Watanabe}, \citenamefont {Taniguchi},
  \citenamefont {Lifshitz}, \citenamefont {Fu}, \citenamefont {Ashoori},\ and\
  \citenamefont {Jarillo-Herrero}}]{Uri2023}%
  \BibitemOpen
  \bibfield  {author} {\bibinfo {author} {\bibfnamefont {Aviram}\ \bibnamefont
  {Uri}}, \bibinfo {author} {\bibfnamefont {Sergio~C.}\ \bibnamefont {de~la
  Barrera}}, \bibinfo {author} {\bibfnamefont {Mallika~T.}\ \bibnamefont
  {Randeria}}, \bibinfo {author} {\bibfnamefont {Daniel}\ \bibnamefont
  {Rodan-Legrain}}, \bibinfo {author} {\bibfnamefont {Trithep}\ \bibnamefont
  {Devakul}}, \bibinfo {author} {\bibfnamefont {Philip J.~D.}\ \bibnamefont
  {Crowley}}, \bibinfo {author} {\bibfnamefont {Nisarga}\ \bibnamefont {Paul}},
  \bibinfo {author} {\bibfnamefont {Kenji}\ \bibnamefont {Watanabe}}, \bibinfo
  {author} {\bibfnamefont {Takashi}\ \bibnamefont {Taniguchi}}, \bibinfo
  {author} {\bibfnamefont {Ron}\ \bibnamefont {Lifshitz}}, \bibinfo {author}
  {\bibfnamefont {Liang}\ \bibnamefont {Fu}}, \bibinfo {author} {\bibfnamefont
  {Raymond~C.}\ \bibnamefont {Ashoori}}, \ and\ \bibinfo {author}
  {\bibfnamefont {Pablo}\ \bibnamefont {Jarillo-Herrero}},\ }\bibfield  {title}
  {\enquote {\bibinfo {title} {Superconductivity and strong interactions in a
  tunable moiré quasicrystal},}\ }\href {\doibase 10.1038/s41586-023-06294-z}
  {\bibfield  {journal} {\bibinfo  {journal} {Nature}\ }\textbf {\bibinfo
  {volume} {620}},\ \bibinfo {pages} {762–767} (\bibinfo {year}
  {2023})}\BibitemShut {NoStop}%
\bibitem [{\citenamefont {Su\'arez~Morell}\ \emph {et~al.}(2010)\citenamefont
  {Su\'arez~Morell}, \citenamefont {Correa}, \citenamefont {Vargas},
  \citenamefont {Pacheco},\ and\ \citenamefont
  {Barticevic}}]{PhysRevB.82.121407}%
  \BibitemOpen
  \bibfield  {author} {\bibinfo {author} {\bibfnamefont {E.}~\bibnamefont
  {Su\'arez~Morell}}, \bibinfo {author} {\bibfnamefont {J.~D.}\ \bibnamefont
  {Correa}}, \bibinfo {author} {\bibfnamefont {P.}~\bibnamefont {Vargas}},
  \bibinfo {author} {\bibfnamefont {M.}~\bibnamefont {Pacheco}}, \ and\
  \bibinfo {author} {\bibfnamefont {Z.}~\bibnamefont {Barticevic}},\ }\bibfield
   {title} {\enquote {\bibinfo {title} {Flat bands in slightly twisted bilayer
  graphene: Tight-binding calculations},}\ }\href {\doibase
  10.1103/PhysRevB.82.121407} {\bibfield  {journal} {\bibinfo  {journal} {Phys.
  Rev. B}\ }\textbf {\bibinfo {volume} {82}},\ \bibinfo {pages} {121407}
  (\bibinfo {year} {2010})}\BibitemShut {NoStop}%
\bibitem [{\citenamefont {Gonzalez-Arraga}\ \emph {et~al.}(2017)\citenamefont
  {Gonzalez-Arraga}, \citenamefont {Lado}, \citenamefont {Guinea},\ and\
  \citenamefont {San-Jose}}]{PhysRevLett.119.107201}%
  \BibitemOpen
  \bibfield  {author} {\bibinfo {author} {\bibfnamefont {Luis~A.}\ \bibnamefont
  {Gonzalez-Arraga}}, \bibinfo {author} {\bibfnamefont {J.~L.}\ \bibnamefont
  {Lado}}, \bibinfo {author} {\bibfnamefont {Francisco}\ \bibnamefont
  {Guinea}}, \ and\ \bibinfo {author} {\bibfnamefont {Pablo}\ \bibnamefont
  {San-Jose}},\ }\bibfield  {title} {\enquote {\bibinfo {title} {Electrically
  controllable magnetism in twisted bilayer graphene},}\ }\href {\doibase
  10.1103/PhysRevLett.119.107201} {\bibfield  {journal} {\bibinfo  {journal}
  {Phys. Rev. Lett.}\ }\textbf {\bibinfo {volume} {119}},\ \bibinfo {pages}
  {107201} (\bibinfo {year} {2017})}\BibitemShut {NoStop}%
\bibitem [{\citenamefont {Long}\ \emph {et~al.}(2022)\citenamefont {Long},
  \citenamefont {Pantaleón}, \citenamefont {Zhan}, \citenamefont {Guinea},
  \citenamefont {Silva-Guillén},\ and\ \citenamefont {Yuan}}]{Long2022}%
  \BibitemOpen
  \bibfield  {author} {\bibinfo {author} {\bibfnamefont {Min}\ \bibnamefont
  {Long}}, \bibinfo {author} {\bibfnamefont {Pierre~A.}\ \bibnamefont
  {Pantaleón}}, \bibinfo {author} {\bibfnamefont {Zhen}\ \bibnamefont {Zhan}},
  \bibinfo {author} {\bibfnamefont {Francisco}\ \bibnamefont {Guinea}},
  \bibinfo {author} {\bibfnamefont {Jose~Angel}\ \bibnamefont
  {Silva-Guillén}}, \ and\ \bibinfo {author} {\bibfnamefont {Shengjun}\
  \bibnamefont {Yuan}},\ }\bibfield  {title} {\enquote {\bibinfo {title} {An
  atomistic approach for the structural and electronic properties of twisted
  bilayer graphene-boron nitride heterostructures},}\ }\href {\doibase
  10.1038/s41524-022-00763-1} {\bibfield  {journal} {\bibinfo  {journal} {npj
  Computational Materials}\ }\textbf {\bibinfo {volume} {8}} (\bibinfo {year}
  {2022}),\ 10.1038/s41524-022-00763-1}\BibitemShut {NoStop}%
\bibitem [{\citenamefont {Julku}\ \emph {et~al.}(2020)\citenamefont {Julku},
  \citenamefont {Peltonen}, \citenamefont {Liang}, \citenamefont {Heikkil\"a},\
  and\ \citenamefont {T\"orm\"a}}]{PhysRevB.101.060505}%
  \BibitemOpen
  \bibfield  {author} {\bibinfo {author} {\bibfnamefont {A.}~\bibnamefont
  {Julku}}, \bibinfo {author} {\bibfnamefont {T.~J.}\ \bibnamefont {Peltonen}},
  \bibinfo {author} {\bibfnamefont {L.}~\bibnamefont {Liang}}, \bibinfo
  {author} {\bibfnamefont {T.~T.}\ \bibnamefont {Heikkil\"a}}, \ and\ \bibinfo
  {author} {\bibfnamefont {P.}~\bibnamefont {T\"orm\"a}},\ }\bibfield  {title}
  {\enquote {\bibinfo {title} {Superfluid weight and
  berezinskii-kosterlitz-thouless transition temperature of twisted bilayer
  graphene},}\ }\href {\doibase 10.1103/PhysRevB.101.060505} {\bibfield
  {journal} {\bibinfo  {journal} {Phys. Rev. B}\ }\textbf {\bibinfo {volume}
  {101}},\ \bibinfo {pages} {060505} (\bibinfo {year} {2020})}\BibitemShut
  {NoStop}%
\bibitem [{\citenamefont {Sboychakov}\ \emph {et~al.}(2015)\citenamefont
  {Sboychakov}, \citenamefont {Rakhmanov}, \citenamefont {Rozhkov},\ and\
  \citenamefont {Nori}}]{PhysRevB.92.075402}%
  \BibitemOpen
  \bibfield  {author} {\bibinfo {author} {\bibfnamefont {A.~O.}\ \bibnamefont
  {Sboychakov}}, \bibinfo {author} {\bibfnamefont {A.~L.}\ \bibnamefont
  {Rakhmanov}}, \bibinfo {author} {\bibfnamefont {A.~V.}\ \bibnamefont
  {Rozhkov}}, \ and\ \bibinfo {author} {\bibfnamefont {Franco}\ \bibnamefont
  {Nori}},\ }\bibfield  {title} {\enquote {\bibinfo {title} {Electronic
  spectrum of twisted bilayer graphene},}\ }\href {\doibase
  10.1103/PhysRevB.92.075402} {\bibfield  {journal} {\bibinfo  {journal} {Phys.
  Rev. B}\ }\textbf {\bibinfo {volume} {92}},\ \bibinfo {pages} {075402}
  (\bibinfo {year} {2015})}\BibitemShut {NoStop}%
\bibitem [{\citenamefont {Ramires}\ and\ \citenamefont
  {Lado}(2021)}]{PhysRevLett.127.026401}%
  \BibitemOpen
  \bibfield  {author} {\bibinfo {author} {\bibfnamefont {Aline}\ \bibnamefont
  {Ramires}}\ and\ \bibinfo {author} {\bibfnamefont {Jose~L.}\ \bibnamefont
  {Lado}},\ }\bibfield  {title} {\enquote {\bibinfo {title} {Emulating heavy
  fermions in twisted trilayer graphene},}\ }\href {\doibase
  10.1103/PhysRevLett.127.026401} {\bibfield  {journal} {\bibinfo  {journal}
  {Phys. Rev. Lett.}\ }\textbf {\bibinfo {volume} {127}},\ \bibinfo {pages}
  {026401} (\bibinfo {year} {2021})}\BibitemShut {NoStop}%
\bibitem [{\citenamefont {Baldo}\ \emph {et~al.}(2023)\citenamefont {Baldo},
  \citenamefont {L\"othman}, \citenamefont {Holmvall},\ and\ \citenamefont
  {Black-Schaffer}}]{PhysRevB.108.125141}%
  \BibitemOpen
  \bibfield  {author} {\bibinfo {author} {\bibfnamefont {Lucas}\ \bibnamefont
  {Baldo}}, \bibinfo {author} {\bibfnamefont {Tomas}\ \bibnamefont
  {L\"othman}}, \bibinfo {author} {\bibfnamefont {Patric}\ \bibnamefont
  {Holmvall}}, \ and\ \bibinfo {author} {\bibfnamefont {Annica~M.}\
  \bibnamefont {Black-Schaffer}},\ }\bibfield  {title} {\enquote {\bibinfo
  {title} {Defect-induced band restructuring and length scales in twisted
  bilayer graphene},}\ }\href {\doibase 10.1103/PhysRevB.108.125141} {\bibfield
   {journal} {\bibinfo  {journal} {Phys. Rev. B}\ }\textbf {\bibinfo {volume}
  {108}},\ \bibinfo {pages} {125141} (\bibinfo {year} {2023})}\BibitemShut
  {NoStop}%
\bibitem [{\citenamefont {Ramzan}\ \emph {et~al.}(2023)\citenamefont {Ramzan},
  \citenamefont {Goodwin}, \citenamefont {Mostofi}, \citenamefont {Kuc},\ and\
  \citenamefont {Lischner}}]{Ramzan2023}%
  \BibitemOpen
  \bibfield  {author} {\bibinfo {author} {\bibfnamefont {Muhammad~Sufyan}\
  \bibnamefont {Ramzan}}, \bibinfo {author} {\bibfnamefont {Zachary A.~H.}\
  \bibnamefont {Goodwin}}, \bibinfo {author} {\bibfnamefont {Arash~A.}\
  \bibnamefont {Mostofi}}, \bibinfo {author} {\bibfnamefont {Agnieszka}\
  \bibnamefont {Kuc}}, \ and\ \bibinfo {author} {\bibfnamefont {Johannes}\
  \bibnamefont {Lischner}},\ }\bibfield  {title} {\enquote {\bibinfo {title}
  {Effect of coulomb impurities on the electronic structure of magic angle
  twisted bilayer graphene},}\ }\href {\doibase 10.1038/s41699-023-00403-2}
  {\bibfield  {journal} {\bibinfo  {journal} {npj 2D Materials and
  Applications}\ }\textbf {\bibinfo {volume} {7}} (\bibinfo {year} {2023}),\
  10.1038/s41699-023-00403-2}\BibitemShut {NoStop}%
\bibitem [{\citenamefont {Ramires}\ and\ \citenamefont
  {Lado}(2018)}]{PhysRevLett.121.146801}%
  \BibitemOpen
  \bibfield  {author} {\bibinfo {author} {\bibfnamefont {Aline}\ \bibnamefont
  {Ramires}}\ and\ \bibinfo {author} {\bibfnamefont {Jose~L.}\ \bibnamefont
  {Lado}},\ }\bibfield  {title} {\enquote {\bibinfo {title} {Electrically
  tunable gauge fields in tiny-angle twisted bilayer graphene},}\ }\href
  {\doibase 10.1103/PhysRevLett.121.146801} {\bibfield  {journal} {\bibinfo
  {journal} {Phys. Rev. Lett.}\ }\textbf {\bibinfo {volume} {121}},\ \bibinfo
  {pages} {146801} (\bibinfo {year} {2018})}\BibitemShut {NoStop}%
\bibitem [{\citenamefont {Mao}\ \emph {et~al.}(2023)\citenamefont {Mao},
  \citenamefont {Guerci},\ and\ \citenamefont {Mora}}]{PhysRevB.107.125423}%
  \BibitemOpen
  \bibfield  {author} {\bibinfo {author} {\bibfnamefont {Yuncheng}\
  \bibnamefont {Mao}}, \bibinfo {author} {\bibfnamefont {Daniele}\ \bibnamefont
  {Guerci}}, \ and\ \bibinfo {author} {\bibfnamefont {Christophe}\ \bibnamefont
  {Mora}},\ }\bibfield  {title} {\enquote {\bibinfo {title} {Supermoir\'e
  low-energy effective theory of twisted trilayer graphene},}\ }\href {\doibase
  10.1103/PhysRevB.107.125423} {\bibfield  {journal} {\bibinfo  {journal}
  {Phys. Rev. B}\ }\textbf {\bibinfo {volume} {107}},\ \bibinfo {pages}
  {125423} (\bibinfo {year} {2023})}\BibitemShut {NoStop}%
\bibitem [{\citenamefont {Carr}\ \emph {et~al.}(2020)\citenamefont {Carr},
  \citenamefont {Fang},\ and\ \citenamefont {Kaxiras}}]{Carr2020}%
  \BibitemOpen
  \bibfield  {author} {\bibinfo {author} {\bibfnamefont {Stephen}\ \bibnamefont
  {Carr}}, \bibinfo {author} {\bibfnamefont {Shiang}\ \bibnamefont {Fang}}, \
  and\ \bibinfo {author} {\bibfnamefont {Efthimios}\ \bibnamefont {Kaxiras}},\
  }\bibfield  {title} {\enquote {\bibinfo {title} {Electronic-structure methods
  for twisted moiré layers},}\ }\href {\doibase 10.1038/s41578-020-0214-0}
  {\bibfield  {journal} {\bibinfo  {journal} {Nature Reviews Materials}\
  }\textbf {\bibinfo {volume} {5}},\ \bibinfo {pages} {748–763} (\bibinfo
  {year} {2020})}\BibitemShut {NoStop}%
\bibitem [{\citenamefont {Orús}(2014)}]{Ors2014}%
  \BibitemOpen
  \bibfield  {author} {\bibinfo {author} {\bibfnamefont {Román}\ \bibnamefont
  {Orús}},\ }\bibfield  {title} {\enquote {\bibinfo {title} {A practical
  introduction to tensor networks: Matrix product states and projected
  entangled pair states},}\ }\href {\doibase 10.1016/j.aop.2014.06.013}
  {\bibfield  {journal} {\bibinfo  {journal} {Annals of Physics}\ }\textbf
  {\bibinfo {volume} {349}},\ \bibinfo {pages} {117–158} (\bibinfo {year}
  {2014})}\BibitemShut {NoStop}%
\bibitem [{\citenamefont {Carleo}\ and\ \citenamefont
  {Troyer}(2017)}]{Carleo2017}%
  \BibitemOpen
  \bibfield  {author} {\bibinfo {author} {\bibfnamefont {Giuseppe}\
  \bibnamefont {Carleo}}\ and\ \bibinfo {author} {\bibfnamefont {Matthias}\
  \bibnamefont {Troyer}},\ }\bibfield  {title} {\enquote {\bibinfo {title}
  {Solving the quantum many-body problem with artificial neural networks},}\
  }\href {\doibase 10.1126/science.aag2302} {\bibfield  {journal} {\bibinfo
  {journal} {Science}\ }\textbf {\bibinfo {volume} {355}},\ \bibinfo {pages}
  {602–606} (\bibinfo {year} {2017})}\BibitemShut {NoStop}%
\bibitem [{\citenamefont {Cerezo}\ \emph {et~al.}(2021)\citenamefont {Cerezo},
  \citenamefont {Arrasmith}, \citenamefont {Babbush}, \citenamefont {Benjamin},
  \citenamefont {Endo}, \citenamefont {Fujii}, \citenamefont {McClean},
  \citenamefont {Mitarai}, \citenamefont {Yuan}, \citenamefont {Cincio},\ and\
  \citenamefont {Coles}}]{Cerezo2021}%
  \BibitemOpen
  \bibfield  {author} {\bibinfo {author} {\bibfnamefont {M.}~\bibnamefont
  {Cerezo}}, \bibinfo {author} {\bibfnamefont {Andrew}\ \bibnamefont
  {Arrasmith}}, \bibinfo {author} {\bibfnamefont {Ryan}\ \bibnamefont
  {Babbush}}, \bibinfo {author} {\bibfnamefont {Simon~C.}\ \bibnamefont
  {Benjamin}}, \bibinfo {author} {\bibfnamefont {Suguru}\ \bibnamefont {Endo}},
  \bibinfo {author} {\bibfnamefont {Keisuke}\ \bibnamefont {Fujii}}, \bibinfo
  {author} {\bibfnamefont {Jarrod~R.}\ \bibnamefont {McClean}}, \bibinfo
  {author} {\bibfnamefont {Kosuke}\ \bibnamefont {Mitarai}}, \bibinfo {author}
  {\bibfnamefont {Xiao}\ \bibnamefont {Yuan}}, \bibinfo {author} {\bibfnamefont
  {Lukasz}\ \bibnamefont {Cincio}}, \ and\ \bibinfo {author} {\bibfnamefont
  {Patrick~J.}\ \bibnamefont {Coles}},\ }\bibfield  {title} {\enquote {\bibinfo
  {title} {Variational quantum algorithms},}\ }\href {\doibase
  10.1038/s42254-021-00348-9} {\bibfield  {journal} {\bibinfo  {journal}
  {Nature Reviews Physics}\ }\textbf {\bibinfo {volume} {3}},\ \bibinfo {pages}
  {625–644} (\bibinfo {year} {2021})}\BibitemShut {NoStop}%
\bibitem [{\citenamefont {Arovas}\ \emph {et~al.}(2022)\citenamefont {Arovas},
  \citenamefont {Berg}, \citenamefont {Kivelson},\ and\ \citenamefont
  {Raghu}}]{Arovas2022}%
  \BibitemOpen
  \bibfield  {author} {\bibinfo {author} {\bibfnamefont {Daniel~P.}\
  \bibnamefont {Arovas}}, \bibinfo {author} {\bibfnamefont {Erez}\ \bibnamefont
  {Berg}}, \bibinfo {author} {\bibfnamefont {Steven~A.}\ \bibnamefont
  {Kivelson}}, \ and\ \bibinfo {author} {\bibfnamefont {Srinivas}\ \bibnamefont
  {Raghu}},\ }\bibfield  {title} {\enquote {\bibinfo {title} {The hubbard
  model},}\ }\href {\doibase 10.1146/annurev-conmatphys-031620-102024}
  {\bibfield  {journal} {\bibinfo  {journal} {Annual Review of Condensed Matter
  Physics}\ }\textbf {\bibinfo {volume} {13}},\ \bibinfo {pages} {239–274}
  (\bibinfo {year} {2022})}\BibitemShut {NoStop}%
\bibitem [{\citenamefont {Qin}\ \emph {et~al.}(2022)\citenamefont {Qin},
  \citenamefont {Sch\"{a}fer}, \citenamefont {Andergassen}, \citenamefont
  {Corboz},\ and\ \citenamefont {Gull}}]{Qin2022}%
  \BibitemOpen
  \bibfield  {author} {\bibinfo {author} {\bibfnamefont {Mingpu}\ \bibnamefont
  {Qin}}, \bibinfo {author} {\bibfnamefont {Thomas}\ \bibnamefont
  {Sch\"{a}fer}}, \bibinfo {author} {\bibfnamefont {Sabine}\ \bibnamefont
  {Andergassen}}, \bibinfo {author} {\bibfnamefont {Philippe}\ \bibnamefont
  {Corboz}}, \ and\ \bibinfo {author} {\bibfnamefont {Emanuel}\ \bibnamefont
  {Gull}},\ }\bibfield  {title} {\enquote {\bibinfo {title} {The hubbard model:
  A computational perspective},}\ }\href {\doibase
  10.1146/annurev-conmatphys-090921-033948} {\bibfield  {journal} {\bibinfo
  {journal} {Annual Review of Condensed Matter Physics}\ }\textbf {\bibinfo
  {volume} {13}},\ \bibinfo {pages} {275–302} (\bibinfo {year}
  {2022})}\BibitemShut {NoStop}%
\bibitem [{\citenamefont {White}(1992)}]{PhysRevLett.69.2863}%
  \BibitemOpen
  \bibfield  {author} {\bibinfo {author} {\bibfnamefont {Steven~R.}\
  \bibnamefont {White}},\ }\bibfield  {title} {\enquote {\bibinfo {title}
  {Density matrix formulation for quantum renormalization groups},}\ }\href
  {\doibase 10.1103/PhysRevLett.69.2863} {\bibfield  {journal} {\bibinfo
  {journal} {Phys. Rev. Lett.}\ }\textbf {\bibinfo {volume} {69}},\ \bibinfo
  {pages} {2863--2866} (\bibinfo {year} {1992})}\BibitemShut {NoStop}%
\bibitem [{\citenamefont {Orús}(2019)}]{Ors2019}%
  \BibitemOpen
  \bibfield  {author} {\bibinfo {author} {\bibfnamefont {Román}\ \bibnamefont
  {Orús}},\ }\bibfield  {title} {\enquote {\bibinfo {title} {Tensor networks
  for complex quantum systems},}\ }\href {\doibase 10.1038/s42254-019-0086-7}
  {\bibfield  {journal} {\bibinfo  {journal} {Nature Reviews Physics}\ }\textbf
  {\bibinfo {volume} {1}},\ \bibinfo {pages} {538–550} (\bibinfo {year}
  {2019})}\BibitemShut {NoStop}%
\bibitem [{\citenamefont {Schollw\"{o}ck}(2011)}]{Schollwck2011}%
  \BibitemOpen
  \bibfield  {author} {\bibinfo {author} {\bibfnamefont {Ulrich}\ \bibnamefont
  {Schollw\"{o}ck}},\ }\bibfield  {title} {\enquote {\bibinfo {title} {The
  density-matrix renormalization group in the age of matrix product states},}\
  }\href {\doibase 10.1016/j.aop.2010.09.012} {\bibfield  {journal} {\bibinfo
  {journal} {Annals of Physics}\ }\textbf {\bibinfo {volume} {326}},\ \bibinfo
  {pages} {96–192} (\bibinfo {year} {2011})}\BibitemShut {NoStop}%
\bibitem [{\citenamefont {Haegeman}\ \emph {et~al.}(2016)\citenamefont
  {Haegeman}, \citenamefont {Lubich}, \citenamefont {Oseledets}, \citenamefont
  {Vandereycken},\ and\ \citenamefont {Verstraete}}]{PhysRevB.94.165116}%
  \BibitemOpen
  \bibfield  {author} {\bibinfo {author} {\bibfnamefont {Jutho}\ \bibnamefont
  {Haegeman}}, \bibinfo {author} {\bibfnamefont {Christian}\ \bibnamefont
  {Lubich}}, \bibinfo {author} {\bibfnamefont {Ivan}\ \bibnamefont
  {Oseledets}}, \bibinfo {author} {\bibfnamefont {Bart}\ \bibnamefont
  {Vandereycken}}, \ and\ \bibinfo {author} {\bibfnamefont {Frank}\
  \bibnamefont {Verstraete}},\ }\bibfield  {title} {\enquote {\bibinfo {title}
  {Unifying time evolution and optimization with matrix product states},}\
  }\href {\doibase 10.1103/PhysRevB.94.165116} {\bibfield  {journal} {\bibinfo
  {journal} {Phys. Rev. B}\ }\textbf {\bibinfo {volume} {94}},\ \bibinfo
  {pages} {165116} (\bibinfo {year} {2016})}\BibitemShut {NoStop}%
\bibitem [{\citenamefont {Schollw\"ock}(2005)}]{RevModPhys.77.259}%
  \BibitemOpen
  \bibfield  {author} {\bibinfo {author} {\bibfnamefont {U.}~\bibnamefont
  {Schollw\"ock}},\ }\bibfield  {title} {\enquote {\bibinfo {title} {The
  density-matrix renormalization group},}\ }\href {\doibase
  10.1103/RevModPhys.77.259} {\bibfield  {journal} {\bibinfo  {journal} {Rev.
  Mod. Phys.}\ }\textbf {\bibinfo {volume} {77}},\ \bibinfo {pages} {259--315}
  (\bibinfo {year} {2005})}\BibitemShut {NoStop}%
\bibitem [{\citenamefont {Cirac}\ \emph {et~al.}(2021)\citenamefont {Cirac},
  \citenamefont {P\'erez-Garc\'{\i}a}, \citenamefont {Schuch},\ and\
  \citenamefont {Verstraete}}]{RevModPhys.93.045003}%
  \BibitemOpen
  \bibfield  {author} {\bibinfo {author} {\bibfnamefont {J.~Ignacio}\
  \bibnamefont {Cirac}}, \bibinfo {author} {\bibfnamefont {David}\ \bibnamefont
  {P\'erez-Garc\'{\i}a}}, \bibinfo {author} {\bibfnamefont {Norbert}\
  \bibnamefont {Schuch}}, \ and\ \bibinfo {author} {\bibfnamefont {Frank}\
  \bibnamefont {Verstraete}},\ }\bibfield  {title} {\enquote {\bibinfo {title}
  {Matrix product states and projected entangled pair states: Concepts,
  symmetries, theorems},}\ }\href {\doibase 10.1103/RevModPhys.93.045003}
  {\bibfield  {journal} {\bibinfo  {journal} {Rev. Mod. Phys.}\ }\textbf
  {\bibinfo {volume} {93}},\ \bibinfo {pages} {045003} (\bibinfo {year}
  {2021})}\BibitemShut {NoStop}%
\bibitem [{\citenamefont {Fishman}\ \emph {et~al.}(2022)\citenamefont
  {Fishman}, \citenamefont {White},\ and\ \citenamefont
  {Stoudenmire}}]{10.21468/SciPostPhysCodeb.4}%
  \BibitemOpen
  \bibfield  {author} {\bibinfo {author} {\bibfnamefont {Matthew}\ \bibnamefont
  {Fishman}}, \bibinfo {author} {\bibfnamefont {Steven~R.}\ \bibnamefont
  {White}}, \ and\ \bibinfo {author} {\bibfnamefont {E.~Miles}\ \bibnamefont
  {Stoudenmire}},\ }\bibfield  {title} {\enquote {\bibinfo {title} {{The
  ITensor Software Library for Tensor Network Calculations}},}\ }\href
  {\doibase 10.21468/SciPostPhysCodeb.4} {\bibfield  {journal} {\bibinfo
  {journal} {SciPost Phys. Codebases}\ ,\ \bibinfo {pages} {4}} (\bibinfo
  {year} {2022})}\BibitemShut {NoStop}%
\bibitem [{\citenamefont {Zheng}\ \emph {et~al.}(2017)\citenamefont {Zheng},
  \citenamefont {Chung}, \citenamefont {Corboz}, \citenamefont {Ehlers},
  \citenamefont {Qin}, \citenamefont {Noack}, \citenamefont {Shi},
  \citenamefont {White}, \citenamefont {Zhang},\ and\ \citenamefont
  {Chan}}]{Zheng_Science_2017}%
  \BibitemOpen
  \bibfield  {author} {\bibinfo {author} {\bibfnamefont {Bo-Xiao}\ \bibnamefont
  {Zheng}}, \bibinfo {author} {\bibfnamefont {Chia-Min}\ \bibnamefont {Chung}},
  \bibinfo {author} {\bibfnamefont {Philippe}\ \bibnamefont {Corboz}}, \bibinfo
  {author} {\bibfnamefont {Georg}\ \bibnamefont {Ehlers}}, \bibinfo {author}
  {\bibfnamefont {Ming-Pu}\ \bibnamefont {Qin}}, \bibinfo {author}
  {\bibfnamefont {Reinhard~M.}\ \bibnamefont {Noack}}, \bibinfo {author}
  {\bibfnamefont {Hao}\ \bibnamefont {Shi}}, \bibinfo {author} {\bibfnamefont
  {Steven~R.}\ \bibnamefont {White}}, \bibinfo {author} {\bibfnamefont
  {Shiwei}\ \bibnamefont {Zhang}}, \ and\ \bibinfo {author} {\bibfnamefont
  {Garnet Kin-Lic}\ \bibnamefont {Chan}},\ }\bibfield  {title} {\enquote
  {\bibinfo {title} {Stripe order in the underdoped region of the
  two-dimensional hubbard model},}\ }\href {\doibase 10.1126/science.aam7127}
  {\bibfield  {journal} {\bibinfo  {journal} {Science}\ }\textbf {\bibinfo
  {volume} {358}},\ \bibinfo {pages} {1155–1160} (\bibinfo {year}
  {2017})}\BibitemShut {NoStop}%
\bibitem [{\citenamefont {Marten}\ \emph {et~al.}(2023)\citenamefont {Marten},
  \citenamefont {Bollmark}, \citenamefont {Köhler}, \citenamefont {Manmana},\
  and\ \citenamefont {Kantian}}]{10.21468/SciPostPhys.15.6.236}%
  \BibitemOpen
  \bibfield  {author} {\bibinfo {author} {\bibfnamefont {Svenja}\ \bibnamefont
  {Marten}}, \bibinfo {author} {\bibfnamefont {Gunnar}\ \bibnamefont
  {Bollmark}}, \bibinfo {author} {\bibfnamefont {Thomas}\ \bibnamefont
  {Köhler}}, \bibinfo {author} {\bibfnamefont {Salvatore~R.}\ \bibnamefont
  {Manmana}}, \ and\ \bibinfo {author} {\bibfnamefont {Adrian}\ \bibnamefont
  {Kantian}},\ }\bibfield  {title} {\enquote {\bibinfo {title} {{Transient
  superconductivity in three-dimensional Hubbard systems by combining
  matrix-product states and self-consistent mean-field theory}},}\ }\href
  {\doibase 10.21468/SciPostPhys.15.6.236} {\bibfield  {journal} {\bibinfo
  {journal} {SciPost Phys.}\ }\textbf {\bibinfo {volume} {15}},\ \bibinfo
  {pages} {236} (\bibinfo {year} {2023})}\BibitemShut {NoStop}%
\bibitem [{\citenamefont {Bollmark}\ \emph {et~al.}(2023)\citenamefont
  {Bollmark}, \citenamefont {K\"ohler}, \citenamefont {Pizzino}, \citenamefont
  {Yang}, \citenamefont {Hofmann}, \citenamefont {Shi}, \citenamefont {Zhang},
  \citenamefont {Giamarchi},\ and\ \citenamefont
  {Kantian}}]{PhysRevX.13.011039}%
  \BibitemOpen
  \bibfield  {author} {\bibinfo {author} {\bibfnamefont {Gunnar}\ \bibnamefont
  {Bollmark}}, \bibinfo {author} {\bibfnamefont {Thomas}\ \bibnamefont
  {K\"ohler}}, \bibinfo {author} {\bibfnamefont {Lorenzo}\ \bibnamefont
  {Pizzino}}, \bibinfo {author} {\bibfnamefont {Yiqi}\ \bibnamefont {Yang}},
  \bibinfo {author} {\bibfnamefont {Johannes~S.}\ \bibnamefont {Hofmann}},
  \bibinfo {author} {\bibfnamefont {Hao}\ \bibnamefont {Shi}}, \bibinfo
  {author} {\bibfnamefont {Shiwei}\ \bibnamefont {Zhang}}, \bibinfo {author}
  {\bibfnamefont {Thierry}\ \bibnamefont {Giamarchi}}, \ and\ \bibinfo {author}
  {\bibfnamefont {Adrian}\ \bibnamefont {Kantian}},\ }\bibfield  {title}
  {\enquote {\bibinfo {title} {Solving 2d and 3d lattice models of correlated
  fermions---combining matrix product states with mean-field theory},}\ }\href
  {\doibase 10.1103/PhysRevX.13.011039} {\bibfield  {journal} {\bibinfo
  {journal} {Phys. Rev. X}\ }\textbf {\bibinfo {volume} {13}},\ \bibinfo
  {pages} {011039} (\bibinfo {year} {2023})}\BibitemShut {NoStop}%
\bibitem [{\citenamefont {Stoudenmire}\ and\ \citenamefont
  {Schwab}(2016)}]{NIPS2016_5314b967}%
  \BibitemOpen
  \bibfield  {author} {\bibinfo {author} {\bibfnamefont {Edwin}\ \bibnamefont
  {Stoudenmire}}\ and\ \bibinfo {author} {\bibfnamefont {David~J}\ \bibnamefont
  {Schwab}},\ }\bibfield  {title} {\enquote {\bibinfo {title} {Supervised
  learning with tensor networks},}\ }in\ \href
  {https://proceedings.neurips.cc/paper_files/paper/2016/file/5314b9674c86e3f9d1ba25ef9bb32895-Paper.pdf}
  {\emph {\bibinfo {booktitle} {Advances in Neural Information Processing
  Systems}}},\ Vol.~\bibinfo {volume} {29},\ \bibinfo {editor} {edited by\
  \bibinfo {editor} {\bibfnamefont {D.}~\bibnamefont {Lee}}, \bibinfo {editor}
  {\bibfnamefont {M.}~\bibnamefont {Sugiyama}}, \bibinfo {editor}
  {\bibfnamefont {U.}~\bibnamefont {Luxburg}}, \bibinfo {editor} {\bibfnamefont
  {I.}~\bibnamefont {Guyon}}, \ and\ \bibinfo {editor} {\bibfnamefont
  {R.}~\bibnamefont {Garnett}}}\ (\bibinfo  {publisher} {Curran Associates,
  Inc.},\ \bibinfo {year} {2016})\BibitemShut {NoStop}%
\bibitem [{\citenamefont {Stoudenmire}(2018)}]{Stoudenmire2018}%
  \BibitemOpen
  \bibfield  {author} {\bibinfo {author} {\bibfnamefont {E~Miles}\ \bibnamefont
  {Stoudenmire}},\ }\bibfield  {title} {\enquote {\bibinfo {title} {Learning
  relevant features of data with multi-scale tensor networks},}\ }\href
  {\doibase 10.1088/2058-9565/aaba1a} {\bibfield  {journal} {\bibinfo
  {journal} {Quantum Science and Technology}\ }\textbf {\bibinfo {volume}
  {3}},\ \bibinfo {pages} {034003} (\bibinfo {year} {2018})}\BibitemShut
  {NoStop}%
\bibitem [{\citenamefont {Dilip}\ \emph {et~al.}(2022)\citenamefont {Dilip},
  \citenamefont {Liu}, \citenamefont {Smith},\ and\ \citenamefont
  {Pollmann}}]{PhysRevResearch.4.043007}%
  \BibitemOpen
  \bibfield  {author} {\bibinfo {author} {\bibfnamefont {Rohit}\ \bibnamefont
  {Dilip}}, \bibinfo {author} {\bibfnamefont {Yu-Jie}\ \bibnamefont {Liu}},
  \bibinfo {author} {\bibfnamefont {Adam}\ \bibnamefont {Smith}}, \ and\
  \bibinfo {author} {\bibfnamefont {Frank}\ \bibnamefont {Pollmann}},\
  }\bibfield  {title} {\enquote {\bibinfo {title} {Data compression for quantum
  machine learning},}\ }\href {\doibase 10.1103/PhysRevResearch.4.043007}
  {\bibfield  {journal} {\bibinfo  {journal} {Phys. Rev. Res.}\ }\textbf
  {\bibinfo {volume} {4}},\ \bibinfo {pages} {043007} (\bibinfo {year}
  {2022})}\BibitemShut {NoStop}%
\bibitem [{\citenamefont {Han}\ \emph {et~al.}(2018)\citenamefont {Han},
  \citenamefont {Wang}, \citenamefont {Fan}, \citenamefont {Wang},\ and\
  \citenamefont {Zhang}}]{PhysRevX.8.031012}%
  \BibitemOpen
  \bibfield  {author} {\bibinfo {author} {\bibfnamefont {Zhao-Yu}\ \bibnamefont
  {Han}}, \bibinfo {author} {\bibfnamefont {Jun}\ \bibnamefont {Wang}},
  \bibinfo {author} {\bibfnamefont {Heng}\ \bibnamefont {Fan}}, \bibinfo
  {author} {\bibfnamefont {Lei}\ \bibnamefont {Wang}}, \ and\ \bibinfo {author}
  {\bibfnamefont {Pan}\ \bibnamefont {Zhang}},\ }\bibfield  {title} {\enquote
  {\bibinfo {title} {Unsupervised generative modeling using matrix product
  states},}\ }\href {\doibase 10.1103/PhysRevX.8.031012} {\bibfield  {journal}
  {\bibinfo  {journal} {Phys. Rev. X}\ }\textbf {\bibinfo {volume} {8}},\
  \bibinfo {pages} {031012} (\bibinfo {year} {2018})}\BibitemShut {NoStop}%
\bibitem [{\citenamefont {Bradley}\ \emph {et~al.}(2020)\citenamefont
  {Bradley}, \citenamefont {Stoudenmire},\ and\ \citenamefont
  {Terilla}}]{Bradley2020}%
  \BibitemOpen
  \bibfield  {author} {\bibinfo {author} {\bibfnamefont {Tai-Danae}\
  \bibnamefont {Bradley}}, \bibinfo {author} {\bibfnamefont {E~M}\ \bibnamefont
  {Stoudenmire}}, \ and\ \bibinfo {author} {\bibfnamefont {John}\ \bibnamefont
  {Terilla}},\ }\bibfield  {title} {\enquote {\bibinfo {title} {Modeling
  sequences with quantum states: a look under the hood},}\ }\href {\doibase
  10.1088/2632-2153/ab8731} {\bibfield  {journal} {\bibinfo  {journal} {Machine
  Learning: Science and Technology}\ }\textbf {\bibinfo {volume} {1}},\
  \bibinfo {pages} {035008} (\bibinfo {year} {2020})}\BibitemShut {NoStop}%
\bibitem [{\citenamefont {Cheng}\ \emph {et~al.}(2019)\citenamefont {Cheng},
  \citenamefont {Wang}, \citenamefont {Xiang},\ and\ \citenamefont
  {Zhang}}]{PhysRevB.99.155131}%
  \BibitemOpen
  \bibfield  {author} {\bibinfo {author} {\bibfnamefont {Song}\ \bibnamefont
  {Cheng}}, \bibinfo {author} {\bibfnamefont {Lei}\ \bibnamefont {Wang}},
  \bibinfo {author} {\bibfnamefont {Tao}\ \bibnamefont {Xiang}}, \ and\
  \bibinfo {author} {\bibfnamefont {Pan}\ \bibnamefont {Zhang}},\ }\bibfield
  {title} {\enquote {\bibinfo {title} {Tree tensor networks for generative
  modeling},}\ }\href {\doibase 10.1103/PhysRevB.99.155131} {\bibfield
  {journal} {\bibinfo  {journal} {Phys. Rev. B}\ }\textbf {\bibinfo {volume}
  {99}},\ \bibinfo {pages} {155131} (\bibinfo {year} {2019})}\BibitemShut
  {NoStop}%
\bibitem [{\citenamefont {Zhou}\ \emph {et~al.}(2020)\citenamefont {Zhou},
  \citenamefont {Stoudenmire},\ and\ \citenamefont
  {Waintal}}]{PhysRevX.10.041038}%
  \BibitemOpen
  \bibfield  {author} {\bibinfo {author} {\bibfnamefont {Yiqing}\ \bibnamefont
  {Zhou}}, \bibinfo {author} {\bibfnamefont {E.~Miles}\ \bibnamefont
  {Stoudenmire}}, \ and\ \bibinfo {author} {\bibfnamefont {Xavier}\
  \bibnamefont {Waintal}},\ }\bibfield  {title} {\enquote {\bibinfo {title}
  {What limits the simulation of quantum computers?}}\ }\href {\doibase
  10.1103/PhysRevX.10.041038} {\bibfield  {journal} {\bibinfo  {journal} {Phys.
  Rev. X}\ }\textbf {\bibinfo {volume} {10}},\ \bibinfo {pages} {041038}
  (\bibinfo {year} {2020})}\BibitemShut {NoStop}%
\bibitem [{\citenamefont {Tindall}\ \emph {et~al.}(2024)\citenamefont
  {Tindall}, \citenamefont {Fishman}, \citenamefont {Stoudenmire},\ and\
  \citenamefont {Sels}}]{PRXQuantum.5.010308}%
  \BibitemOpen
  \bibfield  {author} {\bibinfo {author} {\bibfnamefont {Joseph}\ \bibnamefont
  {Tindall}}, \bibinfo {author} {\bibfnamefont {Matthew}\ \bibnamefont
  {Fishman}}, \bibinfo {author} {\bibfnamefont {E.~Miles}\ \bibnamefont
  {Stoudenmire}}, \ and\ \bibinfo {author} {\bibfnamefont {Dries}\ \bibnamefont
  {Sels}},\ }\bibfield  {title} {\enquote {\bibinfo {title} {Efficient tensor
  network simulation of ibm's eagle kicked ising experiment},}\ }\href
  {\doibase 10.1103/PRXQuantum.5.010308} {\bibfield  {journal} {\bibinfo
  {journal} {PRX Quantum}\ }\textbf {\bibinfo {volume} {5}},\ \bibinfo {pages}
  {010308} (\bibinfo {year} {2024})}\BibitemShut {NoStop}%
\bibitem [{\citenamefont {{Niedermeier}}\ \emph {et~al.}(2024)\citenamefont
  {{Niedermeier}}, \citenamefont {{Nairn}}, \citenamefont {{Flindt}},\ and\
  \citenamefont {{Lado}}}]{2024arXiv240406048N}%
  \BibitemOpen
  \bibfield  {author} {\bibinfo {author} {\bibfnamefont {Marcel}\ \bibnamefont
  {{Niedermeier}}}, \bibinfo {author} {\bibfnamefont {Marc}\ \bibnamefont
  {{Nairn}}}, \bibinfo {author} {\bibfnamefont {Christian}\ \bibnamefont
  {{Flindt}}}, \ and\ \bibinfo {author} {\bibfnamefont {Jose~L.}\ \bibnamefont
  {{Lado}}},\ }\bibfield  {title} {\enquote {\bibinfo {title} {{Quantum
  computing topological invariants of two-dimensional quantum matter}},}\
  }\href {\doibase 10.48550/arXiv.2404.06048} {\bibfield  {journal} {\bibinfo
  {journal} {arXiv e-prints}\ ,\ \bibinfo {eid} {arXiv:2404.06048}} (\bibinfo
  {year} {2024})},\ \Eprint {http://arxiv.org/abs/2404.06048} {arXiv:2404.06048
  [quant-ph]} \BibitemShut {NoStop}%
\bibitem [{\citenamefont {Pan}\ \emph {et~al.}(2022)\citenamefont {Pan},
  \citenamefont {Chen},\ and\ \citenamefont {Zhang}}]{PhysRevLett.129.090502}%
  \BibitemOpen
  \bibfield  {author} {\bibinfo {author} {\bibfnamefont {Feng}\ \bibnamefont
  {Pan}}, \bibinfo {author} {\bibfnamefont {Keyang}\ \bibnamefont {Chen}}, \
  and\ \bibinfo {author} {\bibfnamefont {Pan}\ \bibnamefont {Zhang}},\
  }\bibfield  {title} {\enquote {\bibinfo {title} {Solving the sampling problem
  of the sycamore quantum circuits},}\ }\href {\doibase
  10.1103/PhysRevLett.129.090502} {\bibfield  {journal} {\bibinfo  {journal}
  {Phys. Rev. Lett.}\ }\textbf {\bibinfo {volume} {129}},\ \bibinfo {pages}
  {090502} (\bibinfo {year} {2022})}\BibitemShut {NoStop}%
\bibitem [{\citenamefont {Pan}\ and\ \citenamefont
  {Zhang}(2022)}]{PhysRevLett.128.030501}%
  \BibitemOpen
  \bibfield  {author} {\bibinfo {author} {\bibfnamefont {Feng}\ \bibnamefont
  {Pan}}\ and\ \bibinfo {author} {\bibfnamefont {Pan}\ \bibnamefont {Zhang}},\
  }\bibfield  {title} {\enquote {\bibinfo {title} {Simulation of quantum
  circuits using the big-batch tensor network method},}\ }\href {\doibase
  10.1103/PhysRevLett.128.030501} {\bibfield  {journal} {\bibinfo  {journal}
  {Phys. Rev. Lett.}\ }\textbf {\bibinfo {volume} {128}},\ \bibinfo {pages}
  {030501} (\bibinfo {year} {2022})}\BibitemShut {NoStop}%
\bibitem [{\citenamefont {Niedermeier}\ \emph {et~al.}(2024)\citenamefont
  {Niedermeier}, \citenamefont {Lado},\ and\ \citenamefont
  {Flindt}}]{PhysRevResearch.6.033325}%
  \BibitemOpen
  \bibfield  {author} {\bibinfo {author} {\bibfnamefont {Marcel}\ \bibnamefont
  {Niedermeier}}, \bibinfo {author} {\bibfnamefont {Jose~L.}\ \bibnamefont
  {Lado}}, \ and\ \bibinfo {author} {\bibfnamefont {Christian}\ \bibnamefont
  {Flindt}},\ }\bibfield  {title} {\enquote {\bibinfo {title} {Simulating the
  quantum fourier transform, grover's algorithm, and the quantum counting
  algorithm with limited entanglement using tensor networks},}\ }\href
  {\doibase 10.1103/PhysRevResearch.6.033325} {\bibfield  {journal} {\bibinfo
  {journal} {Phys. Rev. Res.}\ }\textbf {\bibinfo {volume} {6}},\ \bibinfo
  {pages} {033325} (\bibinfo {year} {2024})}\BibitemShut {NoStop}%
\bibitem [{\citenamefont {Zhang}\ \emph {et~al.}(2023)\citenamefont {Zhang},
  \citenamefont {Allcock}, \citenamefont {Wan}, \citenamefont {Liu},
  \citenamefont {Sun}, \citenamefont {Yu}, \citenamefont {Yang}, \citenamefont
  {Qiu}, \citenamefont {Ye}, \citenamefont {Chen}, \citenamefont {Lee},
  \citenamefont {Zheng}, \citenamefont {Jian}, \citenamefont {Yao},
  \citenamefont {Hsieh},\ and\ \citenamefont {Zhang}}]{Zhang2023}%
  \BibitemOpen
  \bibfield  {author} {\bibinfo {author} {\bibfnamefont {Shi-Xin}\ \bibnamefont
  {Zhang}}, \bibinfo {author} {\bibfnamefont {Jonathan}\ \bibnamefont
  {Allcock}}, \bibinfo {author} {\bibfnamefont {Zhou-Quan}\ \bibnamefont
  {Wan}}, \bibinfo {author} {\bibfnamefont {Shuo}\ \bibnamefont {Liu}},
  \bibinfo {author} {\bibfnamefont {Jiace}\ \bibnamefont {Sun}}, \bibinfo
  {author} {\bibfnamefont {Hao}\ \bibnamefont {Yu}}, \bibinfo {author}
  {\bibfnamefont {Xing-Han}\ \bibnamefont {Yang}}, \bibinfo {author}
  {\bibfnamefont {Jiezhong}\ \bibnamefont {Qiu}}, \bibinfo {author}
  {\bibfnamefont {Zhaofeng}\ \bibnamefont {Ye}}, \bibinfo {author}
  {\bibfnamefont {Yu-Qin}\ \bibnamefont {Chen}}, \bibinfo {author}
  {\bibfnamefont {Chee-Kong}\ \bibnamefont {Lee}}, \bibinfo {author}
  {\bibfnamefont {Yi-Cong}\ \bibnamefont {Zheng}}, \bibinfo {author}
  {\bibfnamefont {Shao-Kai}\ \bibnamefont {Jian}}, \bibinfo {author}
  {\bibfnamefont {Hong}\ \bibnamefont {Yao}}, \bibinfo {author} {\bibfnamefont
  {Chang-Yu}\ \bibnamefont {Hsieh}}, \ and\ \bibinfo {author} {\bibfnamefont
  {Shengyu}\ \bibnamefont {Zhang}},\ }\bibfield  {title} {\enquote {\bibinfo
  {title} {Tensorcircuit: a quantum software framework for the nisq era},}\
  }\href {\doibase 10.22331/q-2023-02-02-912} {\bibfield  {journal} {\bibinfo
  {journal} {Quantum}\ }\textbf {\bibinfo {volume} {7}},\ \bibinfo {pages}
  {912} (\bibinfo {year} {2023})}\BibitemShut {NoStop}%
\bibitem [{\citenamefont {Torlai}\ and\ \citenamefont
  {Fishman}(2020)}]{pastaq}%
  \BibitemOpen
  \bibfield  {author} {\bibinfo {author} {\bibfnamefont {Giacomo}\ \bibnamefont
  {Torlai}}\ and\ \bibinfo {author} {\bibfnamefont {Matthew}\ \bibnamefont
  {Fishman}},\ }\href {https://github.com/GTorlai/PastaQ.jl/} {\enquote
  {\bibinfo {title} {\mbox{PastaQ}: A package for simulation, tomography and
  analysis of quantum computers},}\ } (\bibinfo {year} {2020})\BibitemShut
  {NoStop}%
\bibitem [{\citenamefont {Ritter}\ \emph {et~al.}(2024)\citenamefont {Ritter},
  \citenamefont {N\'u\~nez Fern\'andez}, \citenamefont {Wallerberger},
  \citenamefont {von Delft}, \citenamefont {Shinaoka},\ and\ \citenamefont
  {Waintal}}]{PhysRevLett.132.056501}%
  \BibitemOpen
  \bibfield  {author} {\bibinfo {author} {\bibfnamefont {Marc~K.}\ \bibnamefont
  {Ritter}}, \bibinfo {author} {\bibfnamefont {Yuriel}\ \bibnamefont {N\'u\~nez
  Fern\'andez}}, \bibinfo {author} {\bibfnamefont {Markus}\ \bibnamefont
  {Wallerberger}}, \bibinfo {author} {\bibfnamefont {Jan}\ \bibnamefont {von
  Delft}}, \bibinfo {author} {\bibfnamefont {Hiroshi}\ \bibnamefont
  {Shinaoka}}, \ and\ \bibinfo {author} {\bibfnamefont {Xavier}\ \bibnamefont
  {Waintal}},\ }\bibfield  {title} {\enquote {\bibinfo {title} {Quantics tensor
  cross interpolation for high-resolution parsimonious representations of
  multivariate functions},}\ }\href {\doibase 10.1103/PhysRevLett.132.056501}
  {\bibfield  {journal} {\bibinfo  {journal} {Phys. Rev. Lett.}\ }\textbf
  {\bibinfo {volume} {132}},\ \bibinfo {pages} {056501} (\bibinfo {year}
  {2024})}\BibitemShut {NoStop}%
\bibitem [{\citenamefont {Shinaoka}\ \emph {et~al.}(2023)\citenamefont
  {Shinaoka}, \citenamefont {Wallerberger}, \citenamefont {Murakami},
  \citenamefont {Nogaki}, \citenamefont {Sakurai}, \citenamefont {Werner},\
  and\ \citenamefont {Kauch}}]{PhysRevX.13.021015}%
  \BibitemOpen
  \bibfield  {author} {\bibinfo {author} {\bibfnamefont {Hiroshi}\ \bibnamefont
  {Shinaoka}}, \bibinfo {author} {\bibfnamefont {Markus}\ \bibnamefont
  {Wallerberger}}, \bibinfo {author} {\bibfnamefont {Yuta}\ \bibnamefont
  {Murakami}}, \bibinfo {author} {\bibfnamefont {Kosuke}\ \bibnamefont
  {Nogaki}}, \bibinfo {author} {\bibfnamefont {Rihito}\ \bibnamefont
  {Sakurai}}, \bibinfo {author} {\bibfnamefont {Philipp}\ \bibnamefont
  {Werner}}, \ and\ \bibinfo {author} {\bibfnamefont {Anna}\ \bibnamefont
  {Kauch}},\ }\bibfield  {title} {\enquote {\bibinfo {title} {Multiscale
  space-time ansatz for correlation functions of quantum systems based on
  quantics tensor trains},}\ }\href {\doibase 10.1103/PhysRevX.13.021015}
  {\bibfield  {journal} {\bibinfo  {journal} {Phys. Rev. X}\ }\textbf {\bibinfo
  {volume} {13}},\ \bibinfo {pages} {021015} (\bibinfo {year}
  {2023})}\BibitemShut {NoStop}%
\bibitem [{\citenamefont {{Rohshap}}\ \emph {et~al.}(2024)\citenamefont
  {{Rohshap}}, \citenamefont {{Ritter}}, \citenamefont {{Shinaoka}},
  \citenamefont {{von Delft}}, \citenamefont {{Wallerberger}},\ and\
  \citenamefont {{Kauch}}}]{2024arXiv241022975R}%
  \BibitemOpen
  \bibfield  {author} {\bibinfo {author} {\bibfnamefont {Stefan}\ \bibnamefont
  {{Rohshap}}}, \bibinfo {author} {\bibfnamefont {Marc~K.}\ \bibnamefont
  {{Ritter}}}, \bibinfo {author} {\bibfnamefont {Hiroshi}\ \bibnamefont
  {{Shinaoka}}}, \bibinfo {author} {\bibfnamefont {Jan}\ \bibnamefont {{von
  Delft}}}, \bibinfo {author} {\bibfnamefont {Markus}\ \bibnamefont
  {{Wallerberger}}}, \ and\ \bibinfo {author} {\bibfnamefont {Anna}\
  \bibnamefont {{Kauch}}},\ }\bibfield  {title} {\enquote {\bibinfo {title}
  {{Two-particle calculations with quantics tensor trains -- solving the
  parquet equations}},}\ }\href {\doibase 10.48550/arXiv.2410.22975} {\bibfield
   {journal} {\bibinfo  {journal} {arXiv e-prints}\ ,\ \bibinfo {eid}
  {arXiv:2410.22975}} (\bibinfo {year} {2024})},\ \Eprint
  {http://arxiv.org/abs/2410.22975} {arXiv:2410.22975 [cond-mat.str-el]}
  \BibitemShut {NoStop}%
\bibitem [{\citenamefont {{Sakaue}}\ \emph {et~al.}(2024)\citenamefont
  {{Sakaue}}, \citenamefont {{Shinaoka}},\ and\ \citenamefont
  {{Sakurai}}}]{2024arXiv240512730S}%
  \BibitemOpen
  \bibfield  {author} {\bibinfo {author} {\bibfnamefont {Kohtaroh}\
  \bibnamefont {{Sakaue}}}, \bibinfo {author} {\bibfnamefont {Hiroshi}\
  \bibnamefont {{Shinaoka}}}, \ and\ \bibinfo {author} {\bibfnamefont {Rihito}\
  \bibnamefont {{Sakurai}}},\ }\bibfield  {title} {\enquote {\bibinfo {title}
  {{Learning tensor trains from noisy functions with application to quantum
  simulation}},}\ }\href {\doibase 10.48550/arXiv.2405.12730} {\bibfield
  {journal} {\bibinfo  {journal} {arXiv e-prints}\ ,\ \bibinfo {eid}
  {arXiv:2405.12730}} (\bibinfo {year} {2024})},\ \Eprint
  {http://arxiv.org/abs/2405.12730} {arXiv:2405.12730 [quant-ph]} \BibitemShut
  {NoStop}%
\bibitem [{\citenamefont {Wei\ss{}e}\ \emph {et~al.}(2006)\citenamefont
  {Wei\ss{}e}, \citenamefont {Wellein}, \citenamefont {Alvermann},\ and\
  \citenamefont {Fehske}}]{RevModPhys.78.275}%
  \BibitemOpen
  \bibfield  {author} {\bibinfo {author} {\bibfnamefont {Alexander}\
  \bibnamefont {Wei\ss{}e}}, \bibinfo {author} {\bibfnamefont {Gerhard}\
  \bibnamefont {Wellein}}, \bibinfo {author} {\bibfnamefont {Andreas}\
  \bibnamefont {Alvermann}}, \ and\ \bibinfo {author} {\bibfnamefont {Holger}\
  \bibnamefont {Fehske}},\ }\bibfield  {title} {\enquote {\bibinfo {title} {The
  kernel polynomial method},}\ }\href {\doibase 10.1103/RevModPhys.78.275}
  {\bibfield  {journal} {\bibinfo  {journal} {Rev. Mod. Phys.}\ }\textbf
  {\bibinfo {volume} {78}},\ \bibinfo {pages} {275--306} (\bibinfo {year}
  {2006})}\BibitemShut {NoStop}%
\bibitem [{\citenamefont {Nagai}\ \emph {et~al.}(2012)\citenamefont {Nagai},
  \citenamefont {Ota},\ and\ \citenamefont {Machida}}]{Nagai2012}%
  \BibitemOpen
  \bibfield  {author} {\bibinfo {author} {\bibfnamefont {Yuki}\ \bibnamefont
  {Nagai}}, \bibinfo {author} {\bibfnamefont {Yukihiro}\ \bibnamefont {Ota}}, \
  and\ \bibinfo {author} {\bibfnamefont {Masahiko}\ \bibnamefont {Machida}},\
  }\bibfield  {title} {\enquote {\bibinfo {title} {Efficient numerical
  self-consistent mean-field approach for fermionic many-body systems by
  polynomial expansion on spectral density},}\ }\href {\doibase
  10.1143/jpsj.81.024710} {\bibfield  {journal} {\bibinfo  {journal} {Journal
  of the Physical Society of Japan}\ }\textbf {\bibinfo {volume} {81}},\
  \bibinfo {pages} {024710} (\bibinfo {year} {2012})}\BibitemShut {NoStop}%
\bibitem [{\citenamefont {N\'u\~nez Fern\'andez}\ \emph
  {et~al.}(2022)\citenamefont {N\'u\~nez Fern\'andez}, \citenamefont {Jeannin},
  \citenamefont {Dumitrescu}, \citenamefont {Kloss}, \citenamefont {Kaye},
  \citenamefont {Parcollet},\ and\ \citenamefont
  {Waintal}}]{PhysRevX.12.041018}%
  \BibitemOpen
  \bibfield  {author} {\bibinfo {author} {\bibfnamefont {Yuriel}\ \bibnamefont
  {N\'u\~nez Fern\'andez}}, \bibinfo {author} {\bibfnamefont {Matthieu}\
  \bibnamefont {Jeannin}}, \bibinfo {author} {\bibfnamefont {Philipp~T.}\
  \bibnamefont {Dumitrescu}}, \bibinfo {author} {\bibfnamefont {Thomas}\
  \bibnamefont {Kloss}}, \bibinfo {author} {\bibfnamefont {Jason}\ \bibnamefont
  {Kaye}}, \bibinfo {author} {\bibfnamefont {Olivier}\ \bibnamefont
  {Parcollet}}, \ and\ \bibinfo {author} {\bibfnamefont {Xavier}\ \bibnamefont
  {Waintal}},\ }\bibfield  {title} {\enquote {\bibinfo {title} {Learning
  feynman diagrams with tensor trains},}\ }\href {\doibase
  10.1103/PhysRevX.12.041018} {\bibfield  {journal} {\bibinfo  {journal} {Phys.
  Rev. X}\ }\textbf {\bibinfo {volume} {12}},\ \bibinfo {pages} {041018}
  (\bibinfo {year} {2022})}\BibitemShut {NoStop}%
\bibitem [{\citenamefont {{N{\'u}{\~n}ez Fern{\'a}ndez}}\ \emph
  {et~al.}(2024)\citenamefont {{N{\'u}{\~n}ez Fern{\'a}ndez}}, \citenamefont
  {{Ritter}}, \citenamefont {{Jeannin}}, \citenamefont {{Li}}, \citenamefont
  {{Kloss}}, \citenamefont {{Louvet}}, \citenamefont {{Terasaki}},
  \citenamefont {{Parcollet}}, \citenamefont {{von Delft}}, \citenamefont
  {{Shinaoka}},\ and\ \citenamefont {{Waintal}}}]{2024arXiv240702454N}%
  \BibitemOpen
  \bibfield  {author} {\bibinfo {author} {\bibfnamefont {Yuriel}\ \bibnamefont
  {{N{\'u}{\~n}ez Fern{\'a}ndez}}}, \bibinfo {author} {\bibfnamefont {Marc~K.}\
  \bibnamefont {{Ritter}}}, \bibinfo {author} {\bibfnamefont {Matthieu}\
  \bibnamefont {{Jeannin}}}, \bibinfo {author} {\bibfnamefont {Jheng-Wei}\
  \bibnamefont {{Li}}}, \bibinfo {author} {\bibfnamefont {Thomas}\ \bibnamefont
  {{Kloss}}}, \bibinfo {author} {\bibfnamefont {Thibaud}\ \bibnamefont
  {{Louvet}}}, \bibinfo {author} {\bibfnamefont {Satoshi}\ \bibnamefont
  {{Terasaki}}}, \bibinfo {author} {\bibfnamefont {Olivier}\ \bibnamefont
  {{Parcollet}}}, \bibinfo {author} {\bibfnamefont {Jan}\ \bibnamefont {{von
  Delft}}}, \bibinfo {author} {\bibfnamefont {Hiroshi}\ \bibnamefont
  {{Shinaoka}}}, \ and\ \bibinfo {author} {\bibfnamefont {Xavier}\ \bibnamefont
  {{Waintal}}},\ }\bibfield  {title} {\enquote {\bibinfo {title} {{Learning
  tensor networks with tensor cross interpolation: new algorithms and
  libraries}},}\ }\href {\doibase 10.48550/arXiv.2407.02454} {\bibfield
  {journal} {\bibinfo  {journal} {arXiv e-prints}\ ,\ \bibinfo {eid}
  {arXiv:2407.02454}} (\bibinfo {year} {2024})},\ \Eprint
  {http://arxiv.org/abs/2407.02454} {arXiv:2407.02454 [physics.comp-ph]}
  \BibitemShut {NoStop}%
\bibitem [{\citenamefont {Erpenbeck}\ \emph {et~al.}(2023)\citenamefont
  {Erpenbeck}, \citenamefont {Lin}, \citenamefont {Blommel}, \citenamefont
  {Zhang}, \citenamefont {Iskakov}, \citenamefont {Bernheimer}, \citenamefont
  {N\'u\~nez Fern\'andez}, \citenamefont {Cohen}, \citenamefont {Parcollet},
  \citenamefont {Waintal},\ and\ \citenamefont {Gull}}]{PhysRevB.107.245135}%
  \BibitemOpen
  \bibfield  {author} {\bibinfo {author} {\bibfnamefont {A.}~\bibnamefont
  {Erpenbeck}}, \bibinfo {author} {\bibfnamefont {W.-T.}\ \bibnamefont {Lin}},
  \bibinfo {author} {\bibfnamefont {T.}~\bibnamefont {Blommel}}, \bibinfo
  {author} {\bibfnamefont {L.}~\bibnamefont {Zhang}}, \bibinfo {author}
  {\bibfnamefont {S.}~\bibnamefont {Iskakov}}, \bibinfo {author} {\bibfnamefont
  {L.}~\bibnamefont {Bernheimer}}, \bibinfo {author} {\bibfnamefont
  {Y.}~\bibnamefont {N\'u\~nez Fern\'andez}}, \bibinfo {author} {\bibfnamefont
  {G.}~\bibnamefont {Cohen}}, \bibinfo {author} {\bibfnamefont
  {O.}~\bibnamefont {Parcollet}}, \bibinfo {author} {\bibfnamefont
  {X.}~\bibnamefont {Waintal}}, \ and\ \bibinfo {author} {\bibfnamefont
  {E.}~\bibnamefont {Gull}},\ }\bibfield  {title} {\enquote {\bibinfo {title}
  {Tensor train continuous time solver for quantum impurity models},}\ }\href
  {\doibase 10.1103/PhysRevB.107.245135} {\bibfield  {journal} {\bibinfo
  {journal} {Phys. Rev. B}\ }\textbf {\bibinfo {volume} {107}},\ \bibinfo
  {pages} {245135} (\bibinfo {year} {2023})}\BibitemShut {NoStop}%
\bibitem [{\citenamefont {Jeannin}\ \emph {et~al.}(2024)\citenamefont
  {Jeannin}, \citenamefont {N\'u\~nez Fern\'andez}, \citenamefont {Kloss},
  \citenamefont {Parcollet},\ and\ \citenamefont
  {Waintal}}]{PhysRevB.110.035124}%
  \BibitemOpen
  \bibfield  {author} {\bibinfo {author} {\bibfnamefont {Matthieu}\
  \bibnamefont {Jeannin}}, \bibinfo {author} {\bibfnamefont {Yuriel}\
  \bibnamefont {N\'u\~nez Fern\'andez}}, \bibinfo {author} {\bibfnamefont
  {Thomas}\ \bibnamefont {Kloss}}, \bibinfo {author} {\bibfnamefont {Olivier}\
  \bibnamefont {Parcollet}}, \ and\ \bibinfo {author} {\bibfnamefont {Xavier}\
  \bibnamefont {Waintal}},\ }\bibfield  {title} {\enquote {\bibinfo {title}
  {Cross-extrapolation reconstruction of low-rank functions and application to
  quantum many-body observables in the strong coupling regime},}\ }\href
  {\doibase 10.1103/PhysRevB.110.035124} {\bibfield  {journal} {\bibinfo
  {journal} {Phys. Rev. B}\ }\textbf {\bibinfo {volume} {110}},\ \bibinfo
  {pages} {035124} (\bibinfo {year} {2024})}\BibitemShut {NoStop}%
\bibitem [{\citenamefont {{Jolly}}\ \emph {et~al.}(2023)\citenamefont
  {{Jolly}}, \citenamefont {{N{\'u}{\~n}ez Fern{\'a}ndez}},\ and\ \citenamefont
  {{Waintal}}}]{2023arXiv230803508J}%
  \BibitemOpen
  \bibfield  {author} {\bibinfo {author} {\bibfnamefont {Nicolas}\ \bibnamefont
  {{Jolly}}}, \bibinfo {author} {\bibfnamefont {Yuriel}\ \bibnamefont
  {{N{\'u}{\~n}ez Fern{\'a}ndez}}}, \ and\ \bibinfo {author} {\bibfnamefont
  {Xavier}\ \bibnamefont {{Waintal}}},\ }\bibfield  {title} {\enquote {\bibinfo
  {title} {{Tensorized orbitals for computational chemistry}},}\ }\href
  {\doibase 10.48550/arXiv.2308.03508} {\bibfield  {journal} {\bibinfo
  {journal} {arXiv e-prints}\ ,\ \bibinfo {eid} {arXiv:2308.03508}} (\bibinfo
  {year} {2023})},\ \Eprint {http://arxiv.org/abs/2308.03508} {arXiv:2308.03508
  [cond-mat.str-el]} \BibitemShut {NoStop}%
\bibitem [{\citenamefont {Murray}\ \emph {et~al.}(2024)\citenamefont {Murray},
  \citenamefont {Shinaoka},\ and\ \citenamefont
  {Werner}}]{PhysRevB.109.165135}%
  \BibitemOpen
  \bibfield  {author} {\bibinfo {author} {\bibfnamefont {Matthias}\
  \bibnamefont {Murray}}, \bibinfo {author} {\bibfnamefont {Hiroshi}\
  \bibnamefont {Shinaoka}}, \ and\ \bibinfo {author} {\bibfnamefont {Philipp}\
  \bibnamefont {Werner}},\ }\bibfield  {title} {\enquote {\bibinfo {title}
  {Nonequilibrium diagrammatic many-body simulations with quantics tensor
  trains},}\ }\href {\doibase 10.1103/PhysRevB.109.165135} {\bibfield
  {journal} {\bibinfo  {journal} {Phys. Rev. B}\ }\textbf {\bibinfo {volume}
  {109}},\ \bibinfo {pages} {165135} (\bibinfo {year} {2024})}\BibitemShut
  {NoStop}%
\bibitem [{\citenamefont {Wang}\ \emph {et~al.}(2019)\citenamefont {Wang},
  \citenamefont {Zheng}, \citenamefont {Chen}, \citenamefont {Huang},
  \citenamefont {Kartashov}, \citenamefont {Torner}, \citenamefont {Konotop},\
  and\ \citenamefont {Ye}}]{Wang2019}%
  \BibitemOpen
  \bibfield  {author} {\bibinfo {author} {\bibfnamefont {Peng}\ \bibnamefont
  {Wang}}, \bibinfo {author} {\bibfnamefont {Yuanlin}\ \bibnamefont {Zheng}},
  \bibinfo {author} {\bibfnamefont {Xianfeng}\ \bibnamefont {Chen}}, \bibinfo
  {author} {\bibfnamefont {Changming}\ \bibnamefont {Huang}}, \bibinfo {author}
  {\bibfnamefont {Yaroslav~V.}\ \bibnamefont {Kartashov}}, \bibinfo {author}
  {\bibfnamefont {Lluis}\ \bibnamefont {Torner}}, \bibinfo {author}
  {\bibfnamefont {Vladimir~V.}\ \bibnamefont {Konotop}}, \ and\ \bibinfo
  {author} {\bibfnamefont {Fangwei}\ \bibnamefont {Ye}},\ }\bibfield  {title}
  {\enquote {\bibinfo {title} {Localization and delocalization of light in
  photonic moiré lattices},}\ }\href {\doibase 10.1038/s41586-019-1851-6}
  {\bibfield  {journal} {\bibinfo  {journal} {Nature}\ }\textbf {\bibinfo
  {volume} {577}},\ \bibinfo {pages} {42–46} (\bibinfo {year}
  {2019})}\BibitemShut {NoStop}%
\bibitem [{\citenamefont {Fu}\ \emph {et~al.}(2020)\citenamefont {Fu},
  \citenamefont {Wang}, \citenamefont {Huang}, \citenamefont {Kartashov},
  \citenamefont {Torner}, \citenamefont {Konotop},\ and\ \citenamefont
  {Ye}}]{Fu2020}%
  \BibitemOpen
  \bibfield  {author} {\bibinfo {author} {\bibfnamefont {Qidong}\ \bibnamefont
  {Fu}}, \bibinfo {author} {\bibfnamefont {Peng}\ \bibnamefont {Wang}},
  \bibinfo {author} {\bibfnamefont {Changming}\ \bibnamefont {Huang}}, \bibinfo
  {author} {\bibfnamefont {Yaroslav~V.}\ \bibnamefont {Kartashov}}, \bibinfo
  {author} {\bibfnamefont {Lluis}\ \bibnamefont {Torner}}, \bibinfo {author}
  {\bibfnamefont {Vladimir~V.}\ \bibnamefont {Konotop}}, \ and\ \bibinfo
  {author} {\bibfnamefont {Fangwei}\ \bibnamefont {Ye}},\ }\bibfield  {title}
  {\enquote {\bibinfo {title} {Optical soliton formation controlled by angle
  twisting in photonic moiré lattices},}\ }\href {\doibase
  10.1038/s41566-020-0679-9} {\bibfield  {journal} {\bibinfo  {journal} {Nature
  Photonics}\ }\textbf {\bibinfo {volume} {14}},\ \bibinfo {pages} {663–668}
  (\bibinfo {year} {2020})}\BibitemShut {NoStop}%
\bibitem [{\citenamefont {{Caha}}\ \emph {et~al.}(2024)\citenamefont {{Caha}},
  \citenamefont {{Boddapatti}}, \citenamefont {{Ahmad}}, \citenamefont
  {{Banobre}}, \citenamefont {{Costa}}, \citenamefont {{Enyashin}},
  \citenamefont {{Li}}, \citenamefont {{Gargiani}}, \citenamefont
  {{Valvidares}}, \citenamefont {{Fernandez-Rossier}},\ and\ \citenamefont
  {{Deepak}}}]{2024arXiv240514967C}%
  \BibitemOpen
  \bibfield  {author} {\bibinfo {author} {\bibfnamefont {Ihsan}\ \bibnamefont
  {{Caha}}}, \bibinfo {author} {\bibfnamefont {Loukya}\ \bibnamefont
  {{Boddapatti}}}, \bibinfo {author} {\bibfnamefont {Aqrab~ul}\ \bibnamefont
  {{Ahmad}}}, \bibinfo {author} {\bibfnamefont {Manuel}\ \bibnamefont
  {{Banobre}}}, \bibinfo {author} {\bibfnamefont {Antonio~T.}\ \bibnamefont
  {{Costa}}}, \bibinfo {author} {\bibfnamefont {Andrey~N.}\ \bibnamefont
  {{Enyashin}}}, \bibinfo {author} {\bibfnamefont {Weibin}\ \bibnamefont
  {{Li}}}, \bibinfo {author} {\bibfnamefont {Pierluigi}\ \bibnamefont
  {{Gargiani}}}, \bibinfo {author} {\bibfnamefont {Manuel}\ \bibnamefont
  {{Valvidares}}}, \bibinfo {author} {\bibfnamefont {Joaquin}\ \bibnamefont
  {{Fernandez-Rossier}}}, \ and\ \bibinfo {author} {\bibfnamefont
  {Francis~Leonard}\ \bibnamefont {{Deepak}}},\ }\bibfield  {title} {\enquote
  {\bibinfo {title} {{Magnetic single wall CrI3 nanotubes encapsulated within
  multiwall Carbon Nanotubes}},}\ }\href {\doibase 10.48550/arXiv.2405.14967}
  {\bibfield  {journal} {\bibinfo  {journal} {arXiv e-prints}\ ,\ \bibinfo
  {eid} {arXiv:2405.14967}} (\bibinfo {year} {2024})},\ \Eprint
  {http://arxiv.org/abs/2405.14967} {arXiv:2405.14967 [cond-mat.mtrl-sci]}
  \BibitemShut {NoStop}%
\bibitem [{pyq()}]{pyqula}%
  \BibitemOpen
  \href@noop {} {\enquote {\bibinfo {title} {pyqula library,
  {\url{https://github.com/joselado/pyqula}}},}\ }\BibitemShut {NoStop}%
\bibitem [{qtc()}]{qtcipy}%
  \BibitemOpen
  \href@noop {} {\enquote {\bibinfo {title} {qtcipy library,
  {\url{https://github.com/joselado/qtcipy}}},}\ }\BibitemShut {NoStop}%
\bibitem [{\citenamefont {Mao}\ \emph {et~al.}(2020)\citenamefont {Mao},
  \citenamefont {Milovanović}, \citenamefont {Andelković}, \citenamefont
  {Lai}, \citenamefont {Cao}, \citenamefont {Watanabe}, \citenamefont
  {Taniguchi}, \citenamefont {Covaci}, \citenamefont {Peeters}, \citenamefont
  {Geim}, \citenamefont {Jiang},\ and\ \citenamefont {Andrei}}]{Mao2020}%
  \BibitemOpen
  \bibfield  {author} {\bibinfo {author} {\bibfnamefont {Jinhai}\ \bibnamefont
  {Mao}}, \bibinfo {author} {\bibfnamefont {Slaviša~P.}\ \bibnamefont
  {Milovanović}}, \bibinfo {author} {\bibfnamefont {Miša}\ \bibnamefont
  {Andelković}}, \bibinfo {author} {\bibfnamefont {Xinyuan}\ \bibnamefont
  {Lai}}, \bibinfo {author} {\bibfnamefont {Yang}\ \bibnamefont {Cao}},
  \bibinfo {author} {\bibfnamefont {Kenji}\ \bibnamefont {Watanabe}}, \bibinfo
  {author} {\bibfnamefont {Takashi}\ \bibnamefont {Taniguchi}}, \bibinfo
  {author} {\bibfnamefont {Lucian}\ \bibnamefont {Covaci}}, \bibinfo {author}
  {\bibfnamefont {Francois~M.}\ \bibnamefont {Peeters}}, \bibinfo {author}
  {\bibfnamefont {Andre~K.}\ \bibnamefont {Geim}}, \bibinfo {author}
  {\bibfnamefont {Yuhang}\ \bibnamefont {Jiang}}, \ and\ \bibinfo {author}
  {\bibfnamefont {Eva~Y.}\ \bibnamefont {Andrei}},\ }\bibfield  {title}
  {\enquote {\bibinfo {title} {Evidence of flat bands and correlated states in
  buckled graphene superlattices},}\ }\href {\doibase
  10.1038/s41586-020-2567-3} {\bibfield  {journal} {\bibinfo  {journal}
  {Nature}\ }\textbf {\bibinfo {volume} {584}},\ \bibinfo {pages} {215–220}
  (\bibinfo {year} {2020})}\BibitemShut {NoStop}%
\bibitem [{\citenamefont {Manesco}\ and\ \citenamefont
  {Lado}(2021)}]{Manesco2021}%
  \BibitemOpen
  \bibfield  {author} {\bibinfo {author} {\bibfnamefont {Antonio L~R}\
  \bibnamefont {Manesco}}\ and\ \bibinfo {author} {\bibfnamefont {Jose~L}\
  \bibnamefont {Lado}},\ }\bibfield  {title} {\enquote {\bibinfo {title}
  {Correlation-induced valley topology in buckled graphene superlattices},}\
  }\href {\doibase 10.1088/2053-1583/ac0b48} {\bibfield  {journal} {\bibinfo
  {journal} {2D Materials}\ }\textbf {\bibinfo {volume} {8}},\ \bibinfo {pages}
  {035057} (\bibinfo {year} {2021})}\BibitemShut {NoStop}%
\bibitem [{\citenamefont {Gao}\ \emph {et~al.}(2023)\citenamefont {Gao},
  \citenamefont {Dong}, \citenamefont {Ledwith}, \citenamefont {Parker},\ and\
  \citenamefont {Khalaf}}]{PhysRevLett.131.096401}%
  \BibitemOpen
  \bibfield  {author} {\bibinfo {author} {\bibfnamefont {Qiang}\ \bibnamefont
  {Gao}}, \bibinfo {author} {\bibfnamefont {Junkai}\ \bibnamefont {Dong}},
  \bibinfo {author} {\bibfnamefont {Patrick}\ \bibnamefont {Ledwith}}, \bibinfo
  {author} {\bibfnamefont {Daniel}\ \bibnamefont {Parker}}, \ and\ \bibinfo
  {author} {\bibfnamefont {Eslam}\ \bibnamefont {Khalaf}},\ }\bibfield  {title}
  {\enquote {\bibinfo {title} {Untwisting moire physics: Almost ideal bands and
  fractional chern insulators in periodically strained monolayer graphene},}\
  }\href {\doibase 10.1103/PhysRevLett.131.096401} {\bibfield  {journal}
  {\bibinfo  {journal} {Phys. Rev. Lett.}\ }\textbf {\bibinfo {volume} {131}},\
  \bibinfo {pages} {096401} (\bibinfo {year} {2023})}\BibitemShut {NoStop}%
\bibitem [{\citenamefont {Wan}\ \emph {et~al.}(2023)\citenamefont {Wan},
  \citenamefont {Sarkar}, \citenamefont {Lin},\ and\ \citenamefont
  {Sun}}]{PhysRevLett.130.216401}%
  \BibitemOpen
  \bibfield  {author} {\bibinfo {author} {\bibfnamefont {Xiaohan}\ \bibnamefont
  {Wan}}, \bibinfo {author} {\bibfnamefont {Siddhartha}\ \bibnamefont
  {Sarkar}}, \bibinfo {author} {\bibfnamefont {Shi-Zeng}\ \bibnamefont {Lin}},
  \ and\ \bibinfo {author} {\bibfnamefont {Kai}\ \bibnamefont {Sun}},\
  }\bibfield  {title} {\enquote {\bibinfo {title} {Topological exact flat bands
  in two-dimensional materials under periodic strain},}\ }\href {\doibase
  10.1103/PhysRevLett.130.216401} {\bibfield  {journal} {\bibinfo  {journal}
  {Phys. Rev. Lett.}\ }\textbf {\bibinfo {volume} {130}},\ \bibinfo {pages}
  {216401} (\bibinfo {year} {2023})}\BibitemShut {NoStop}%
\bibitem [{\citenamefont {Phong}\ and\ \citenamefont
  {Mele}(2022)}]{PhysRevLett.128.176406}%
  \BibitemOpen
  \bibfield  {author} {\bibinfo {author} {\bibfnamefont {Vo~Tien}\ \bibnamefont
  {Phong}}\ and\ \bibinfo {author} {\bibfnamefont {E.~J.}\ \bibnamefont
  {Mele}},\ }\bibfield  {title} {\enquote {\bibinfo {title} {Boundary modes
  from periodic magnetic and pseudomagnetic fields in graphene},}\ }\href
  {\doibase 10.1103/PhysRevLett.128.176406} {\bibfield  {journal} {\bibinfo
  {journal} {Phys. Rev. Lett.}\ }\textbf {\bibinfo {volume} {128}},\ \bibinfo
  {pages} {176406} (\bibinfo {year} {2022})}\BibitemShut {NoStop}%
\bibitem [{\citenamefont {Nakatsuji}\ \emph {et~al.}(2023)\citenamefont
  {Nakatsuji}, \citenamefont {Kawakami},\ and\ \citenamefont
  {Koshino}}]{PhysRevX.13.041007}%
  \BibitemOpen
  \bibfield  {author} {\bibinfo {author} {\bibfnamefont {Naoto}\ \bibnamefont
  {Nakatsuji}}, \bibinfo {author} {\bibfnamefont {Takuto}\ \bibnamefont
  {Kawakami}}, \ and\ \bibinfo {author} {\bibfnamefont {Mikito}\ \bibnamefont
  {Koshino}},\ }\bibfield  {title} {\enquote {\bibinfo {title} {Multiscale
  lattice relaxation in general twisted trilayer graphenes},}\ }\href {\doibase
  10.1103/PhysRevX.13.041007} {\bibfield  {journal} {\bibinfo  {journal} {Phys.
  Rev. X}\ }\textbf {\bibinfo {volume} {13}},\ \bibinfo {pages} {041007}
  (\bibinfo {year} {2023})}\BibitemShut {NoStop}%
\bibitem [{\citenamefont {Craig}\ \emph {et~al.}(2024)\citenamefont {Craig},
  \citenamefont {Van~Winkle}, \citenamefont {Groschner}, \citenamefont {Zhang},
  \citenamefont {Dowlatshahi}, \citenamefont {Zhu}, \citenamefont {Taniguchi},
  \citenamefont {Watanabe}, \citenamefont {Griffin},\ and\ \citenamefont
  {Bediako}}]{Craig2024}%
  \BibitemOpen
  \bibfield  {author} {\bibinfo {author} {\bibfnamefont {Isaac~M.}\
  \bibnamefont {Craig}}, \bibinfo {author} {\bibfnamefont {Madeline}\
  \bibnamefont {Van~Winkle}}, \bibinfo {author} {\bibfnamefont {Catherine}\
  \bibnamefont {Groschner}}, \bibinfo {author} {\bibfnamefont {Kaidi}\
  \bibnamefont {Zhang}}, \bibinfo {author} {\bibfnamefont {Nikita}\
  \bibnamefont {Dowlatshahi}}, \bibinfo {author} {\bibfnamefont {Ziyan}\
  \bibnamefont {Zhu}}, \bibinfo {author} {\bibfnamefont {Takashi}\ \bibnamefont
  {Taniguchi}}, \bibinfo {author} {\bibfnamefont {Kenji}\ \bibnamefont
  {Watanabe}}, \bibinfo {author} {\bibfnamefont {Sinéad~M.}\ \bibnamefont
  {Griffin}}, \ and\ \bibinfo {author} {\bibfnamefont {D.~Kwabena}\
  \bibnamefont {Bediako}},\ }\bibfield  {title} {\enquote {\bibinfo {title}
  {Local atomic stacking and symmetry in twisted graphene trilayers},}\ }\href
  {\doibase 10.1038/s41563-023-01783-y} {\bibfield  {journal} {\bibinfo
  {journal} {Nature Materials}\ }\textbf {\bibinfo {volume} {23}},\ \bibinfo
  {pages} {323–330} (\bibinfo {year} {2024})}\BibitemShut {NoStop}%
\bibitem [{\citenamefont {Engelke}\ \emph {et~al.}(2023)\citenamefont
  {Engelke}, \citenamefont {Yoo}, \citenamefont {Carr}, \citenamefont {Xu},
  \citenamefont {Cazeaux}, \citenamefont {Allen}, \citenamefont {Valdivia},
  \citenamefont {Luskin}, \citenamefont {Kaxiras}, \citenamefont {Kim},
  \citenamefont {Han},\ and\ \citenamefont {Kim}}]{PhysRevB.107.125413}%
  \BibitemOpen
  \bibfield  {author} {\bibinfo {author} {\bibfnamefont {Rebecca}\ \bibnamefont
  {Engelke}}, \bibinfo {author} {\bibfnamefont {Hyobin}\ \bibnamefont {Yoo}},
  \bibinfo {author} {\bibfnamefont {Stephen}\ \bibnamefont {Carr}}, \bibinfo
  {author} {\bibfnamefont {Kevin}\ \bibnamefont {Xu}}, \bibinfo {author}
  {\bibfnamefont {Paul}\ \bibnamefont {Cazeaux}}, \bibinfo {author}
  {\bibfnamefont {Richard}\ \bibnamefont {Allen}}, \bibinfo {author}
  {\bibfnamefont {Andres~Mier}\ \bibnamefont {Valdivia}}, \bibinfo {author}
  {\bibfnamefont {Mitchell}\ \bibnamefont {Luskin}}, \bibinfo {author}
  {\bibfnamefont {Efthimios}\ \bibnamefont {Kaxiras}}, \bibinfo {author}
  {\bibfnamefont {Minhyong}\ \bibnamefont {Kim}}, \bibinfo {author}
  {\bibfnamefont {Jung~Hoon}\ \bibnamefont {Han}}, \ and\ \bibinfo {author}
  {\bibfnamefont {Philip}\ \bibnamefont {Kim}},\ }\bibfield  {title} {\enquote
  {\bibinfo {title} {Topological nature of dislocation networks in
  two-dimensional moir\'e materials},}\ }\href {\doibase
  10.1103/PhysRevB.107.125413} {\bibfield  {journal} {\bibinfo  {journal}
  {Phys. Rev. B}\ }\textbf {\bibinfo {volume} {107}},\ \bibinfo {pages}
  {125413} (\bibinfo {year} {2023})}\BibitemShut {NoStop}%
\bibitem [{\citenamefont {Kreutzer}\ \emph {et~al.}(2015)\citenamefont
  {Kreutzer}, \citenamefont {Pieper}, \citenamefont {Hager}, \citenamefont
  {Wellein}, \citenamefont {Alvermann},\ and\ \citenamefont
  {Fehske}}]{Kreutzer2015}%
  \BibitemOpen
  \bibfield  {author} {\bibinfo {author} {\bibfnamefont {Moritz}\ \bibnamefont
  {Kreutzer}}, \bibinfo {author} {\bibfnamefont {Andreas}\ \bibnamefont
  {Pieper}}, \bibinfo {author} {\bibfnamefont {Georg}\ \bibnamefont {Hager}},
  \bibinfo {author} {\bibfnamefont {Gerhard}\ \bibnamefont {Wellein}}, \bibinfo
  {author} {\bibfnamefont {Andreas}\ \bibnamefont {Alvermann}}, \ and\ \bibinfo
  {author} {\bibfnamefont {Holger}\ \bibnamefont {Fehske}},\ }\bibfield
  {title} {\enquote {\bibinfo {title} {Performance engineering of the kernel
  polynomal method on large-scale cpu-gpu systems},}\ }in\ \href {\doibase
  10.1109/ipdps.2015.76} {\emph {\bibinfo {booktitle} {2015 IEEE International
  Parallel and Distributed Processing Symposium}}},\ Vol.~\bibinfo {volume}
  {78}\ (\bibinfo  {publisher} {IEEE},\ \bibinfo {year} {2015})\ p.\ \bibinfo
  {pages} {417–426}\BibitemShut {NoStop}%
\bibitem [{\citenamefont {Kronik}\ \emph {et~al.}(2006)\citenamefont {Kronik},
  \citenamefont {Makmal}, \citenamefont {Tiago}, \citenamefont {Alemany},
  \citenamefont {Jain}, \citenamefont {Huang}, \citenamefont {Saad},\ and\
  \citenamefont {Chelikowsky}}]{Kronik2006}%
  \BibitemOpen
  \bibfield  {author} {\bibinfo {author} {\bibfnamefont {Leeor}\ \bibnamefont
  {Kronik}}, \bibinfo {author} {\bibfnamefont {Adi}\ \bibnamefont {Makmal}},
  \bibinfo {author} {\bibfnamefont {Murilo~L.}\ \bibnamefont {Tiago}}, \bibinfo
  {author} {\bibfnamefont {M.~M.~G.}\ \bibnamefont {Alemany}}, \bibinfo
  {author} {\bibfnamefont {Manish}\ \bibnamefont {Jain}}, \bibinfo {author}
  {\bibfnamefont {Xiangyang}\ \bibnamefont {Huang}}, \bibinfo {author}
  {\bibfnamefont {Yousef}\ \bibnamefont {Saad}}, \ and\ \bibinfo {author}
  {\bibfnamefont {James~R.}\ \bibnamefont {Chelikowsky}},\ }\bibfield  {title}
  {\enquote {\bibinfo {title} {Parsec – the pseudopotential algorithm for
  real‐space electronic structure calculations: recent advances and novel
  applications to nano‐structures},}\ }\href {\doibase
  10.1002/pssb.200541463} {\bibfield  {journal} {\bibinfo  {journal} {physica
  status solidi (b)}\ }\textbf {\bibinfo {volume} {243}},\ \bibinfo {pages}
  {1063–1079} (\bibinfo {year} {2006})}\BibitemShut {NoStop}%
\bibitem [{\citenamefont {Prentice}\ \emph {et~al.}(2020)\citenamefont
  {Prentice}, \citenamefont {Aarons}, \citenamefont {Womack}, \citenamefont
  {Allen}, \citenamefont {Andrinopoulos}, \citenamefont {Anton}, \citenamefont
  {Bell}, \citenamefont {Bhandari}, \citenamefont {Bramley}, \citenamefont
  {Charlton}, \citenamefont {Clements}, \citenamefont {Cole}, \citenamefont
  {Constantinescu}, \citenamefont {Corsetti}, \citenamefont {Dubois},
  \citenamefont {Duff}, \citenamefont {Escartín}, \citenamefont {Greco},
  \citenamefont {Hill}, \citenamefont {Lee}, \citenamefont {Linscott},
  \citenamefont {O’Regan}, \citenamefont {Phipps}, \citenamefont {Ratcliff},
  \citenamefont {Serrano}, \citenamefont {Tait}, \citenamefont {Teobaldi},
  \citenamefont {Vitale}, \citenamefont {Yeung}, \citenamefont {Zuehlsdorff},
  \citenamefont {Dziedzic}, \citenamefont {Haynes}, \citenamefont {Hine},
  \citenamefont {Mostofi}, \citenamefont {Payne},\ and\ \citenamefont
  {Skylaris}}]{Prentice2020}%
  \BibitemOpen
  \bibfield  {author} {\bibinfo {author} {\bibfnamefont {Joseph C.~A.}\
  \bibnamefont {Prentice}}, \bibinfo {author} {\bibfnamefont {Jolyon}\
  \bibnamefont {Aarons}}, \bibinfo {author} {\bibfnamefont {James~C.}\
  \bibnamefont {Womack}}, \bibinfo {author} {\bibfnamefont {Alice E.~A.}\
  \bibnamefont {Allen}}, \bibinfo {author} {\bibfnamefont {Lampros}\
  \bibnamefont {Andrinopoulos}}, \bibinfo {author} {\bibfnamefont {Lucian}\
  \bibnamefont {Anton}}, \bibinfo {author} {\bibfnamefont {Robert~A.}\
  \bibnamefont {Bell}}, \bibinfo {author} {\bibfnamefont {Arihant}\
  \bibnamefont {Bhandari}}, \bibinfo {author} {\bibfnamefont {Gabriel~A.}\
  \bibnamefont {Bramley}}, \bibinfo {author} {\bibfnamefont {Robert~J.}\
  \bibnamefont {Charlton}}, \bibinfo {author} {\bibfnamefont {Rebecca~J.}\
  \bibnamefont {Clements}}, \bibinfo {author} {\bibfnamefont {Daniel~J.}\
  \bibnamefont {Cole}}, \bibinfo {author} {\bibfnamefont {Gabriel}\
  \bibnamefont {Constantinescu}}, \bibinfo {author} {\bibfnamefont {Fabiano}\
  \bibnamefont {Corsetti}}, \bibinfo {author} {\bibfnamefont {Simon M.-M.}\
  \bibnamefont {Dubois}}, \bibinfo {author} {\bibfnamefont {Kevin K.~B.}\
  \bibnamefont {Duff}}, \bibinfo {author} {\bibfnamefont {Jose~María}\
  \bibnamefont {Escartín}}, \bibinfo {author} {\bibfnamefont {Andrea}\
  \bibnamefont {Greco}}, \bibinfo {author} {\bibfnamefont {Quintin}\
  \bibnamefont {Hill}}, \bibinfo {author} {\bibfnamefont {Louis~P.}\
  \bibnamefont {Lee}}, \bibinfo {author} {\bibfnamefont {Edward}\ \bibnamefont
  {Linscott}}, \bibinfo {author} {\bibfnamefont {David~D.}\ \bibnamefont
  {O’Regan}}, \bibinfo {author} {\bibfnamefont {Maximillian J.~S.}\
  \bibnamefont {Phipps}}, \bibinfo {author} {\bibfnamefont {Laura~E.}\
  \bibnamefont {Ratcliff}}, \bibinfo {author} {\bibfnamefont {Alvaro~Ruiz}\
  \bibnamefont {Serrano}}, \bibinfo {author} {\bibfnamefont {Edward~W.}\
  \bibnamefont {Tait}}, \bibinfo {author} {\bibfnamefont {Gilberto}\
  \bibnamefont {Teobaldi}}, \bibinfo {author} {\bibfnamefont {Valerio}\
  \bibnamefont {Vitale}}, \bibinfo {author} {\bibfnamefont {Nelson}\
  \bibnamefont {Yeung}}, \bibinfo {author} {\bibfnamefont {Tim~J.}\
  \bibnamefont {Zuehlsdorff}}, \bibinfo {author} {\bibfnamefont {Jacek}\
  \bibnamefont {Dziedzic}}, \bibinfo {author} {\bibfnamefont {Peter~D.}\
  \bibnamefont {Haynes}}, \bibinfo {author} {\bibfnamefont {Nicholas D.~M.}\
  \bibnamefont {Hine}}, \bibinfo {author} {\bibfnamefont {Arash~A.}\
  \bibnamefont {Mostofi}}, \bibinfo {author} {\bibfnamefont {Mike~C.}\
  \bibnamefont {Payne}}, \ and\ \bibinfo {author} {\bibfnamefont
  {Chris-Kriton}\ \bibnamefont {Skylaris}},\ }\bibfield  {title} {\enquote
  {\bibinfo {title} {The onetep linear-scaling density functional theory
  program},}\ }\href {\doibase 10.1063/5.0004445} {\bibfield  {journal}
  {\bibinfo  {journal} {The Journal of Chemical Physics}\ }\textbf {\bibinfo
  {volume} {152}} (\bibinfo {year} {2020}),\ 10.1063/5.0004445}\BibitemShut
  {NoStop}%
\bibitem [{\citenamefont {Ivanov}\ \emph {et~al.}(2021)\citenamefont {Ivanov},
  \citenamefont {Levi}, \citenamefont {Jónsson},\ and\ \citenamefont
  {Jónsson}}]{Ivanov2021}%
  \BibitemOpen
  \bibfield  {author} {\bibinfo {author} {\bibfnamefont {Aleksei~V.}\
  \bibnamefont {Ivanov}}, \bibinfo {author} {\bibfnamefont {Gianluca}\
  \bibnamefont {Levi}}, \bibinfo {author} {\bibfnamefont {Elvar~\"{O}.}\
  \bibnamefont {Jónsson}}, \ and\ \bibinfo {author} {\bibfnamefont {Hannes}\
  \bibnamefont {Jónsson}},\ }\bibfield  {title} {\enquote {\bibinfo {title}
  {Method for calculating excited electronic states using density functionals
  and direct orbital optimization with real space grid or plane-wave basis
  set},}\ }\href {\doibase 10.1021/acs.jctc.1c00157} {\bibfield  {journal}
  {\bibinfo  {journal} {Journal of Chemical Theory and Computation}\ }\textbf
  {\bibinfo {volume} {17}},\ \bibinfo {pages} {5034–5049} (\bibinfo {year}
  {2021})}\BibitemShut {NoStop}%
\bibitem [{\citenamefont {Mortensen}\ \emph {et~al.}(2024)\citenamefont
  {Mortensen}, \citenamefont {Larsen}, \citenamefont {Kuisma}, \citenamefont
  {Ivanov}, \citenamefont {Taghizadeh}, \citenamefont {Peterson}, \citenamefont
  {Haldar}, \citenamefont {Dohn}, \citenamefont {Sch\"{a}fer}, \citenamefont
  {Jónsson}, \citenamefont {Hermes}, \citenamefont {Nilsson}, \citenamefont
  {Kastlunger}, \citenamefont {Levi}, \citenamefont {Jónsson}, \citenamefont
  {H\"{a}kkinen}, \citenamefont {Fojt}, \citenamefont {Kangsabanik},
  \citenamefont {Sødequist}, \citenamefont {Lehtom\"{a}ki}, \citenamefont
  {Heske}, \citenamefont {Enkovaara}, \citenamefont {Winther}, \citenamefont
  {Dulak}, \citenamefont {Melander}, \citenamefont {Ovesen}, \citenamefont
  {Louhivuori}, \citenamefont {Walter}, \citenamefont {Gjerding}, \citenamefont
  {Lopez-Acevedo}, \citenamefont {Erhart}, \citenamefont {Warmbier},
  \citenamefont {W\"{u}rdemann}, \citenamefont {Kaappa}, \citenamefont
  {Latini}, \citenamefont {Boland}, \citenamefont {Bligaard}, \citenamefont
  {Skovhus}, \citenamefont {Susi}, \citenamefont {Maxson}, \citenamefont
  {Rossi}, \citenamefont {Chen}, \citenamefont {Schmerwitz}, \citenamefont
  {Schiøtz}, \citenamefont {Olsen}, \citenamefont {Jacobsen},\ and\
  \citenamefont {Thygesen}}]{Mortensen2024}%
  \BibitemOpen
  \bibfield  {author} {\bibinfo {author} {\bibfnamefont {Jens~Jørgen}\
  \bibnamefont {Mortensen}}, \bibinfo {author} {\bibfnamefont {Ask~Hjorth}\
  \bibnamefont {Larsen}}, \bibinfo {author} {\bibfnamefont {Mikael}\
  \bibnamefont {Kuisma}}, \bibinfo {author} {\bibfnamefont {Aleksei~V.}\
  \bibnamefont {Ivanov}}, \bibinfo {author} {\bibfnamefont {Alireza}\
  \bibnamefont {Taghizadeh}}, \bibinfo {author} {\bibfnamefont {Andrew}\
  \bibnamefont {Peterson}}, \bibinfo {author} {\bibfnamefont {Anubhab}\
  \bibnamefont {Haldar}}, \bibinfo {author} {\bibfnamefont {Asmus~Ougaard}\
  \bibnamefont {Dohn}}, \bibinfo {author} {\bibfnamefont {Christian}\
  \bibnamefont {Sch\"{a}fer}}, \bibinfo {author} {\bibfnamefont
  {Elvar~\"{O}rn}\ \bibnamefont {Jónsson}}, \bibinfo {author} {\bibfnamefont
  {Eric~D.}\ \bibnamefont {Hermes}}, \bibinfo {author} {\bibfnamefont
  {Fredrik~Andreas}\ \bibnamefont {Nilsson}}, \bibinfo {author} {\bibfnamefont
  {Georg}\ \bibnamefont {Kastlunger}}, \bibinfo {author} {\bibfnamefont
  {Gianluca}\ \bibnamefont {Levi}}, \bibinfo {author} {\bibfnamefont {Hannes}\
  \bibnamefont {Jónsson}}, \bibinfo {author} {\bibfnamefont {Hannu}\
  \bibnamefont {H\"{a}kkinen}}, \bibinfo {author} {\bibfnamefont {Jakub}\
  \bibnamefont {Fojt}}, \bibinfo {author} {\bibfnamefont {Jiban}\ \bibnamefont
  {Kangsabanik}}, \bibinfo {author} {\bibfnamefont {Joachim}\ \bibnamefont
  {Sødequist}}, \bibinfo {author} {\bibfnamefont {Jouko}\ \bibnamefont
  {Lehtom\"{a}ki}}, \bibinfo {author} {\bibfnamefont {Julian}\ \bibnamefont
  {Heske}}, \bibinfo {author} {\bibfnamefont {Jussi}\ \bibnamefont
  {Enkovaara}}, \bibinfo {author} {\bibfnamefont {Kirsten~Trøstrup}\
  \bibnamefont {Winther}}, \bibinfo {author} {\bibfnamefont {Marcin}\
  \bibnamefont {Dulak}}, \bibinfo {author} {\bibfnamefont {Marko~M.}\
  \bibnamefont {Melander}}, \bibinfo {author} {\bibfnamefont {Martin}\
  \bibnamefont {Ovesen}}, \bibinfo {author} {\bibfnamefont {Martti}\
  \bibnamefont {Louhivuori}}, \bibinfo {author} {\bibfnamefont {Michael}\
  \bibnamefont {Walter}}, \bibinfo {author} {\bibfnamefont {Morten}\
  \bibnamefont {Gjerding}}, \bibinfo {author} {\bibfnamefont {Olga}\
  \bibnamefont {Lopez-Acevedo}}, \bibinfo {author} {\bibfnamefont {Paul}\
  \bibnamefont {Erhart}}, \bibinfo {author} {\bibfnamefont {Robert}\
  \bibnamefont {Warmbier}}, \bibinfo {author} {\bibfnamefont {Rolf}\
  \bibnamefont {W\"{u}rdemann}}, \bibinfo {author} {\bibfnamefont {Sami}\
  \bibnamefont {Kaappa}}, \bibinfo {author} {\bibfnamefont {Simone}\
  \bibnamefont {Latini}}, \bibinfo {author} {\bibfnamefont {Tara~Maria}\
  \bibnamefont {Boland}}, \bibinfo {author} {\bibfnamefont {Thomas}\
  \bibnamefont {Bligaard}}, \bibinfo {author} {\bibfnamefont {Thorbjørn}\
  \bibnamefont {Skovhus}}, \bibinfo {author} {\bibfnamefont {Toma}\
  \bibnamefont {Susi}}, \bibinfo {author} {\bibfnamefont {Tristan}\
  \bibnamefont {Maxson}}, \bibinfo {author} {\bibfnamefont {Tuomas}\
  \bibnamefont {Rossi}}, \bibinfo {author} {\bibfnamefont {Xi}~\bibnamefont
  {Chen}}, \bibinfo {author} {\bibfnamefont {Yorick Leonard~A.}\ \bibnamefont
  {Schmerwitz}}, \bibinfo {author} {\bibfnamefont {Jakob}\ \bibnamefont
  {Schiøtz}}, \bibinfo {author} {\bibfnamefont {Thomas}\ \bibnamefont
  {Olsen}}, \bibinfo {author} {\bibfnamefont {Karsten~Wedel}\ \bibnamefont
  {Jacobsen}}, \ and\ \bibinfo {author} {\bibfnamefont {Kristian~Sommer}\
  \bibnamefont {Thygesen}},\ }\bibfield  {title} {\enquote {\bibinfo {title}
  {Gpaw: An open python package for electronic structure calculations},}\
  }\href {\doibase 10.1063/5.0182685} {\bibfield  {journal} {\bibinfo
  {journal} {The Journal of Chemical Physics}\ }\textbf {\bibinfo {volume}
  {160}} (\bibinfo {year} {2024}),\ 10.1063/5.0182685}\BibitemShut {NoStop}%
\bibitem [{\citenamefont {Soler}\ \emph {et~al.}(2002)\citenamefont {Soler},
  \citenamefont {Artacho}, \citenamefont {Gale}, \citenamefont {García},
  \citenamefont {Junquera}, \citenamefont {Ordejón},\ and\ \citenamefont
  {Sánchez-Portal}}]{Soler2002}%
  \BibitemOpen
  \bibfield  {author} {\bibinfo {author} {\bibfnamefont {José~M}\ \bibnamefont
  {Soler}}, \bibinfo {author} {\bibfnamefont {Emilio}\ \bibnamefont {Artacho}},
  \bibinfo {author} {\bibfnamefont {Julian~D}\ \bibnamefont {Gale}}, \bibinfo
  {author} {\bibfnamefont {Alberto}\ \bibnamefont {García}}, \bibinfo {author}
  {\bibfnamefont {Javier}\ \bibnamefont {Junquera}}, \bibinfo {author}
  {\bibfnamefont {Pablo}\ \bibnamefont {Ordejón}}, \ and\ \bibinfo {author}
  {\bibfnamefont {Daniel}\ \bibnamefont {Sánchez-Portal}},\ }\bibfield
  {title} {\enquote {\bibinfo {title} {The siesta method for ab initio order-n
  materials simulation},}\ }\href {\doibase 10.1088/0953-8984/14/11/302}
  {\bibfield  {journal} {\bibinfo  {journal} {Journal of Physics: Condensed
  Matter}\ }\textbf {\bibinfo {volume} {14}},\ \bibinfo {pages} {2745–2779}
  (\bibinfo {year} {2002})}\BibitemShut {NoStop}%
\bibitem [{\citenamefont {Park}\ \emph {et~al.}(2024)\citenamefont {Park},
  \citenamefont {Son}, \citenamefont {Kim}, \citenamefont {Chang},
  \citenamefont {Zhang}, \citenamefont {Kim}, \citenamefont {Lee},\ and\
  \citenamefont {Park}}]{Park_2024_2D}%
  \BibitemOpen
  \bibfield  {author} {\bibinfo {author} {\bibfnamefont {Giung}\ \bibnamefont
  {Park}}, \bibinfo {author} {\bibfnamefont {Suhan}\ \bibnamefont {Son}},
  \bibinfo {author} {\bibfnamefont {Jongchan}\ \bibnamefont {Kim}}, \bibinfo
  {author} {\bibfnamefont {Yunyeong}\ \bibnamefont {Chang}}, \bibinfo {author}
  {\bibfnamefont {Kaixuan}\ \bibnamefont {Zhang}}, \bibinfo {author}
  {\bibfnamefont {Miyoung}\ \bibnamefont {Kim}}, \bibinfo {author}
  {\bibfnamefont {Jieun}\ \bibnamefont {Lee}}, \ and\ \bibinfo {author}
  {\bibfnamefont {Je-Geun}\ \bibnamefont {Park}},\ }\bibfield  {title}
  {\enquote {\bibinfo {title} {New twisted van der waals fabrication method
  based on strongly adhesive polymer},}\ }\href {\doibase
  10.1088/2053-1583/ad2524} {\bibfield  {journal} {\bibinfo  {journal} {2D
  Materials}\ }\textbf {\bibinfo {volume} {11}},\ \bibinfo {pages} {025021}
  (\bibinfo {year} {2024})}\BibitemShut {NoStop}%
\bibitem [{\citenamefont {Son}\ \emph {et~al.}(2020)\citenamefont {Son},
  \citenamefont {Shin}, \citenamefont {Zhang}, \citenamefont {Shin},
  \citenamefont {Lee}, \citenamefont {Idzuchi}, \citenamefont {Coak},
  \citenamefont {Kim}, \citenamefont {Kim}, \citenamefont {Kim}, \citenamefont
  {Kim}, \citenamefont {Kim}, \citenamefont {Kim},\ and\ \citenamefont
  {Park}}]{Son20202D}%
  \BibitemOpen
  \bibfield  {author} {\bibinfo {author} {\bibfnamefont {Suhan}\ \bibnamefont
  {Son}}, \bibinfo {author} {\bibfnamefont {Young~Jae}\ \bibnamefont {Shin}},
  \bibinfo {author} {\bibfnamefont {Kaixuan}\ \bibnamefont {Zhang}}, \bibinfo
  {author} {\bibfnamefont {Jeacheol}\ \bibnamefont {Shin}}, \bibinfo {author}
  {\bibfnamefont {Sungmin}\ \bibnamefont {Lee}}, \bibinfo {author}
  {\bibfnamefont {Hiroshi}\ \bibnamefont {Idzuchi}}, \bibinfo {author}
  {\bibfnamefont {Matthew~J}\ \bibnamefont {Coak}}, \bibinfo {author}
  {\bibfnamefont {Hwangsun}\ \bibnamefont {Kim}}, \bibinfo {author}
  {\bibfnamefont {Jangwon}\ \bibnamefont {Kim}}, \bibinfo {author}
  {\bibfnamefont {Jae~Hoon}\ \bibnamefont {Kim}}, \bibinfo {author}
  {\bibfnamefont {Miyoung}\ \bibnamefont {Kim}}, \bibinfo {author}
  {\bibfnamefont {Dohun}\ \bibnamefont {Kim}}, \bibinfo {author} {\bibfnamefont
  {Philip}\ \bibnamefont {Kim}}, \ and\ \bibinfo {author} {\bibfnamefont
  {Je-Geun}\ \bibnamefont {Park}},\ }\bibfield  {title} {\enquote {\bibinfo
  {title} {Strongly adhesive dry transfer technique for van der waals
  heterostructure},}\ }\href {\doibase 10.1088/2053-1583/abad0b} {\bibfield
  {journal} {\bibinfo  {journal} {2D Materials}\ }\textbf {\bibinfo {volume}
  {7}},\ \bibinfo {pages} {041005} (\bibinfo {year} {2020})}\BibitemShut
  {NoStop}%
\bibitem [{\citenamefont {Nuckolls}\ \emph {et~al.}(2023)\citenamefont
  {Nuckolls}, \citenamefont {Lee}, \citenamefont {Oh}, \citenamefont {Wong},
  \citenamefont {Soejima}, \citenamefont {Hong}, \citenamefont {Călugăru},
  \citenamefont {Herzog-Arbeitman}, \citenamefont {Bernevig}, \citenamefont
  {Watanabe}, \citenamefont {Taniguchi}, \citenamefont {Regnault},
  \citenamefont {Zaletel},\ and\ \citenamefont {Yazdani}}]{Nuckolls2023}%
  \BibitemOpen
  \bibfield  {author} {\bibinfo {author} {\bibfnamefont {Kevin~P.}\
  \bibnamefont {Nuckolls}}, \bibinfo {author} {\bibfnamefont {Ryan~L.}\
  \bibnamefont {Lee}}, \bibinfo {author} {\bibfnamefont {Myungchul}\
  \bibnamefont {Oh}}, \bibinfo {author} {\bibfnamefont {Dillon}\ \bibnamefont
  {Wong}}, \bibinfo {author} {\bibfnamefont {Tomohiro}\ \bibnamefont
  {Soejima}}, \bibinfo {author} {\bibfnamefont {Jung~Pyo}\ \bibnamefont
  {Hong}}, \bibinfo {author} {\bibfnamefont {Dumitru}\ \bibnamefont
  {Călugăru}}, \bibinfo {author} {\bibfnamefont {Jonah}\ \bibnamefont
  {Herzog-Arbeitman}}, \bibinfo {author} {\bibfnamefont {B.~Andrei}\
  \bibnamefont {Bernevig}}, \bibinfo {author} {\bibfnamefont {Kenji}\
  \bibnamefont {Watanabe}}, \bibinfo {author} {\bibfnamefont {Takashi}\
  \bibnamefont {Taniguchi}}, \bibinfo {author} {\bibfnamefont {Nicolas}\
  \bibnamefont {Regnault}}, \bibinfo {author} {\bibfnamefont {Michael~P.}\
  \bibnamefont {Zaletel}}, \ and\ \bibinfo {author} {\bibfnamefont {Ali}\
  \bibnamefont {Yazdani}},\ }\bibfield  {title} {\enquote {\bibinfo {title}
  {Quantum textures of the many-body wavefunctions in magic-angle graphene},}\
  }\href {\doibase 10.1038/s41586-023-06226-x} {\bibfield  {journal} {\bibinfo
  {journal} {Nature}\ }\textbf {\bibinfo {volume} {620}},\ \bibinfo {pages}
  {525–532} (\bibinfo {year} {2023})}\BibitemShut {NoStop}%
\bibitem [{\citenamefont {Nuckolls}\ and\ \citenamefont
  {Yazdani}(2024)}]{Nuckolls2024}%
  \BibitemOpen
  \bibfield  {author} {\bibinfo {author} {\bibfnamefont {Kevin~P.}\
  \bibnamefont {Nuckolls}}\ and\ \bibinfo {author} {\bibfnamefont {Ali}\
  \bibnamefont {Yazdani}},\ }\bibfield  {title} {\enquote {\bibinfo {title} {A
  microscopic perspective on moiré materials},}\ }\href {\doibase
  10.1038/s41578-024-00682-1} {\bibfield  {journal} {\bibinfo  {journal}
  {Nature Reviews Materials}\ }\textbf {\bibinfo {volume} {9}},\ \bibinfo
  {pages} {460–480} (\bibinfo {year} {2024})}\BibitemShut {NoStop}%
\bibitem [{\citenamefont {Kim}\ \emph {et~al.}(2022)\citenamefont {Kim},
  \citenamefont {Choi}, \citenamefont {Lewandowski}, \citenamefont {Thomson},
  \citenamefont {Zhang}, \citenamefont {Polski}, \citenamefont {Watanabe},
  \citenamefont {Taniguchi}, \citenamefont {Alicea},\ and\ \citenamefont
  {Nadj-Perge}}]{Kim2022}%
  \BibitemOpen
  \bibfield  {author} {\bibinfo {author} {\bibfnamefont {Hyunjin}\ \bibnamefont
  {Kim}}, \bibinfo {author} {\bibfnamefont {Youngjoon}\ \bibnamefont {Choi}},
  \bibinfo {author} {\bibfnamefont {Cyprian}\ \bibnamefont {Lewandowski}},
  \bibinfo {author} {\bibfnamefont {Alex}\ \bibnamefont {Thomson}}, \bibinfo
  {author} {\bibfnamefont {Yiran}\ \bibnamefont {Zhang}}, \bibinfo {author}
  {\bibfnamefont {Robert}\ \bibnamefont {Polski}}, \bibinfo {author}
  {\bibfnamefont {Kenji}\ \bibnamefont {Watanabe}}, \bibinfo {author}
  {\bibfnamefont {Takashi}\ \bibnamefont {Taniguchi}}, \bibinfo {author}
  {\bibfnamefont {Jason}\ \bibnamefont {Alicea}}, \ and\ \bibinfo {author}
  {\bibfnamefont {Stevan}\ \bibnamefont {Nadj-Perge}},\ }\bibfield  {title}
  {\enquote {\bibinfo {title} {Evidence for unconventional superconductivity in
  twisted trilayer graphene},}\ }\href {\doibase 10.1038/s41586-022-04715-z}
  {\bibfield  {journal} {\bibinfo  {journal} {Nature}\ }\textbf {\bibinfo
  {volume} {606}},\ \bibinfo {pages} {494–500} (\bibinfo {year}
  {2022})}\BibitemShut {NoStop}%
\bibitem [{\citenamefont {Frano}\ \emph {et~al.}(2016)\citenamefont {Frano},
  \citenamefont {Blanco-Canosa}, \citenamefont {Schierle}, \citenamefont {Lu},
  \citenamefont {Wu}, \citenamefont {Bluschke}, \citenamefont {Minola},
  \citenamefont {Christiani}, \citenamefont {Habermeier}, \citenamefont
  {Logvenov}, \citenamefont {Wang}, \citenamefont {van Aken}, \citenamefont
  {Benckiser}, \citenamefont {Weschke}, \citenamefont {Le~Tacon},\ and\
  \citenamefont {Keimer}}]{Frano2016}%
  \BibitemOpen
  \bibfield  {author} {\bibinfo {author} {\bibfnamefont {A.}~\bibnamefont
  {Frano}}, \bibinfo {author} {\bibfnamefont {S.}~\bibnamefont
  {Blanco-Canosa}}, \bibinfo {author} {\bibfnamefont {E.}~\bibnamefont
  {Schierle}}, \bibinfo {author} {\bibfnamefont {Y.}~\bibnamefont {Lu}},
  \bibinfo {author} {\bibfnamefont {M.}~\bibnamefont {Wu}}, \bibinfo {author}
  {\bibfnamefont {M.}~\bibnamefont {Bluschke}}, \bibinfo {author}
  {\bibfnamefont {M.}~\bibnamefont {Minola}}, \bibinfo {author} {\bibfnamefont
  {G.}~\bibnamefont {Christiani}}, \bibinfo {author} {\bibfnamefont {H.~U.}\
  \bibnamefont {Habermeier}}, \bibinfo {author} {\bibfnamefont
  {G.}~\bibnamefont {Logvenov}}, \bibinfo {author} {\bibfnamefont
  {Y.}~\bibnamefont {Wang}}, \bibinfo {author} {\bibfnamefont {P.~A.}\
  \bibnamefont {van Aken}}, \bibinfo {author} {\bibfnamefont {E.}~\bibnamefont
  {Benckiser}}, \bibinfo {author} {\bibfnamefont {E.}~\bibnamefont {Weschke}},
  \bibinfo {author} {\bibfnamefont {M.}~\bibnamefont {Le~Tacon}}, \ and\
  \bibinfo {author} {\bibfnamefont {B.}~\bibnamefont {Keimer}},\ }\bibfield
  {title} {\enquote {\bibinfo {title} {Long-range charge-density-wave proximity
  effect at cuprate/manganate interfaces},}\ }\href {\doibase 10.1038/nmat4682}
  {\bibfield  {journal} {\bibinfo  {journal} {Nature Materials}\ }\textbf
  {\bibinfo {volume} {15}},\ \bibinfo {pages} {831–834} (\bibinfo {year}
  {2016})}\BibitemShut {NoStop}%
\bibitem [{\citenamefont {Miao}\ \emph {et~al.}(2019)\citenamefont {Miao},
  \citenamefont {Fumagalli}, \citenamefont {Rossi}, \citenamefont {Lorenzana},
  \citenamefont {Seibold}, \citenamefont {Yakhou-Harris}, \citenamefont
  {Kummer}, \citenamefont {Brookes}, \citenamefont {Gu}, \citenamefont
  {Braicovich}, \citenamefont {Ghiringhelli},\ and\ \citenamefont
  {Dean}}]{PhysRevX.9.031042}%
  \BibitemOpen
  \bibfield  {author} {\bibinfo {author} {\bibfnamefont {H.}~\bibnamefont
  {Miao}}, \bibinfo {author} {\bibfnamefont {R.}~\bibnamefont {Fumagalli}},
  \bibinfo {author} {\bibfnamefont {M.}~\bibnamefont {Rossi}}, \bibinfo
  {author} {\bibfnamefont {J.}~\bibnamefont {Lorenzana}}, \bibinfo {author}
  {\bibfnamefont {G.}~\bibnamefont {Seibold}}, \bibinfo {author} {\bibfnamefont
  {F.}~\bibnamefont {Yakhou-Harris}}, \bibinfo {author} {\bibfnamefont
  {K.}~\bibnamefont {Kummer}}, \bibinfo {author} {\bibfnamefont {N.~B.}\
  \bibnamefont {Brookes}}, \bibinfo {author} {\bibfnamefont {G.~D.}\
  \bibnamefont {Gu}}, \bibinfo {author} {\bibfnamefont {L.}~\bibnamefont
  {Braicovich}}, \bibinfo {author} {\bibfnamefont {G.}~\bibnamefont
  {Ghiringhelli}}, \ and\ \bibinfo {author} {\bibfnamefont {M.~P.~M.}\
  \bibnamefont {Dean}},\ }\bibfield  {title} {\enquote {\bibinfo {title}
  {Formation of incommensurate charge density waves in cuprates},}\ }\href
  {\doibase 10.1103/PhysRevX.9.031042} {\bibfield  {journal} {\bibinfo
  {journal} {Phys. Rev. X}\ }\textbf {\bibinfo {volume} {9}},\ \bibinfo {pages}
  {031042} (\bibinfo {year} {2019})}\BibitemShut {NoStop}%
\bibitem [{\citenamefont {Fleming}\ \emph {et~al.}(1980)\citenamefont
  {Fleming}, \citenamefont {Moncton}, \citenamefont {McWhan},\ and\
  \citenamefont {DiSalvo}}]{PhysRevLett.45.576}%
  \BibitemOpen
  \bibfield  {author} {\bibinfo {author} {\bibfnamefont {R.~M.}\ \bibnamefont
  {Fleming}}, \bibinfo {author} {\bibfnamefont {D.~E.}\ \bibnamefont
  {Moncton}}, \bibinfo {author} {\bibfnamefont {D.~B.}\ \bibnamefont {McWhan}},
  \ and\ \bibinfo {author} {\bibfnamefont {F.~J.}\ \bibnamefont {DiSalvo}},\
  }\bibfield  {title} {\enquote {\bibinfo {title} {Broken hexagonal symmetry in
  the incommensurate charge-density wave structure of
  $2h$-ta${\mathrm{se}}_{2}$},}\ }\href {\doibase 10.1103/PhysRevLett.45.576}
  {\bibfield  {journal} {\bibinfo  {journal} {Phys. Rev. Lett.}\ }\textbf
  {\bibinfo {volume} {45}},\ \bibinfo {pages} {576--579} (\bibinfo {year}
  {1980})}\BibitemShut {NoStop}%
\bibitem [{\citenamefont {Yan}\ \emph {et~al.}(2017)\citenamefont {Yan},
  \citenamefont {Iaia}, \citenamefont {Morosan}, \citenamefont {Fradkin},
  \citenamefont {Abbamonte},\ and\ \citenamefont
  {Madhavan}}]{PhysRevLett.118.106405}%
  \BibitemOpen
  \bibfield  {author} {\bibinfo {author} {\bibfnamefont {Shichao}\ \bibnamefont
  {Yan}}, \bibinfo {author} {\bibfnamefont {Davide}\ \bibnamefont {Iaia}},
  \bibinfo {author} {\bibfnamefont {Emilia}\ \bibnamefont {Morosan}}, \bibinfo
  {author} {\bibfnamefont {Eduardo}\ \bibnamefont {Fradkin}}, \bibinfo {author}
  {\bibfnamefont {Peter}\ \bibnamefont {Abbamonte}}, \ and\ \bibinfo {author}
  {\bibfnamefont {Vidya}\ \bibnamefont {Madhavan}},\ }\bibfield  {title}
  {\enquote {\bibinfo {title} {Influence of domain walls in the incommensurate
  charge density wave state of cu intercalated
  $1t\text{\ensuremath{-}}{\mathrm{tise}}_{2}$},}\ }\href {\doibase
  10.1103/PhysRevLett.118.106405} {\bibfield  {journal} {\bibinfo  {journal}
  {Phys. Rev. Lett.}\ }\textbf {\bibinfo {volume} {118}},\ \bibinfo {pages}
  {106405} (\bibinfo {year} {2017})}\BibitemShut {NoStop}%
\bibitem [{\citenamefont {Fang}\ \emph {et~al.}(2007)\citenamefont {Fang},
  \citenamefont {Ru}, \citenamefont {Fisher},\ and\ \citenamefont
  {Kapitulnik}}]{PhysRevLett.99.046401}%
  \BibitemOpen
  \bibfield  {author} {\bibinfo {author} {\bibfnamefont {A.}~\bibnamefont
  {Fang}}, \bibinfo {author} {\bibfnamefont {N.}~\bibnamefont {Ru}}, \bibinfo
  {author} {\bibfnamefont {I.~R.}\ \bibnamefont {Fisher}}, \ and\ \bibinfo
  {author} {\bibfnamefont {A.}~\bibnamefont {Kapitulnik}},\ }\bibfield  {title}
  {\enquote {\bibinfo {title} {Stm studies of ${\mathrm{tbte}}_{3}$: Evidence
  for a fully incommensurate charge density wave},}\ }\href {\doibase
  10.1103/PhysRevLett.99.046401} {\bibfield  {journal} {\bibinfo  {journal}
  {Phys. Rev. Lett.}\ }\textbf {\bibinfo {volume} {99}},\ \bibinfo {pages}
  {046401} (\bibinfo {year} {2007})}\BibitemShut {NoStop}%
\end{thebibliography}%

\end{document}